\documentclass[12pt,nofootinbib,preprint,superscriptaddress]{revtex4}
\pdfoutput=1

\usepackage{amsmath}
\usepackage{amssymb}
\usepackage{anysize}
\usepackage{bold-extra}
\usepackage{color}
\usepackage{enumerate}
\usepackage{fancyhdr}
\usepackage{graphicx}
\usepackage[linktocpage,colorlinks,urlcolor=blue]{hyperref}
\usepackage[final]{pdfpages}
\usepackage{rotating}
\usepackage{subfig}
\usepackage{textcomp}
\usepackage{url}
\usepackage{verbatim}
\usepackage{wrapfig}
\usepackage{amsfonts}
\usepackage{comment}
\usepackage{epigraph}
\usepackage[utf8]{inputenc}
\usepackage{slashed}
\usepackage{bbm}
\usepackage{cancel}
\usepackage{epsf, array, color}

\def\lsim{\mbox{\raisebox{-.6ex}{~$\stackrel{<}{\sim}$~}}}
\def\gsim{\mbox{\raisebox{-.6ex}{~$\stackrel{>}{\sim}$~}}}

\def\beq{\begin{equation}}
\def\eeq{\end{equation}}

\def\beqn{\begin{eqnarray}}

\def\eeqn{\end{eqnarray}}

\def\lsim{\mbox{\raisebox{-.6ex}{~$\stackrel{<}{\sim}$~}}}
\def\gsim{\mbox{\raisebox{-.6ex}{~$\stackrel{>}{\sim}$~}}}

\newcommand{\mhsm}{m_h}
\newcommand{\hsm}{h}
\newcommand{\tev}{\rm TeV}
\newcommand{\gev}{\rm GeV}
\newcommand{\mt}{m_t}

\newcommand{\mh}{m_h}
\newcommand{\mf}{m_f}

\begin{document}
\title{TASI 2016 Lectures:  Electroweak Symmetry Breaking and Effective Field Theory
}
\author{ S.~Dawson}
\address{  Physics Department,\\
               Brookhaven National Laboratory,\\  Upton, NY 11973}

\begin{abstract} 
I give a pedagogical introduction to the physics of electroweak
symmetry breaking and the uses of effective field theory in the
context of Higgs physics.
Higgs boson production and decay at the LHC and  the consistency  of the Higgs measurements with
triviality arguments, vacuum
stability,  and precision electroweak  measurements are discussed.  
Effective Lagrangian techniques are used to understand  potential  deviations
from the Standard Model (SM) predictions. Finally, I end with a brief discussion of the future of Higgs physics. 
\end{abstract}

\maketitle

\section{Introduction}
The experimental discovery of the Higgs boson \cite{Aad:2012tfa,Chatrchyan:2012xdj} implies that the Weinberg Salam Standard
Model (SM) is a valid low energy theory at the weak scale.
  All current measurements are consistent with this statement
and physics  in the electroweak symmetry breaking (EWSB) sector beyond 
that predicted by the SM is highly constrained by  experimental results, both
at the LHC and from precision electroweak measurements.    These lectures
summarize the underlying theoretical framework of the SM and its experimental predictions
 and discuss 
possible high scale 
extensions of the theory in terms of 
an effective field theory.  

Section \ref{sec:smintro}  contains an introduction to the SM and  Section \ref{sec:theory}
discusses theoretical restrictions on the EWSB sector with an aside on unitarity.  Section \ref{sec:prod}
 presents the basics of Higgs production and decay,
along with a summary of current experimental results.   Pedagogical discussions of  the
gluon fusion production rate at leading order, low energy theorems that can be used
to approximate the gluon fusion rate in Beyond the SM scenarios (BSM) and the determination of the Higgs
width are also found in Section \ref{sec:prod}.   Extensions 
of the SM in terms of an effective field theory are presented in Section \ref{sec:eftch} and Section \ref{sec:conc}
contains some conclusions and a personal view on the future of Higgs physics. 
There are many excellent reviews of Higgs physics and the reader is referred to them for additional details and further 
references\cite{Logan:2014jla,Reina:2012fs,Djouadi:2005gi,Spira:2016ztx,Quigg:2013ufa,Dawson:1998yi,Gunion:1989we}.

\section{Introduction to the Standard Model}
\label{sec:smintro}

\subsection{The Higgs Mechanism}
\label{sec:abh}
We begin by discussing a simplified version of the SM with a $U(1)$ symmetry,
the Abelian Higgs model, that illustrates a basic version of EWSB and the
motivation for the $SU(2)_L\times U(1)_Y$ SM.
   To
   understand  the problem of gauge boson masses, consider 
a $U(1)$ gauge theory with a single gauge field, the photon, $A_\mu$,
\beq
{\cal L}=-{1\over 4} F_{\mu \nu}F^{\mu\nu},
\eeq
where
\beq F_{\mu\nu}=\partial_\nu A_\mu-\partial _\mu A_\nu.
\eeq
Local $U(1)$ gauge invariance requires  that the Lagrangian
be invariant under the transformation: $ A_\mu(x)
\rightarrow A_\mu(x)-
\partial_\mu \eta(x)$ for any $\eta$ and $x$.   If we add a
mass term for the photon to the Lagrangian,
\beq
{\cal L}=-{1\over 4} F_{\mu \nu}F^{\mu\nu}+{1\over 2}m^2 A_\mu A^\mu
,
\eeq
it is easy to see that the mass term violates the local gauge invariance.

The model can be extended  by adding a single complex scalar field
with charge $-e$\footnote{My conventions follow \cite{Quigg:2013ufa} and have $e>0$ and $Q_e=-1$.}
that  couples to the photon,
\beq
{\cal L}=-{1\over 4} F_{\mu\nu} F^{\mu\nu}+\mid D_\mu\phi\mid^2
-V(\phi),
\eeq
where,
\beqn D_\mu & =&\partial_\mu -i e A_\mu \nonumber \\
V(\hsm) &=& \mu^2 \mid \hsm\mid^2+\lambda(\mid \hsm\mid^2)^2 \, . 
\label{eq:smpot}
\eeqn
$V(\phi)$ is the most general renormalizable potential allowed by
the $U(1)$  gauge invariance.

This Lagrangian is invariant under global $U(1)$ rotations, $\phi
\rightarrow e^{i\theta}\phi$ and also  under local gauge transformations:
\beqn
A_\mu(x) &\rightarrow & A_\mu(x)- \partial _\mu \eta(x) \nonumber \\
\phi(x) &\rightarrow & e^{-i e \eta(x)} \phi(x).\\
\eeqn
There are now two possibilities for the theory.\footnote{We
assume $\lambda>0$. If $\lambda<0$, the potential is unbounded
from below and has no state of minimum energy.}  If $\mu^2>0$, the
potential preserves the symmetries
of the Lagrangian and
the state of lowest energy is that with $\phi=0$, the vacuum state.
This theory is quantum electrodynamics
with a massless photon and a charged
scalar field $\phi$ with mass $\mu$.

 In the alternative scenario,  $\mu^2<0$ and the 
potential is, 
\beq V(\phi)=-\mid \mu^2\mid \mid \phi\mid^2 +
\lambda (\mid \phi\mid^2)^2.
\eeq
 In  this case
the minimum energy state is not at $\phi=0$, but rather at
\beq
\langle \phi\rangle=\sqrt{-{\mu^2\over 2 \lambda}}
\equiv {v\over\sqrt{2}}.
\eeq
$\langle \phi\rangle$ is called the vacuum expectation value (VEV)
of $\phi$.
The direction in which the vacuum is chosen is 
arbitrary, but it is conventional to choose it to lie along
the direction of the real part of $\phi$.
The VEV breaks the global $U(1)$ symmetry.  
 
It is convenient to rewrite $\phi$ as
\beq \phi\equiv {1\over \sqrt{2}}e^{i {\chi\over v}} 
\biggl(v+h\biggr),
\label{phidef}
\eeq
where $\chi$ and $h$ are real fields that have no VEVs.
If we substitute Eq.~\ref{phidef}  back
 into the original Lagrangian, 
the interactions  are,
\beqn
{\cal L} & = &
-{1\over 4} F_{\mu\nu}F^{\mu\nu} - e v A_{\mu}\partial^{\mu}\chi
+{e^2 v^2\over 2} A_{\mu} A^\mu \nonumber +{1\over 2} \biggl( \partial_\mu h \partial^\mu h
+ 2 \mu^2 h^2\biggr)
\nonumber \\ &&
+{1\over 2} \partial_\mu\chi\partial^\mu\chi
+(h,~\chi {\rm ~interactions}).
\label{eq:ints}
\eeqn
Eq.~\ref{eq:ints}
 describes a theory with a photon of mass $M_A=ev$, a physical scalar field
$h$ with mass-squared $-2 \mu^2>0$, and a massless scalar field $\chi$.
The mixed $\chi-A$ propagator can be
removed by making a gauge transformation,
\beq
A^\prime_\mu\equiv A_\mu -{1\over e v} \partial_\mu\chi.
\label{eq:gauget}
\eeq
After making the gauge transformation of Eq.~\ref{eq:gauget},
 the $\chi$ field disappears from the theory.
  In the gauge of Eq. \ref{eq:gauget}
the particle content of the theory
 is  apparent: a massive photon and
a scalar field $h$, which we call
a Higgs boson. Clearly, the choice $\mu^2<0$ leads to very different physical consequences from $\mu^2>0$. 

Now consider the gauge dependance of these
results.  The gauge choice above with the transformation 
$A_\mu^\prime=A_\mu -{1\over e v }\partial_\mu\chi$ is called the
unitary gauge.  This gauge has the advantage that the particle
spectrum is obvious and there is no $\chi$ field.
The unitary gauge, however, has the disadvantage that the 
photon 
propagator, $\Delta_{\mu\nu}(k)$,
 has bad high energy behaviour,
\beq
{\text{Unitary~gauge}}:\qquad
\Delta_{\mu\nu}(k)=-{i\over k^2-M_A^2}\biggl( g_{\mu\nu}-{k^{\mu}k^{\nu}
\over M_A^2}\biggr).
\eeq
In the unitary gauge, scattering cross sections have contributions that
grow with powers of $k^2$ (such as $k^4$, $k^6$, etc.) that  cannot
be removed by the conventional mass, coupling constant, and wavefunction
renormalizations.
More convenient gauges are the $R_{\xi}$ gauges
that are obtained by adding the gauge fixing term to
the Lagrangian\cite{Abers:1973qs},
\beq {\cal L}_{GF}=-{1\over 2 \xi}\biggl(\partial_\mu A^\mu+\xi
e v \chi\biggl)^2.
\label{eq:rtsi}
\eeq
After integration by parts,
 the cross term in Eq. ~\ref{eq:rtsi}  exactly cancels the mixed
$\chi \partial_\mu A^\mu$ term of Eq. \ref{eq:ints}.
Different choices for $\xi$ correspond to different gauges and
in the limit $\xi\rightarrow \infty$, the unitary gauge is recovered.

The gauge boson propagator in $R_\xi$ gauge is given by
\beq \Delta_{\mu\nu}(k)=-{i\over k^2-M_A^2}
\biggl( g_{\mu\nu}-{(1-\xi)k_{\mu}k_{\nu}\over k^2-\xi M_A^2}\biggr).
\eeq
 In the $R_\xi$ gauges the $\chi$ field is part of the spectrum
and has mass $M_{\chi}^2=\xi M_A^2$.  The field $\chi$ is called 
a Goldstone boson.   Feynman gauge corresponds to the
choice $\xi=1$  and has a massive Goldstone boson, $\chi$, 
while Landau gauge has $\xi=0$ and  the Goldstone boson
$\chi$ is massless.  We see that the particle spectrum and the mass of the Goldstone
boson are gauge dependent. 

\subsection{Weinberg-Salam Model}
 
It  straightforward to obtain the Weinberg-Salam model
of electroweak interactions by generalizing the results of the previous section\cite{Quigg:2013ufa}.
The Weinberg- Salam model is an $SU(2)_L \times U(1)_Y$ gauge theory containing
three $SU(2)_L$ gauge bosons, $W_\mu^I$, $I=1,2,3$, and one $U(1)_Y$
gauge boson, $B_\mu$, with  kinetic energy terms,
\beq
{\cal L}_{\rm KE} =-{1\over 4}W_{\mu\nu}^I W^{\mu\nu I}
-{1\over 4} B_{\mu\nu} B^{\mu\nu}\, ,
\eeq
where the index $I$ is summed over and,
\beqn
W_{\mu\nu}^I&=& \partial_\nu W_\mu^I-\partial _\mu W_\nu^I
+g \epsilon^{IJK}W_\mu^J W_\nu^K \, ,
\nonumber \\
B_{\mu\nu}&=&\partial_\nu B_\mu-\partial_\mu B_\nu\quad .
\eeqn
The $SU(2)_L$ and $U(1)_Y$ coupling constants are $g$ and $g^\prime$, respectively. 
Coupled to the gauge fields is a complex scalar $SU(2)$
doublet, $\Phi$,
\beq
\Phi
= \left(\begin{array}{c}
 \phi^+   \\
 \phi^0   \end{array}\right)\, .
\eeq
The scalar potential is  given by,
\beq
 V(\Phi)=\mu^2 \mid \Phi^\dagger\Phi\mid +\lambda
\biggl(\mid \Phi^\dagger \Phi\mid\biggr)^2\, ,
\label{eq:wspot}
\eeq
where $\lambda>0$.

Just as in the Abelian model of Section \ref{sec:abh}, the state of minimum
energy for $\mu^2<0$ is not at $\phi^0=0$ and the scalar field develops
a VEV\footnote{As in the Abelian model, there is no mechanism or
motivation
for determining the sign$(\mu^2)$ in the SM.}.
The direction of the minimum in $SU(2)_L$ space is not determined,
since the potential depends only on 
the combination 
$\Phi^\dagger \Phi$
and we arbitrarily choose
\beq
\langle \Phi\rangle
\equiv {1\over\sqrt{2}} \left(\begin{array}{c}
 0   \\
 v   \end{array}\right)\, .
\label{EQ:vevdef}
\eeq
With this choice, the 
 electromagnetic charge is,\footnote{The $\tau_I$ are
 the Pauli matrices with $Tr(\tau_I\tau_J)
=2\delta_{IJ}$.}
\beq
Q={(\tau_3 +Y)\over 2} \,,
\label{eq:qdef}
\eeq
where we assign hypercharge $Y=1$ to $\Phi$. 

  Therefore,
\beq 
Q \langle \Phi\rangle
= 0
\eeq
and electromagnetism is unbroken by the scalar VEV.
The VEV of Eq. \ref{EQ:vevdef}  yields the desired symmetry breaking
pattern,
\beq
SU(2)_L\times U(1)_Y\rightarrow U(1)_{EM}.
\eeq
The scalar contribution to the Lagrangian is,
\beq
{\cal L}_s=(D^\mu \Phi)^\dagger (D_\mu \Phi)-V(\Phi)\, ,
\label{scalepot}
\eeq
where\footnote{ Different choices for the gauge kinetic energy
and the covariant derivative depend on whether $g$ and $g^\prime$
are chosen positive or negative.  There are no physical consequences
of this choice.}
\beq
D_\mu=\partial_\mu +i {g\over 2}\tau\cdot W_\mu+i{g^\prime\over 2}
B_\mu Y.
\eeq
In unitary gauge there are no Goldstone
bosons and only the physical Higgs scalar remains in the spectrum
after the spontaneous symmetry breaking has occurred.
In unitary gauge,
\beq
\Phi={1\over \sqrt{2}}\left(\begin{array}{c}  0 \\
 v+h\end{array}\right)                  \, ,
\eeq
which gives the contribution to the gauge boson masses
from the scalar kinetic energy term of Eq. \ref{scalepot},
\beq
 M^2\sim 
{1\over 2} (0 ,v )
\biggl({1\over 2}g \tau\cdot W_\mu
+{1\over 2} g^\prime B_\mu
\biggr)^2 \left(\begin{array}{c}  0 \\  v \end{array}
\right).
\label{eq:wmasgen}
\eeq
The physical gauge fields are two
charged fields, $W^\pm$, and two  neutral gauge
bosons, $Z$ and $\gamma$.  
\beqn
W^{\pm}_\mu&=&
{1\over \sqrt{2}}(W_\mu^1 \mp i W_\mu^2)\nonumber \\
Z^\mu&=& {-g^\prime B_\mu+ g W_\mu^3\over \sqrt{g^2+g^{\prime~2}}}\equiv -\sin\theta_W B_\mu+\cos\theta_W W_\mu^3
\nonumber \\
A^\mu&=& {g B_\mu+ g^{\prime} W_\mu^3\over \sqrt{g^2+g^{\prime~2}}}\equiv \cos\theta_W B_\mu+\sin\theta_WW_\mu^3.
\label{eq:masseig}
\eeqn
Eq. \ref{eq:masseig} defines a mixing angle,
\beq
\sin\theta_W\equiv {g^\prime \over \sqrt{g^2+g^{\prime~2}}}\,.
\eeq
Since the massless photon must couple with electromagnetic
strength, $e$, the coupling constants 
define the weak mixing angle $\theta_W$,
\beqn
e&=& g \sin\theta_W \equiv gs_W\nonumber \\
e&=& g^\prime \cos\theta_W\equiv g^\prime c_W
\, .
\eeqn
The gauge bosons   obtain  masses
 from the Higgs mechanism, as demonstrated in Eq. \ref{eq:wmasgen}:
\beq
M_W^2 = {1\over 4} g^2 v^2, \qquad \qquad
M_Z^2 = {1\over 4} (g^2 + g^{\prime~2})v^2,\qquad\qquad
M_A =  0\, .
\eeq

Just as in the case of the Abelian Higgs model, if we go to
a gauge other than unitary gauge, there are Goldstone
bosons in the spectrum and the scalar field can be parameterized,
\beq
\Phi={1\over \sqrt{2}}
 e^{i{\omega\cdot\tau\over 2 v}}
\left(\begin{array}{c}  0 \\
 v+h\end{array}\right).
\eeq 
In the Standard Model, there are three Goldstone bosons,
${\vec \omega}=(\omega^\pm,z)$, with masses $M_W$ and $M_Z$ in
the Feynman gauge.  

Fermions can easily be included in the theory. 
We write the fermions in terms of their left-
and right-handed projections,
\beq
\psi_{L,R}={1\over 2}(1\mp\gamma_5)\psi
\,\, .
\eeq 
From the four-Fermi theory of weak interactions\cite{Quigg:2013ufa}, we know experimentally that the
$W$-boson couples only to left-handed fermions and so we construct the 
$SU(2)_L$ doublet, 
\beq
L_L=
\left(\begin{array}{c}
\nu_L\\ e_L \end{array}\right)
.
\eeq
From Eq. \ref{eq:qdef}, the hypercharge of the lepton doublet
must be $Y_L=-1$. 
In the limit where  the neutrino is massless, it can have only
one helicity state which is taken to be $\nu_L$.   Including neutrino
masses requires interactions beyond the standard construction
of the Weinberg-Salam model\footnote{A pedagogical introduction to $\nu$ masses
can be found in Ref. \cite{deGouvea:2004gd}.}.  The SM is  therefore constructed 
with no right-handed neutrinos.  Further, we assume that
right-handed fields do not interact
with the $W$ boson, and so the right-handed electron, $e_R$, must be an
$SU(2)_L$ singlet with  $Y_{e_R}=-2$.  Using these hypercharge
assignments, the leptons can be coupled
in a gauge invariant manner
 to the $SU(2)_L\times
U(1)_Y$ gauge fields,
\beq
{\cal L}_{lepton}
=i {\overline e}_R \gamma^\mu
\biggl(\partial_\mu+i{g^\prime\over 2}Y_e B_\mu\biggr)e_R+
i{\overline L}_L\gamma^\mu
\biggl(\partial_\mu+i {g\over 2}\tau\cdot W_\mu
+i{g^\prime\over 2}Y_L
B_\mu\biggr)L_L\,\,.
\eeq
All of the known fermions can be accommodated in the Standard
Model in this fashion. 
The $SU(2)_L$ and $U(1)_Y$ charge assignments of
the first generation of fermions 
are given in Table \ref{tab:ferm}.  The quantum numbers of the $2^{nd}$ and $3^{rd}$ generation are
identical to those of first generation. 
\begin{table}[t]
\begin{center}
\vskip6pt
\renewcommand\arraystretch{1.2}
\begin{tabular}{|lccr|}
\hline
\multicolumn{1}{|c}{Field}& SU(3)& $SU(2)_L$& $U(1)_Y$
\\
\hline\hline
$Q_L=\left(\begin{array}{c}
u_L\\ d_L \end{array}\right)$
   &    $3$          & $2$&  $~{1\over 3}$
\\
\hline
$u_R$ & $3$ & $1$& ${4\over 3}$
\\\hline
$ d_R$ & $ 3$ & $1$&  $~-{2\over 3}$
\\
\hline
$L_L=\left(\begin{array}{c}
\nu_L\\ e_L \end{array}\right)
$  & $1$             & $2$& $~-1$
\\
\hline
$e_R$ & $1$             & $1$& $~-2$ 
\\ \hline
$\Phi= \left(\begin{array}{c}
\phi^+\\ \phi^0 \end{array}\right)
$  & $1$             & $2$& $1$ 
\\
\hline
\end{tabular}
\caption{Quantum numbers of the  SM fermions.\label{tab:ferm}}
\end{center}
\end{table}

A fermion mass term takes the form
\beq
{\cal L}_{mass}=-m{\overline{\psi}}\psi=-m
\biggl({\overline{\psi}}_L\psi_R+
{\overline{\psi}}_R\psi_L\biggr) 
\,\, .
\label{eq:massterm}
\eeq
As is obvious from Table \ref{tab:ferm}, the left-and right-handed
fermions transform differently under $SU(2)_L$ and $U(1)_Y$ gauge
transformations and so
gauge invariance forbids a term like Eq. ~\ref{eq:massterm}.  
 The Higgs
boson, however,  can couple in a gauge invariant fashion to the down quarks,
\beq
{\cal L}_d=-Y_d {\overline Q}_L \Phi d_R + h.c.\, ,
\label{eq:dyuk}
\eeq
After the Higgs obtains a VEV, we have the effective coupling,
\beq
-Y_d {1\over\sqrt{2}}
({\overline u}_L,~ {\overline d}_L)\left(
\begin{array}{c}  0 \\
v+ \hsm \end{array} \right) d_R + h.c.
\eeq
which can be seen to yield a mass term for the down quark,
\beq 
Y_d = {m_d \sqrt{2}\over v}.
\eeq
In order to generate a mass term for the up-type  quarks we use the 
fact that 
\begin{equation}
{\tilde{\Phi}} \equiv  i \tau_2 \Phi^*
=\left(
\begin{array}{c}\phi^{0} \\
-\phi^-\end{array}\right)
\end{equation}
is an $SU(2)_L$
doublet, and  write the $SU(2)_L$ invariant coupling
\beq
{\cal L}_u=-Y_u {\overline Q}_L {\tilde{\Phi}} u_R + h.c.
\label{eq:uyuk}
\eeq
which generates a mass term for the up quark.  Similar couplings
can be used to generate mass terms for the charged leptons.
Since the neutrino has no right handed partner in the SM, it remains 
massless.  

For the multi-family case, the Yukawa couplings, $
Y_d$ and $Y_u$, become $N_F \times N_F$ matrices
(where $N_F$ is the number of families).  Since the  fermion
mass matrices
and Yukawa matrices  are proportional, the interactions of the
Higgs boson with the fermion mass eigenstates are flavor diagonal
and
 the Higgs boson does not mediate flavor changing interactions.  This is
 an important prediction of the SM.  

The parameter $v$ can be found  from 
the charged current for $\mu$ decay,
$\mu\rightarrow e {\overline \nu}_e \nu_\mu$, which is measured
very accurately to be $G_F=1.16638\times 10^{-5}~GeV^{-2}$.
Since the momentum carried by the $W$ boson is of order $m_\mu$ it
can be neglected in comparison with $M_W$ and we make the identification,
\beq
{G_F\over \sqrt{2}}={g^2\over 8 M_W^2}={1\over 2 v^2},
\label{eq:gdef}
\eeq
which gives the result
\beq
v=(\sqrt{2} G_F)^{-1/2} = 246~GeV\,\,.
\eeq

One of the most important points about the Higgs mechanism is
that all of the couplings of the Higgs boson to fermions and
gauge bosons are completely determined in terms of coupling
constants and fermion masses. A complete set of Feynman
rules can be found in Ref. \cite{Gunion:1989we}. The potential of Eq. \ref{eq:wspot}
had two free parameters, $\mu$ and $\lambda$, which can
be traded for,
\beqn
v^2&=&-{\mu^2\over 2 \lambda}\nonumber \\
m_h^2&=& 2 v^2 \lambda \, .
\label{mhdef}
\eeqn
The scalar potential is now,
\begin{equation}
V={\mhsm^2\over2}\hsm^2+{\mhsm^2\over 2 v}\hsm^3+{\mhsm^2\over 8 v^2}\hsm^4\, .
\label{eq:potfin}
\end{equation}
The self-interactions of the Higgs boson are determined in terms of the Higgs mass. 
  There are no remaining adjustable
parameters and so Higgs production  and decay processes
can be computed unambiguously in terms of the Higgs mass.

\subsection{Aside: Anomaly Cancellation}

The requirement of gauge anomaly
cancellation  puts restrictions on the couplings of the fermions to
vector and axial gauge bosons, denoted here by $V^\mu$ and
$A^\mu$.
The fermions of the Standard Model
have couplings to the gauge bosons of the general form:
\begin{equation}
{\cal L}\sim g_A {\overline \psi}
 T^\alpha \gamma_\mu\gamma_5\psi A^{\alpha\mu}
+g_V {\overline \psi} T^\alpha \gamma_\mu\psi V^{\alpha\mu},
\end{equation}
\begin{figure}
\centering
\includegraphics[width=.4\textwidth]{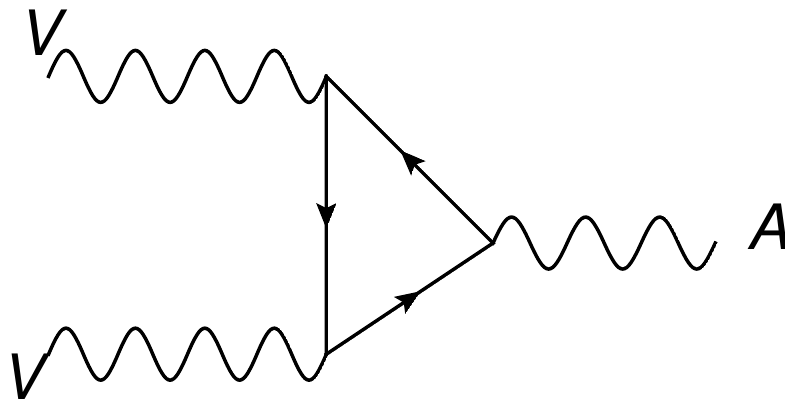}
\caption{\label{fg:anomfig}Contribution to gauge anomalies.}
\end{figure}
where $T^\alpha$ is the gauge generator in the adjunct representation.
These fermion-gauge boson couplings contribute to
triangle graphs (Fig. \ref{fg:anomfig}) that diverge at high energy, 
\begin{equation}
T^{abc}
\sim Tr [\eta_i T^a \{ T^b, T^c\}] \int{d^nk\over (2\pi)^n}{1\over k^3},
\label{eq:anomdef}
\end{equation}
where $\eta_i =\mp 1$ for left- and right-handed
fermions, $\psi_{L,R}={1\over 2} (1\mp \gamma_5)\psi$.
This divergence  is independent of the
fermion mass and depends only on the fermion couplings to
the gauge bosons.  Such
divergences cannot exist in a physical theory.
The theory can be anomaly free in a vector-like model where the left-
and right-handed particles have identical couplings to gauge bosons and
the contribution to Eq. \ref{eq:anomdef} cancels for each pair of
particles.
From Table \ref{tab:ferm}, however, it is clear that the Standard Model is
not vector-like since left- and right-handed fermions transform differently under $SU(2)_L$. The anomaly, $T^{abc}$,
must therefore
be cancelled by a judicious choice
of fermion representations under the various gauge groups.

The only non-vanishing contribution to the
anomaly in the Standard Model is from
\begin{equation}
\Sigma Tr [Y \{T^a, T^b\}],
\label{anom_sm}
\end{equation}
where $T^a$ are the $SU(2)_L$ generators and the sum is over all
fermions in the theory.
Eq. \ref{anom_sm} vanishes for the hypercharge assignments given in Table \ref{tab:ferm}. 
Note that the anomaly cancels separately for each generation of fermions and the SM thus requires complete
generations of chiral fermions.

\section{Theoretical Constraints}
\label{sec:theory}

\subsection{Bounds from Precision Measurements}

The Higgs boson enters into one loop
radiative  corrections in the Standard 
Model and precision electroweak measurements
test the consistency of the theory\footnote{An introductory review of precision measurements
in the SM can be found in Ref. \cite{Wells:2005vk}.} .
In the electroweak sector of the SM, there are four fundamental
parameters, the $SU(2)_{L} \times U(1)_{Y}$ gauge coupling constants,
$g$ and $g^\prime$, as well as the two parameters of the Higgs potential,
which are usually taken to be the  vacuum expectation value of the
Higgs boson,  $v$, and the Higgs mass, $\mhsm$. Once these  parameters are fixed,
all other physical quantities  can be derived in terms
of them (and of course the fermion masses and
CKM mixing parameters, along with the strong coupling constant
$\alpha_s$).
Equivalently, the muon decay constant, $G_{\mu}$,
the Z-boson mass, $M_{Z}$, and the fine structure constant,
 $\alpha$, can be used as input parameters.
Experimentally, the measured values for these input parameters
are\cite{Patrignani:2016xqp,Aad:2015zhl},
\begin{eqnarray}
G_{\mu} & = & 1.16638(1) \times 10^{-5} \; GeV^{-2}\nonumber \\
M_{Z} & = & 91.1876(21) \; GeV\nonumber \\
\alpha^{-1} & = & 137.035 999 679 (94)
\nonumber \\
\mh&=& 125.09\pm.21(stat)\pm.11(syst) ~GeV
\, .
\end{eqnarray}
The $W$ boson mass
is thus a prediction of the theory and is defined through muon decay,
\begin{eqnarray}
M_W^2 &=& \frac{\pi \alpha}{\sqrt{2} G_\mu (1-M_W^2/M_Z^2)}\nonumber \\
M_W^2 &=&{M_Z^2\over 2}\biggl\{ 1+\sqrt{1-{4\pi \alpha\over \sqrt{2}G_\mu M_Z^2}}
\biggr\}
\, .
\label{eq:mwtree}
\end{eqnarray}
At tree level, the SM prediction from Eq. \ref{eq:mwtree} is,
\begin{equation}
M_W(tree)=79.829~GeV\, ,
\end{equation}
in  slight disagreement with the measured value\cite{Patrignani:2016xqp},
\begin{equation}
M_W(experiment)=80.379\pm0.012~GeV \, .
\end{equation}

In order to obtain good agreement between theory and the experimental data, it
is crucial to include radiative corrections.
The prediction
for $M_W$ can be written as\cite{Marciano:1983wwa},
\begin{equation}
 M_{W}^{2} = \frac{\pi \alpha}{\sqrt{2}G_{\mu} s_W^{2}}
\biggl[1 + \Delta r_{SM} \biggr]\, ,
\label{eq:rdef}
\end{equation}
where $\Delta r_{SM}$ summarizes the radiative corrections.
The dependence on the top quark mass, $m_t$, is particularly significant as
$\Delta r_{SM}$ depends on $m_{t}$ quadratically,
\begin{eqnarray}
\Delta r_{SM}^t& = &
 - {G_\mu\over \sqrt{2}}{N_c\over 8 \pi^2}
\biggl( {c_W^2\over s_W^2}
\biggr) m_t^2+\log(m_t)~{\rm{terms}}\, ,
\end{eqnarray}
where $N_c=3$ is the number of colors. The dependence on  $\mh$ is 
logarithmic,
\begin{eqnarray}
\Delta r_{SM}^h&\sim &{\alpha\over \pi
s_W^2}{11\over 48} \log\biggl({\mh^2\over M_Z^2}\biggr) + {\cal {O}}\biggl({m_h^2\over M_Z^2}, {v^4\over\Lambda^4}\biggr)\, .
\label{eq:drsm}
\end{eqnarray}
The top quark does not decouple from the theory even at energies far above
the top quark mass.  This is  because the top quark coupling to the Higgs boson 
is proportional to $m_t$.

\begin{figure}
\begin{center}
\includegraphics[width=0.5\textwidth]{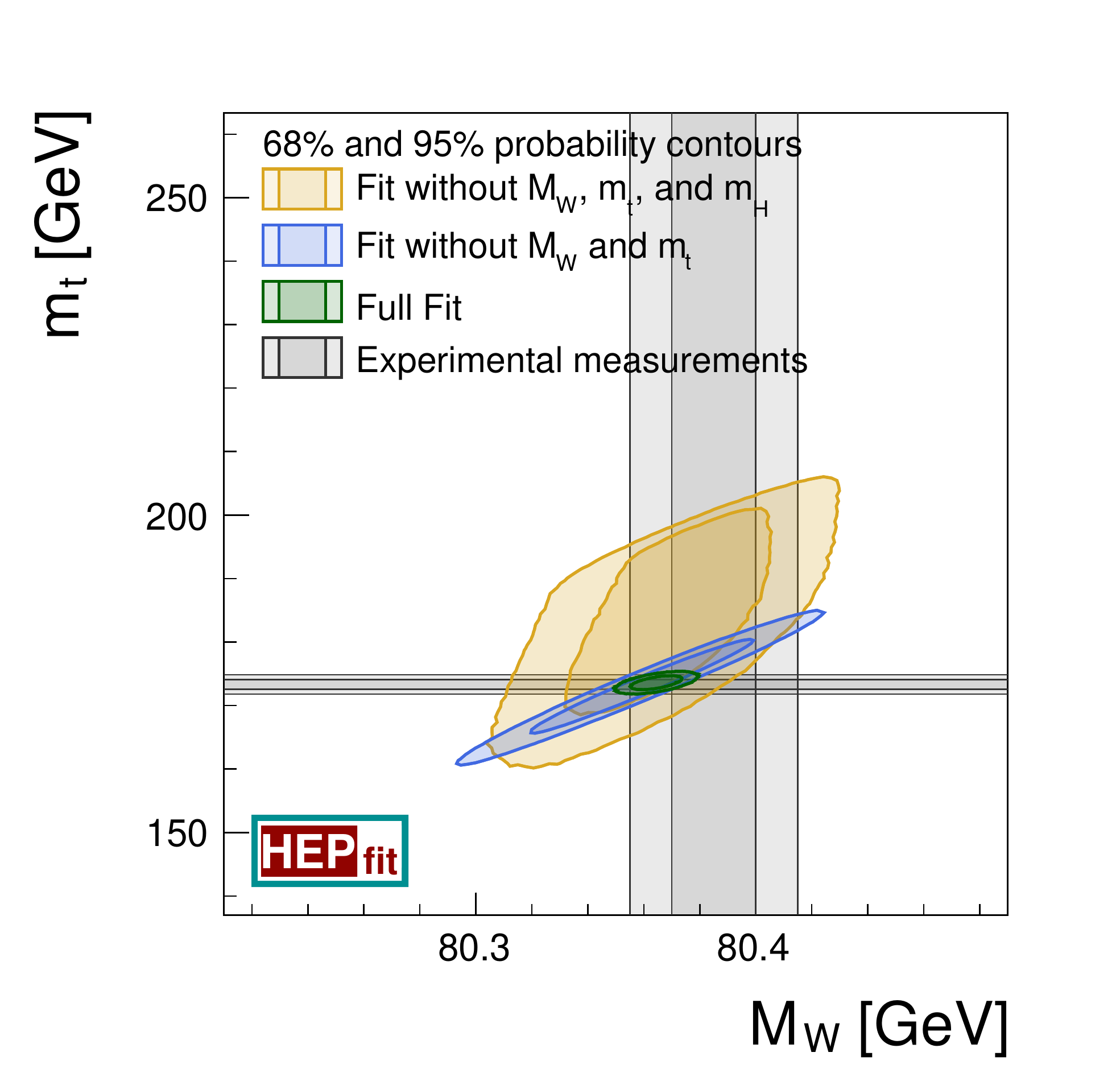}
\caption{\label{fg:ewprec}Experimental limits on $M_W$ and $m_t$ from precision electroweak measurements.
The straight bands are the direct measurements of $M_W$ and $m_t$\cite{deBlas:2016ojx}.}
\end{center}
\end{figure}

The agreement between the radiatively corrected prediction for the $W$ mass given by
Eq. \ref{eq:rdef} with the measured value is a strong test
of the theory.  In a similar fashion,
the full set of electroweak data can be used  to test the self consistency of the theory, 
as demonstrated in Fig. \ref{fg:ewprec}\cite{deBlas:2016ojx}.  Similar
studies have been performed by the GFITTER collaboration\cite{Baak:2012kk}.   (The most restrictive data points are the measurements
of the $Zb{\overline b}$ coupling and the $W$ boson mass.) When the experimental values of $M_W$,  $m_t$, and 
$\mh$ are omitted, the fit is in good agreement with the directly measured values of the masses.  Note that
the fit excludes a large ($\sim 100's$ of $GeV$) value of $\mh$ and so even before the Higgs boson was discovered,
we knew that  if there were no new physics contributions  to the predictions for electroweak quantities such as $M_W$,
the Higgs boson could not be too heavy. 

\subsection{Oblique Parameters}

Extensions of the SM with modified Higgs sectors are significantly restricted by
the requirement of consistency with the electroweak measurements.  A simple way to examine
whether a theory with a complicated Higgs sector is consistent with electroweak
experiments is to use the oblique parameters. Using the oblique parameters
to obtain limits on BSM physics
assumes that the dominant contributions
resulting from  the
 expanded theory 
 are to the gauge
boson 2-point functions\cite{Peskin:1991sw,Altarelli:1990zd},
\begin{equation}
\Pi_{XY}^{\mu\nu}(p^2)=\Pi_{XY}(p^2)g^{\mu\nu}+B_{XY}(p^2)p^\mu p^\nu\, ,
\end{equation}
with $XY=\gamma\gamma, \gamma Z, ZZ$ and $W^+W^-$.
We define the $S$,$T$ and $U$ functions
following the notation of Peskin and
Takeuchi\cite{Peskin:1991sw},
\begin{eqnarray}
\alpha S&=&
\biggl({4 s_W^2 c_W^2\over M_Z^2}\biggr)
\biggl\{ \Pi_{ZZ}(M_Z^2)- \Pi_{ZZ}(0)-
\Pi_{\gamma\gamma}(M_Z^2)
\nonumber \\ &&
-{c_W^2-s_W^2\over c_W s_W}\biggl(
\Pi_{\gamma Z}(M_Z^2)
-\Pi_{\gamma Z}(0)\biggr)\biggr\}\nonumber \\
\alpha  T &=&
 \biggl({ \Pi_{WW}(0)\over M_W^2}-{
\Pi_{ZZ}(0)\over M_Z^2}-{2 s_W
\over c_W }{\Pi_{\gamma Z}(0) \over M_Z^2}
\biggr)
\nonumber \\
\alpha U&=& 4 s_W^2\biggl\{
{ \Pi_{WW}(M_W^2)-\Pi_{WW}(0)\over M_W^2} -c_W^2
\biggl({ \Pi_{ZZ}(M_Z^2)-\Pi_{ZZ}(0)\over M_Z^2}\biggr)
\nonumber \\
&&-2 s_W c_W
\biggl(
{ \Pi_{\gamma Z}(M_Z^2)-\Pi_{\gamma Z}(0)
\over M_Z^2}\biggr)
-s_W^2
{ \Pi_{\gamma \gamma}(M_Z^2)\over M_Z^2}\biggr\} \, .
\label{sdef}
\end{eqnarray}
New physics effects are then determined by subtracting the SM contribution, e.g. $\Delta S \equiv S_{BSM}-S_{SM}$.

\subsubsection{Aside: Restrictions on New Physics from Oblique Parameters with a Scalar Singlet}
The simplest possible extension of the SM in the Higgs sector is to add 
a real scalar singlet, $S$,  with hypercharge $Y=0$.
After imposing a $Z_2$ symmetry under which $S\rightarrow -S$, 
the most general scalar potential is\cite{OConnell:2006rsp}
\begin{eqnarray}
V &=& - \mu^2 \, \Phi^\dagger \Phi -m^2 S^2 +\lambda (\Phi^\dagger \Phi)^2
   + \frac{a_2}{2} \, \Phi^\dagger \Phi \, S^2   + \frac{b_4}{4} S^4.
\label{potential}
\end{eqnarray}
After spontaneous symmetry breaking,  both $\Phi$ and $S$ obtain VEVs and 
the mass eigenstates $h$ and $H$ are a mixture of $S$ and $\Phi$ ($s\equiv \langle S\rangle$),
\begin{equation}
\left(
\begin{array}{c}
h \\ H\end{array}
\right)=\left(\begin{array}{cc} 
\cos\alpha & -\sin\alpha \\
\sin\alpha & \cos \alpha
\end{array}
\right)
\left( 
\begin{array}{c}
\sqrt{2}\phi_0-v\\
S-s
\end{array}
\right)\, ,
\label{eq:singdef}
\end{equation}
with physical masses, $m_h$ and $M_H$.
The singlet cannot couple  directly
to fermions or gauge bosons, so the only physical effect 
on single Higgs production is through the mixing of Eq. \ref{eq:singdef}. The mixing affects the SM-like
Higgs couplings to both fermions and gauge bosons in an identical fashion and all SM couplings are suppressed
by the factor $\cos\alpha$.
This model is particularly simple since it can be studied in terms of $M_H$ and the mixing angle $\alpha$.
 For $m_h, M_{H} >> M_W, M_Z$,
 the contributions to the oblique parameters are,
\begin{eqnarray}
\Delta S&=&{1\over 12 \pi}
\sin^2\alpha
\log\biggl({M_H^2\over m_h^2}\biggr)
\nonumber \\
\Delta T&=& -{3\over 16 \pi c_W^2}
\sin^2\alpha \log\biggl({M_H^2\over {m_h}^2}\biggr)
\nonumber \\
\Delta U&=& 0
\, .
\label{eq:loglim}
\end{eqnarray}
and for any given value of $M_H$, an upper limit on $\sin\alpha$ can be determined\cite{Pruna:2013bma}.
Limits from the oblique parameters are an important tool in understanding what BSM models are allowed
experimentally and in restricting the parameters of the models.

\subsection{Restrictions from Unitarity}

\label{sec:unitsec}
Scattering amplitudes in a weakly interacting theory are required to obey perturbative 
unitarity\cite{Lee:1977eg}. Cross sections that grow with energy  will eventually
violate perturbative unitarity, a simple result derived from the optical theorem.
 The simplest version of the unitarity requirement 
 is,
\beq
\mid Re(a_0^0)\mid < {1\over 2}\, ,
\eeq
where $a_0^0$ is the $J=0$ partial wave.  For a $2\rightarrow 2$ scattering
process with massless particles,
\begin{equation}
a_0={1\over 16\pi s}\int_{-s}^0\mid A\mid\, .
\end{equation}

The interesting physics is in the longitudinal gauge boson sector, since the longitudinal 
polarizations are the result of the electroweak symmetry breaking.  For a gauge boson $V=(W,Z)$ 
with momentum in the $z-$ direction,
\begin{equation}
p_V=(E_V,0,0,p),
\end{equation}
the longitudinal polarization vector is,
\begin{equation}
\epsilon_L^\mu={1\over M_V}(p,0,0,E_V)\rightarrow_{E_V>>M_V} {p^\mu_V\over M_V}+{\cal{O}}({M_V^2\over E_V^2})
\label{eq:long}
\, .
\end{equation}
Eq. \ref{eq:long}  makes it clear that longitudinally polarized gauge bosons can give potentially dangerous
contributions  to scattering processes at high energies.
\vskip 1in

\begin{figure}
\begin{centering}
\includegraphics[width=0.6\textwidth]{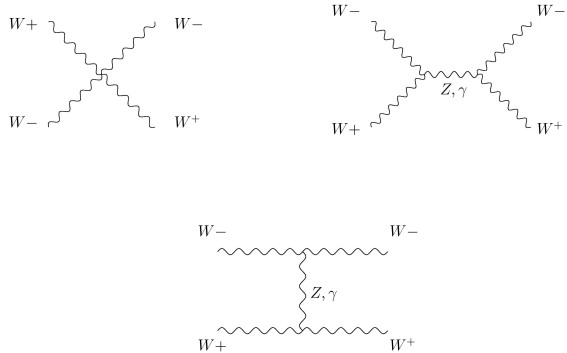}
\par\end{centering}
\vskip -.5in
\caption{\label{fig:wwun1}Diagrams contributing to $W^+_LW^-_L\rightarrow W^+_L W^-_L$.}
\end{figure} 

\vskip -.5in 
\begin{figure}
\begin{centering}
\includegraphics[width=0.6\textwidth]{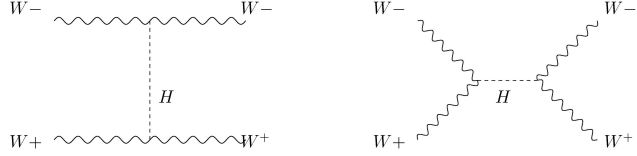}
\par\end{centering}
\vskip -1.in
\caption{\label{fig:wwun2}Diagrams contributing to $W^+_LW^-_L\rightarrow W^+_L W^-_L$. In this figure, $H$ is the SM Higgs boson. } 
\end{figure}

The elastic scattering of longitudinally polarized gauge bosons,
 $W^+_LW^-_L\rightarrow W^+_L W^-_L$, illustrates this point\cite{Duncan:1985vj}.
 The $s-$ and $t-$ channel diagrams containing the exchange of a $Z$ boson or $\gamma$, combined with the
$4-$point interaction shown in Fig. \ref{fig:wwun1}  give the contribution to the scattering amplitude in
 the limit $s>>M_W^2,M_Z^2$,
\beq
 {\cal A}(W^+_LW^-_L\rightarrow W^+_LW^-_L)_V\sim 
-{1\over v^2}\biggl\{
-s-t
\biggr\}
\quad .
\label{eq:www}
\eeq
This contribution grows with $s$.
However, the $s-$ and $t-$ channel contributions from Higgs exchange shown in Fig. \ref{fig:wwun2} 
 contribute,
\beq
 {\cal A}(W^+_LW^-_L\rightarrow W^+_LW^-_L)_h\sim 
-{1\over v^2}\biggl\{
{s^2\over s-\mh^2}+{t^2\over t-\mh^2}
\biggr\}
\quad .
\label{eq:wwh}
\eeq
Combining Eqs. \ref{eq:www} and \ref{eq:wwh}, we find
the high  energy limit, $M_W^2<<s$, of the $J=0$ partial wave,
\beqn
a_0^0(W_L^+W_L^-\rightarrow W_L^+W_L^-)& \equiv &{1\over 16 \pi s}
\int^0_{-s}\mid {\cal A}\mid dt +{\cal{O}}\biggl({s\over M_W^2}\biggr)
\nonumber \\
&=&- {\mh^2\over 16 \pi v^2 }
\biggl[2 + {\mh^2\over s-\mh^2}-{\mh^2\over s}\log
\biggl(1+{s\over \mh^2}\biggr)\biggr]\, .
\label{eq:wwscat}
\eeqn
This limit is well behaved in the high energy limit, due the the cancellations
between the different contributions.  Note that this cancellation requires the SM Higgs
boson contribution with SM couplings to the gauge bosons. 

In the high energy limit,
 $s >>\mh^2$,  Eq. ~\ref{eq:wwscat}
has the limit,
\beq
\mid a_0^0(W_L^+W_L^-\rightarrow W_L^+W_L^-)\mid 
\longrightarrow_{s>>\mh^2}
{\mh^2\over 8 \pi v^2}.
\label{eq:unitlim}
\eeq
Eq. \ref{eq:unitlim} gives 
the upper bound on the SM Higgs mass from perturbative unitarity of $\mh \lsim 870~GeV$.

Although individual contributions in Eqs. \ref{eq:www} and \ref{eq:wwh} grow with energy, in the SM they combine in just the right fashion
to preserve perturbative unitarity.  This is another strong constraint on BSM models.  If the $WWZ$
coupling were altered from the SM value, the high energy cancellation of Eq. \ref{eq:wwscat} would
not occur and partial wave unitarity would be violated. 
The Higgs boson  plays a fundamental role in the
theory since it cuts off the growth of the partial wave
amplitudes and makes the theory obey perturbative unitarity.  

\subsection{Restrictions from Triviality}

\begin{figure}
\begin{center}
\includegraphics[width=0.5\textwidth]{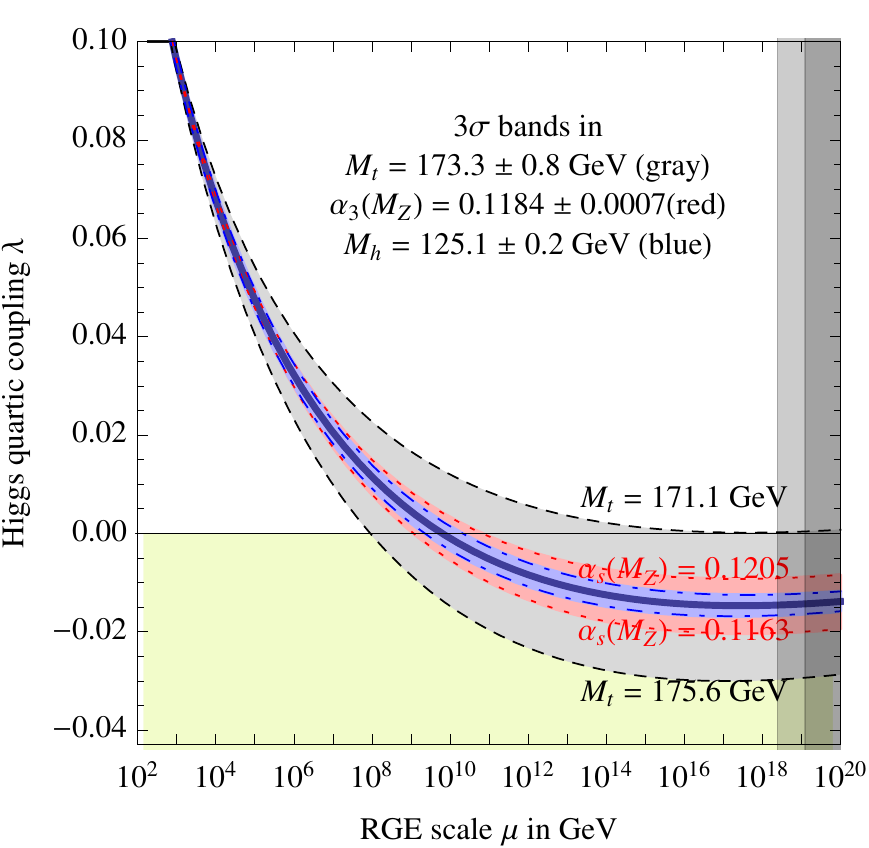}
\caption{\label{fg:lamsc}Dependence of the Higgs quartic coupling on the  renormalization scale \cite{Buttazzo:2013uya}.}
\end{center}

\end{figure}

Theoretical bounds on the Higgs boson  mass can be  deduced on the grounds
of {\it triviality},
which can be summarized
as the requirement that the Higgs quartic coupling remain finite
at high energy scales.  If the quartic coupling becomes infinite, the theory is no longer
perturbative, while if the quartic coupling goes to zero, the theory is non-interacting.
The Higgs quartic coupling, $\lambda$,
changes with the effective energy scale, $\Lambda$, due to
the self interactions of the scalar field:
\beq
{d \lambda \over dt}={3 \lambda^2\over 4 \pi^2},
\label{eq:lams}
\eeq
where $t\equiv \log(\Lambda^2/v^2)$.
In the SM, however, there are also contributions due to gauge boson and fermion loops\footnote{We neglect
the gauge contributions here.}.  Including the 
top quark contribution, Eq. \ref{eq:lams} becomes,
\beq
{d \lambda \over dt}={3 \over 4 \pi^2}\biggl\{\lambda^2-Y_t^2\lambda-Y_t^4\biggr\}\, ,
\label{eq:lams}
\eeq
where $Y_t=m_t/v$.
For small $\lambda$ ( small $\mhsm$), the $Y_t^4$ term dominates and the quartic coupling decreases with energy,
\begin{equation}
\lambda(\Lambda)\sim \lambda(v)-{3Y_t^4\over 4\pi^2}\log\biggl({\Lambda^2\over v^2}\biggr)\, .
\end{equation}
The scaling of $\lambda$ has been performed to $2-$ loops\cite{Buttazzo:2013uya},
 including contributions from gauge and Yukawa couplings
and the result is shown in Fig. \ref{fg:lamsc}.  The quartic coupling becomes negative at a high scale that
is quite sensitive to $m_t$ and $\alpha_s$, suggesting that at this scale some new physics is required to force $\lambda$ to be
positive which is need in order  for the potential to be bounded from below.

\section{Higgs Production and Decay}
\label{sec:prod}

In this section we review the SM rates for Higgs production and decay.  Numerical values, including the most precisely
known higher order calculations, have been tabulated by the LHC Higgs cross section working group\cite{Dittmaier:2011ti}.

\subsection{Higgs Decays}
Expressions for the SM Higgs decay widths at leading order can be found in Ref. \cite{Gunion:1989we}, and the QCD
corrected rates, with references to the original literature, are given in Refs. \cite{Djouadi:2005gi,Spira:2016ztx}.  The QCD 
NLO corrected decay rates can be found using the public code, HDECAY\cite{Djouadi:1997yw}. 

\subsubsection{$h\rightarrow f {\overline{f}}$}
The Higgs couplings  to fermions are proportional to fermion mass and the lowest order
width for the Higgs decay to fermions 
 of mass $\mf$ is, 
\beq
\Gamma(\hsm\rightarrow f {\overline f} )={G_F \mf^2 N_{ci}\over 
4\sqrt{2}\pi} \mhsm \beta_F^3
\, ,
\eeq
where
$\beta_F\equiv \sqrt{1-4\mf^2/\mhsm^2}$ is the velocity of the
final state fermions and $N_{ci}=1 (3)$ for charged leptons (fermions).  
 The largest fermion decay channel is $\hsm \rightarrow b {\overline{b}}$, which
 receives large  QCD corrections.
 A
 significant portion of the QCD corrections can be accounted for by
expressing the decay width in terms of a
running  quark mass, $\mf(\mu)$, evaluated at the
scale $\mu=\mhsm$.
  The QCD
corrected decay width can then be approximated as\cite{Drees:1990dq,Braaten:1980yq},
\beq
\Gamma(\hsm\rightarrow q {\overline q})=
{3G_F\over 4 \sqrt{2} \pi} m_q^2(\mh^2)
\mh \beta_q^3 \biggl(1+5.67{\alpha_s(\mh^2)\over \pi}+\cdots
\biggr)                             ,
\eeq
where $\alpha_s(\mh^2)$ is defined in the ${\overline{MS}}$ scheme with
$5$ flavors.
In
leading log QCD, the running of the $b$ quark mass is,
\beq
m_{b}(\mu^2)=m \biggl[{\alpha_s(m^2)
\over \alpha_s(\mu^2)}
\biggr]^{(-12/23)}
\biggl\{1+{\cal O}(\alpha_s^2)\biggr\} ,
\label{bscale}
\eeq
where $m_b(m^2)\equiv m $ implies that the running mass at the
position of the propagator pole is equal to the location of the pole.
For $m_b(m_b^2)=4.18~GeV$, this yields an
effective value $m_b(\mh=125~GeV)\mid_{LL} = 2.8~GeV$
(at NLL, $m_b(\mh=125~GeV)\mid_{NLL} = 2.7~GeV$).
 Inserting the QCD corrected mass into  the expression
for the width thus leads to a suppression of the width by $\sim .4$.
Using the running $b$ mass absorbs the large logarithms of the form $\log(m_h^2/m_b^2)$
and is important for numerical accuracy. 
The electroweak radiative corrections to $\hsm \rightarrow f {\overline f}$
 amount to only a few
percent correction\cite{Kniehl:1993ay}.

\subsubsection{$h\rightarrow WW, ZZ$}
The Higgs boson can also decay to gauge boson pairs.
At tree level, the decays $\hsm\rightarrow WW^*$ and $\hsm\rightarrow
ZZ^*$ are possible (with one of the gauge bosons off-shell), while at one-loop 
the decays $\hsm\rightarrow gg,\gamma\gamma$, and 
$\gamma Z$
occur. 

The decay width for  the off-shell decay, $h\rightarrow Z Z^*\rightarrow  f_1(p_1) f_2(p_2) Z(p_3)$, is,
\begin{eqnarray}
\Gamma&=&\int_0^{(m_h-M_Z)^2} dq^2 \int dm_{23}^2\,{\mid A\mid^2\over 256\pi^3 m_h^3}\, ,
\end{eqnarray}
where $m_{ij}=(p_i+p_j)^2$, $m_{12}^2\equiv q^2$, and $m_{12}^2+m_{23}^2+m_{13}^2=m_h^2+M_Z^2$,
$\lambda(m_h^2,M_Z^2,q^2)\equiv q^4-2q^2(m_h^2+M_Z^2)+(m_h^2-M_Z^2)^2$, and 
$m_{23}^2\mid_{max,min}\equiv {1\over 2}
\biggl(m_h^2+M_Z^2-q^2\pm \sqrt{\lambda}\biggr)$.
The amplitude-squared is, 
\begin{eqnarray}
\mid A (h\rightarrow Z f {\overline {f}})\mid^2&=&
32\,(g_L^{\, 2}+g_R^{2})\,G_F^2\,
M_Z^4
\nonumber \\ && \cdot
\biggl [
{2 M_Z^2q^2 -m_{13}^2 q^2 
-m_h^2 M_Z^2+m_{13}^2M_Z^2
 +m_{13}^2\,m_h^2- m_{13}^4\over
 (q^2-M_Z^2)^2+\Gamma_Z^2M_Z^2}\biggr]\, ,
   \label{eq:treeamp}
\end{eqnarray}
with $g_{Lf}=T_{3f}-Q_fs_W^2$, $g_{Rf}=-Q_fs_W^2$, and  $T_3=\pm {1\over 2}$.  
 We see that the amplitude is peaked at  $q^2=M_Z^2$. 
Integrating over $dm_{23}^2$, 
\begin{eqnarray}
{d\Gamma\over dq^2}(h\rightarrow Z f {\overline {f}})&=&
(g_L^{\, 2}+g_R^{2})\,G_F^2\,\sqrt{\lambda(m_h^2,M_Z^2,q^2)}{M_Z^4\over 48 \pi^3 m_h^3}
\nonumber \\ && \cdot 
\biggl[{(12 M_Z^2 q^2+\lambda(m_h^2,M_Z^2,q^2))
  \over
 (q^2-M_Z^2)^2+\Gamma_Z^2M_Z^2}
\biggr]\, .
\end{eqnarray}
The result for $h\rightarrow W f {\overline {f^\prime}}$ can  be found 
by making the appropriate redefinitions of the fermion 
- gauge boson couplings. 
 
 Performing the $q^2$ integral and
summing over the final state fermions\cite{Keung:1984hn}, 
\begin{eqnarray}
\Gamma(\hsm\rightarrow WW^*)&=&{3g^4\mhsm\over 512 \pi^3} F\biggl({M_W\over \mhsm}\biggr)\nonumber \\
\Gamma(\hsm\rightarrow ZZ^*)&=&{g^4 \mhsm\over 2048  \cos_W^4 \pi^3}\biggl(7-{40\over 3}s_W^2
+{160\over 9}s_W^4\biggr)F\biggl({M_Z\over\mhsm}\biggr)\, ,
\end{eqnarray}
where
\begin{eqnarray}
F(x)&=&- \mid 1-x^2\mid 
\biggl(
{47\over 2}x^2-{13\over 2} +{1\over x^2}\biggr)\nonumber \\
&&+3(1-6x^2+4x^4)\mid \ln x\mid 
+{3(1-8x^2+20x^4)\over \sqrt{4x^2-1}} 
\cos^{-1}\biggl({3x^2-1\over 2 x^3}\biggr)\, . 
\end{eqnarray}
The NLO QCD and electroweak corrections to the off-shell decays, $h\rightarrow V^* V^*\rightarrow $4-fermions , $V=(W,Z)$, are
implemented in the public code, PROPHECY4f\cite{Bredenstein:2007ec}. 

\subsubsection{$h\rightarrow gg$}
The decay of the Higgs boson to gluons  only arises through
fermion loops in the SM  and is sensitive to new colored particles that interact with the Higgs,
\beq
\Gamma(\hsm\rightarrow gg)={ G_F \alpha_s^2 \mhsm^3
\over 64 \sqrt{2}\pi^3}
\mid \sum_q F_{1/2}(\tau_q)\mid^2\,  ,
\eeq
where $\tau_q\equiv                                                             
4 m_q^2/\mhsm^2$ and  $F_{1/2}(\tau_q)$ is defined to be,
\beq
F_{1/2}(\tau_q) \equiv -2\tau_q\biggl[1+(1-\tau_q)f(\tau_q)\biggr]
\, .
\label{eq:etadef}
\eeq
The function $f(\tau_q)$ is given by,
\beq
f(\tau_q)=\left\{\begin{array}{ll}
\biggl[\sin^{-1}\biggl(\sqrt{1/\tau_q}\biggr)\biggr]^2,&\hbox{if~}
\tau_q\ge 1
\\
-{1\over 4}\biggl[\log\biggl({x_+\over x_-}\biggr)
-i\pi\biggr]^2,
&\hbox{ if~}\tau_q<1,
\end{array}
\right .
\label{fundef}
\eeq
with
\beq
x_{\pm}=1\pm\sqrt{1-\tau_q}
.
\eeq
In the limit in which the quark mass is much less than the Higgs boson mass,
\beq
F_{1/2}\rightarrow {2 m_q^2\over \mhsm^2}\log^2\biggl(
{m_q\over m_h}\biggr)
.
\label{eq:lighth}
\eeq
On
the other hand, for a heavy quark, $\tau_q\rightarrow\infty$,
 and $F_{1/2}(\tau_q)$ approaches
a constant,
\beq
F_{1/2}\rightarrow -{4\over 3}
.
\label{eq:f12}
\eeq
Eqs. \ref{eq:lighth} and \ref{eq:f12} make it clear that the top quark loop is the dominant contribution. 
QCD corrections to the decay $h\rightarrow gg $ are known at NLO for a finite top quark mass
 and increase the rate by roughly $60\%$\cite{Spira:1995rr}. 

\subsubsection{$h\rightarrow \gamma \gamma$}

The decay $h\rightarrow \gamma \gamma$  arises  from fermion and $W$ loops and is an
important mode for Higgs measurements at the LHC, despite the smallness of the branching ratio.
At lowest order the width is, \cite{Gunion:1989we}
\beq
\Gamma(\hsm\rightarrow \gamma\gamma)={\alpha^2 G_F\over 128\sqrt{2} \pi^3}
m_h^3\mid \sum_i N_{ci} Q_i^2 F_i(\tau_i)\mid^2\, ,
\eeq
where the sum is over fermions and  $W^\pm$ bosons with
$F_{1/2}(\tau_q)$ given in Eq. \ref{eq:etadef}, and
\beq
F_W(\tau_W)= 2+3\tau_W[1+(2-\tau_W)f(\tau_W)]
\, ,
\label{fdef}
\eeq
with $\tau_W=4 M_W^2/m_h^2$, $N_{ci}=1 (3)$ for leptons (quarks),
and $Q_i$ is the electric charge in units of $e$.  In the (unphysical) limit $\tau_W\rightarrow\infty$,
$F_W\rightarrow 7$ and we see that the top quark and $W$ contributions have opposite signs.  The decay $\hsm\rightarrow
\gamma\gamma$ is therefore sensitive to the sign of the top quark Yukawa coupling through the interference of the $W$ and $t$ loops.  
Similarly, the rate for $\hsm\rightarrow Z\gamma$ receives contributions from both fermions and the
$W$ boson.  The analytic formula is given in \cite{Gunion:1989we} and the $Z\gamma$ width is quite small. 

\begin{figure}
\begin{centering}
\includegraphics[width=0.4\textwidth]{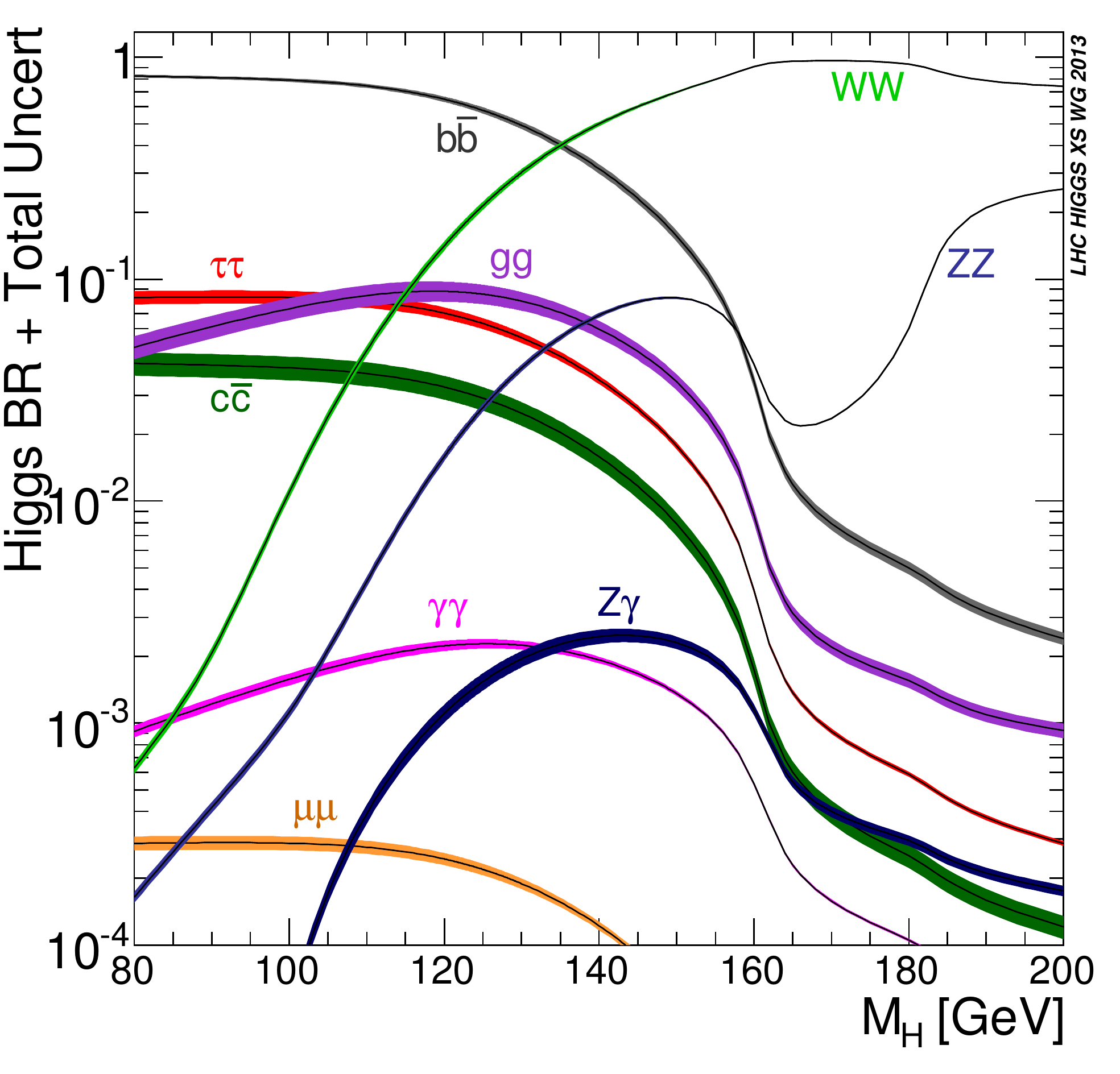}
\includegraphics[width=0.4\textwidth]{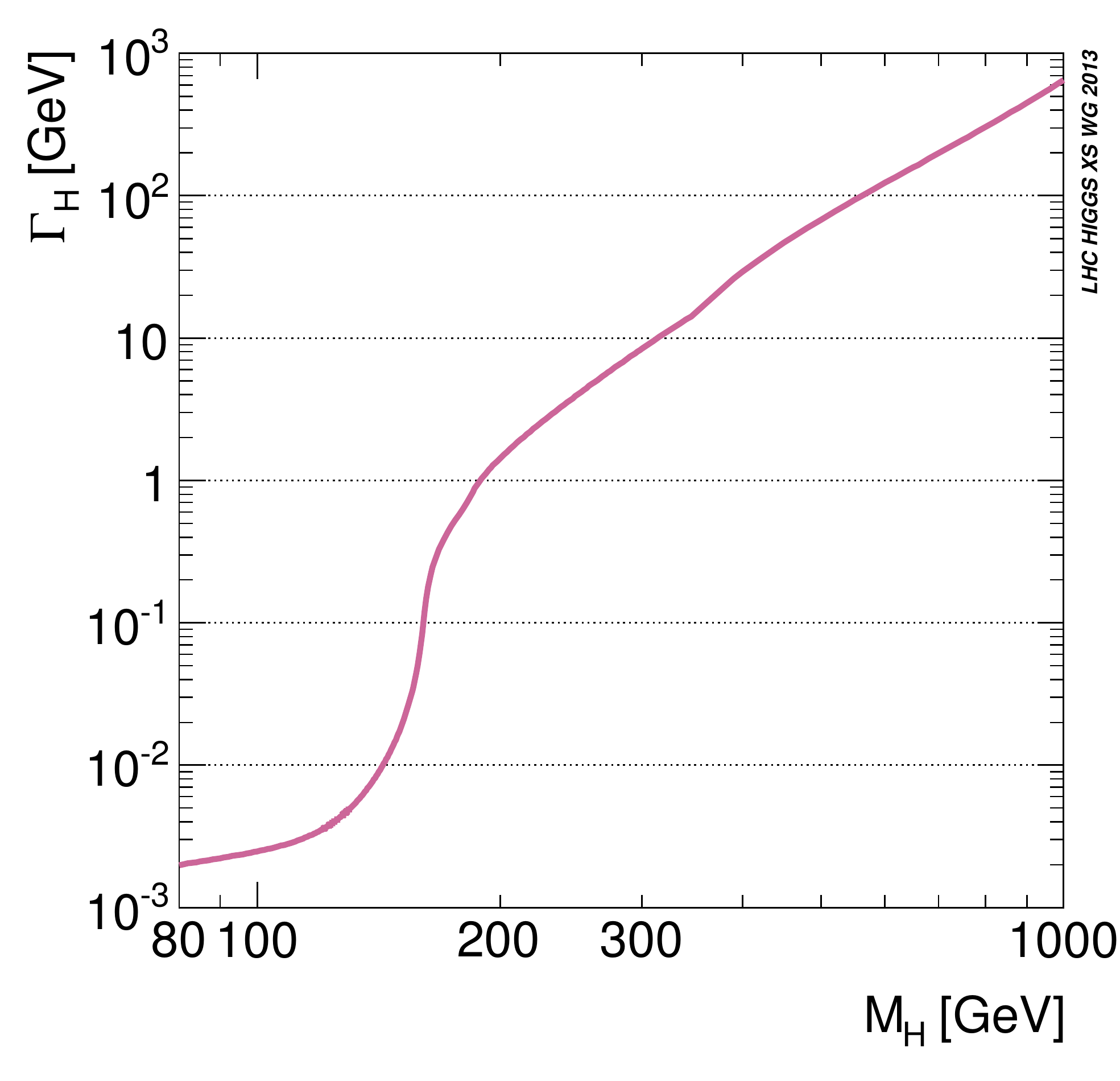}
\par\end{centering}
\caption{\label{fig:hbrs}
SM Higgs Branching ratios   (LHS) and total width for a SM-like Higgs boson of arbitrary mass 
(RHS)\cite{Dittmaier:2011ti}.  In this figure, $H$ is the SM Higgs boson. }
\end{figure}

The Higgs branching ratios are shown in Fig. \ref{fig:hbrs} for a SM Higgs boson of
arbitrary mass\cite{Dittmaier:2011ti}.  The
width of the curves is an estimate of the theoretical uncertainties on the branching ratios. 
The branching ratios assume SM couplings and
no new decay channels and include all known radiative corrections\cite{Dittmaier:2011ti}. 
 Also shown
in Fig. \ref{fig:hbrs} is the Higgs total decay width as a function of Higgs mass.  For $\mhsm=125~GeV$,
the total width is very narrow, $\Gamma_\hsm=4~MeV$.  

\subsection{Higgs Production in Hadronic Collisions}
\begin{figure}
\begin{centering}
\includegraphics[width=0.35\textwidth]{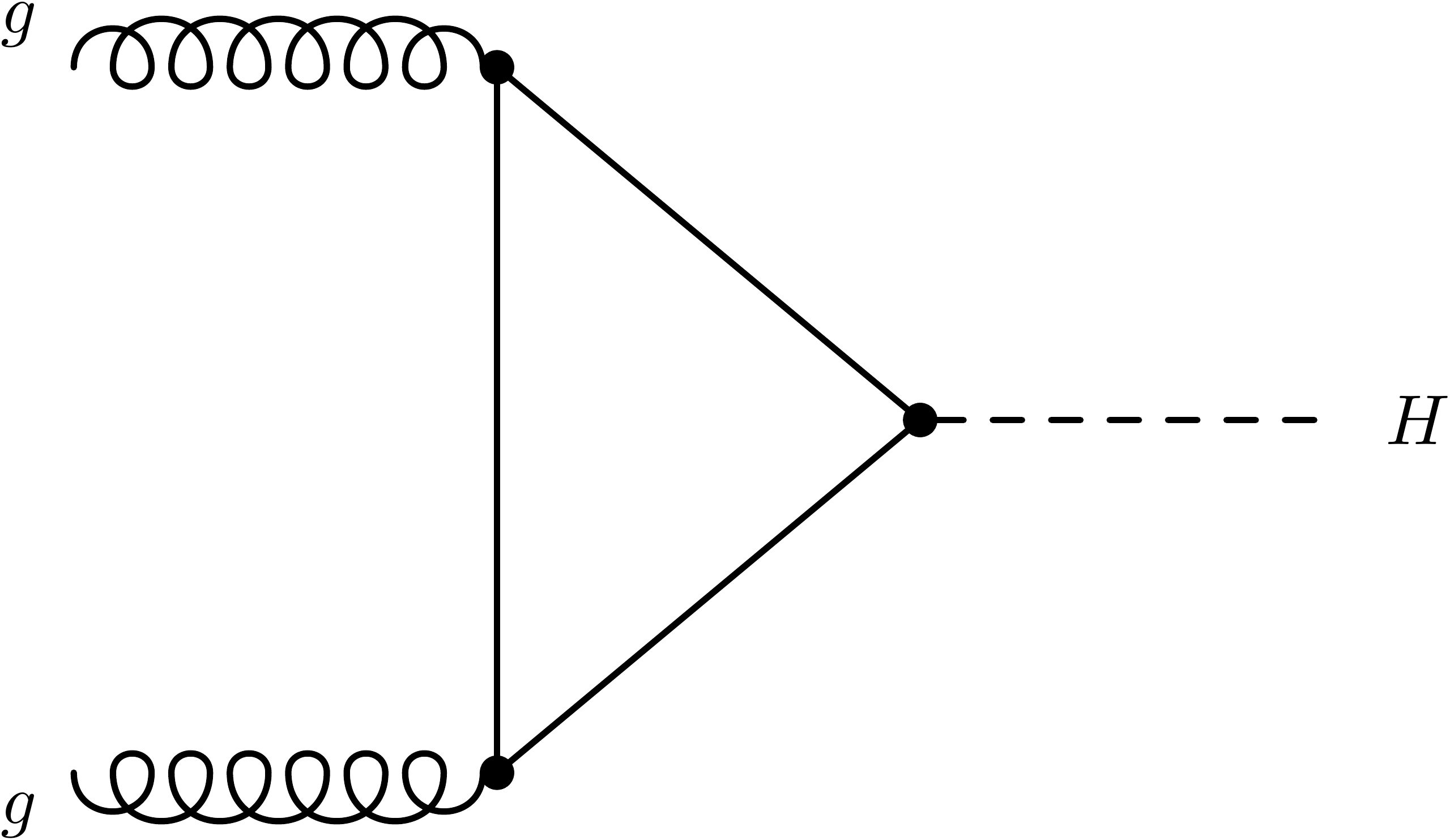}
\hskip .2in
\includegraphics[width=0.35\textwidth]{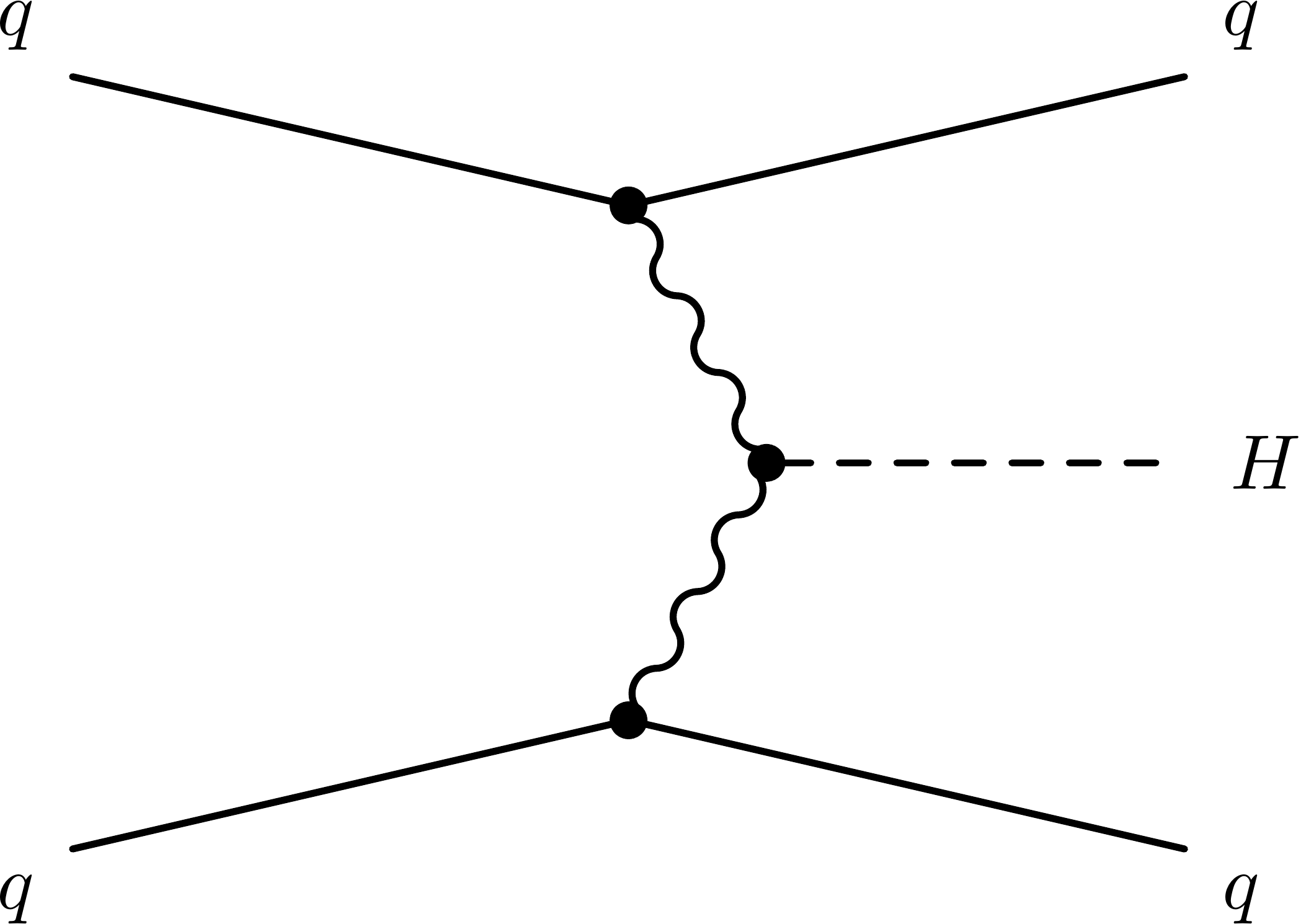}
\par\end{centering}
\caption{\label{fig:prodgg}Contribution to Higgs boson production from
(LHS) gluon fusion and (RHS) vector boson scattering. In this figure, $H$ is the SM Higgs boson. }
\end{figure}

\begin{figure}
\begin{centering}
\includegraphics[width=0.35\textwidth]{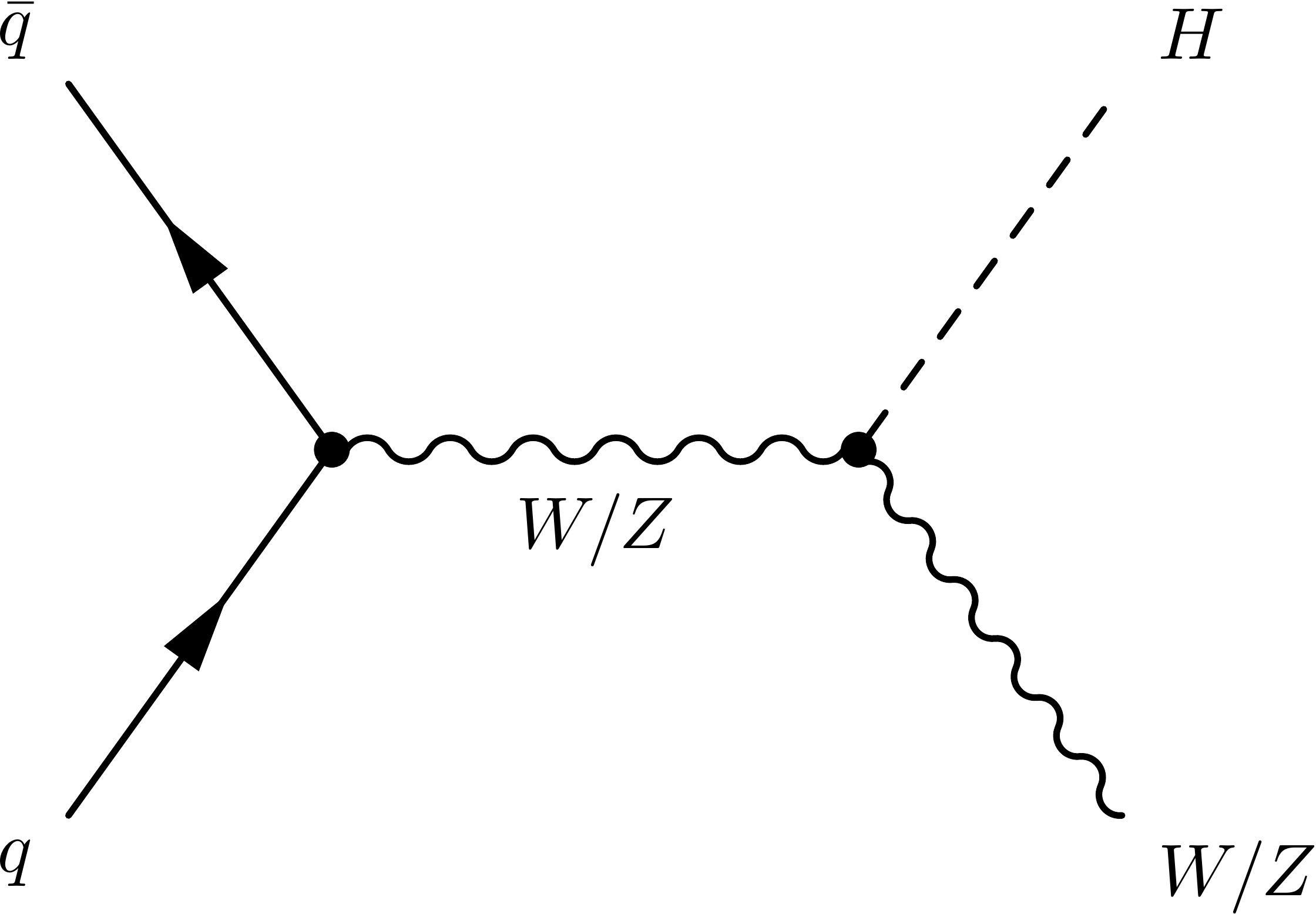}
\hskip .2in
\includegraphics[width=0.35\textwidth]{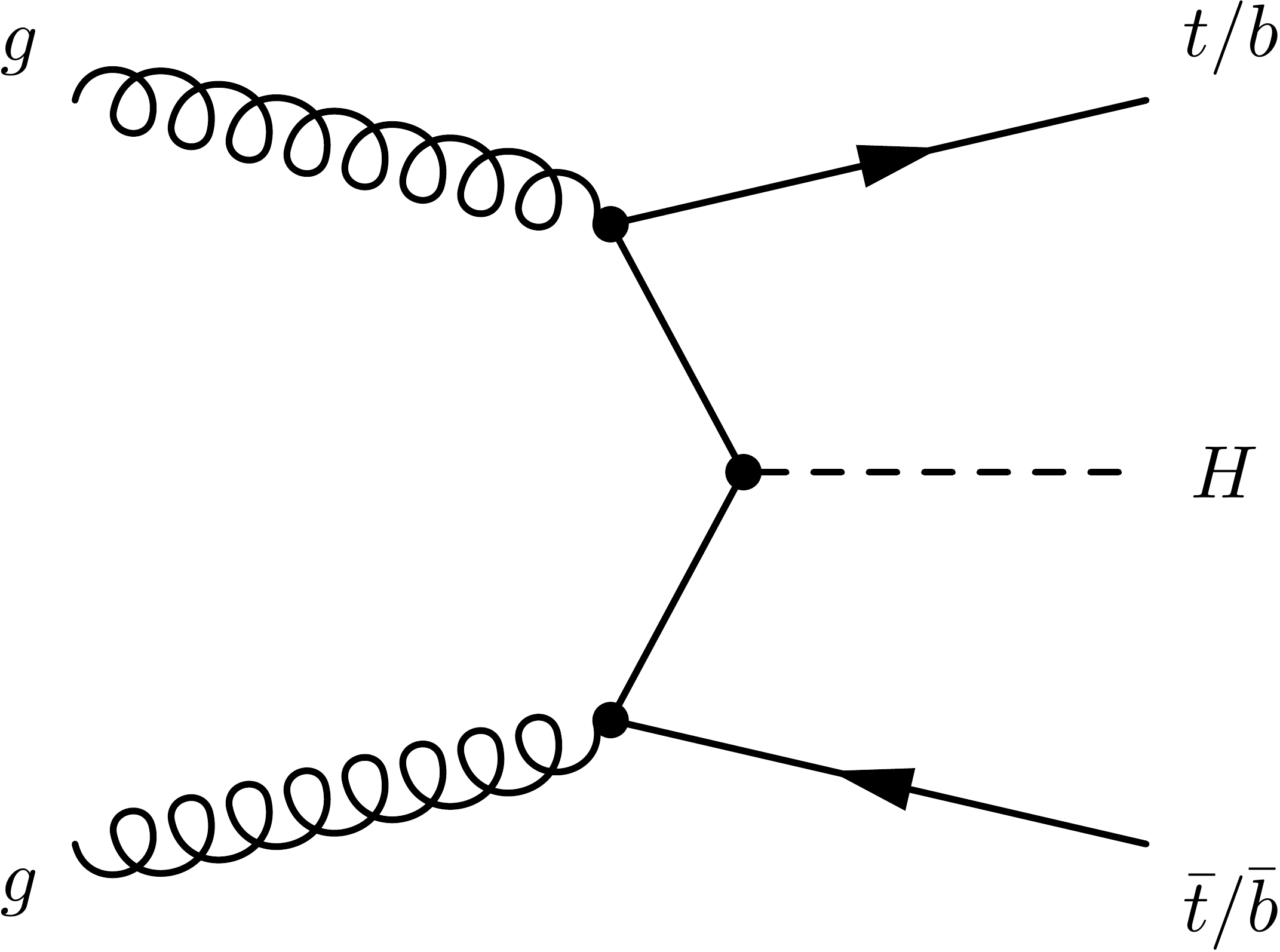}
\par\end{centering}
\caption{\label{fig:prodtt}Contribution to Higgs boson production from
(LHS)   associated  $Vh$ production and (RHS) $t {\overline t} h$ production.  In this figure, $H$ is the SM Higgs boson. }
\label{fig:hprodtt}
\end{figure}

At the LHC, the dominant production mechanisms are gluon fusion, followed by vector boson fusion,
shown in Fig. \ref{fig:prodgg}.  The associated production mechanisms of the Higgs with vector bosons or top quarks
have smaller rates, but  these channels are theoretically important and are shown in Fig. \ref{fig:prodtt}.
It is immediately apparent that gluon fusion and $t {\overline t}\hsm$  production are sensitive to the top quark Yukawa
coupling, while vector boson fusion and associated $\hsm V$, $V=(W,Z)$, production probe the gauge-Higgs couplings. 
 
\begin{figure}
\begin{centering}
\includegraphics[width=0.35\textwidth]{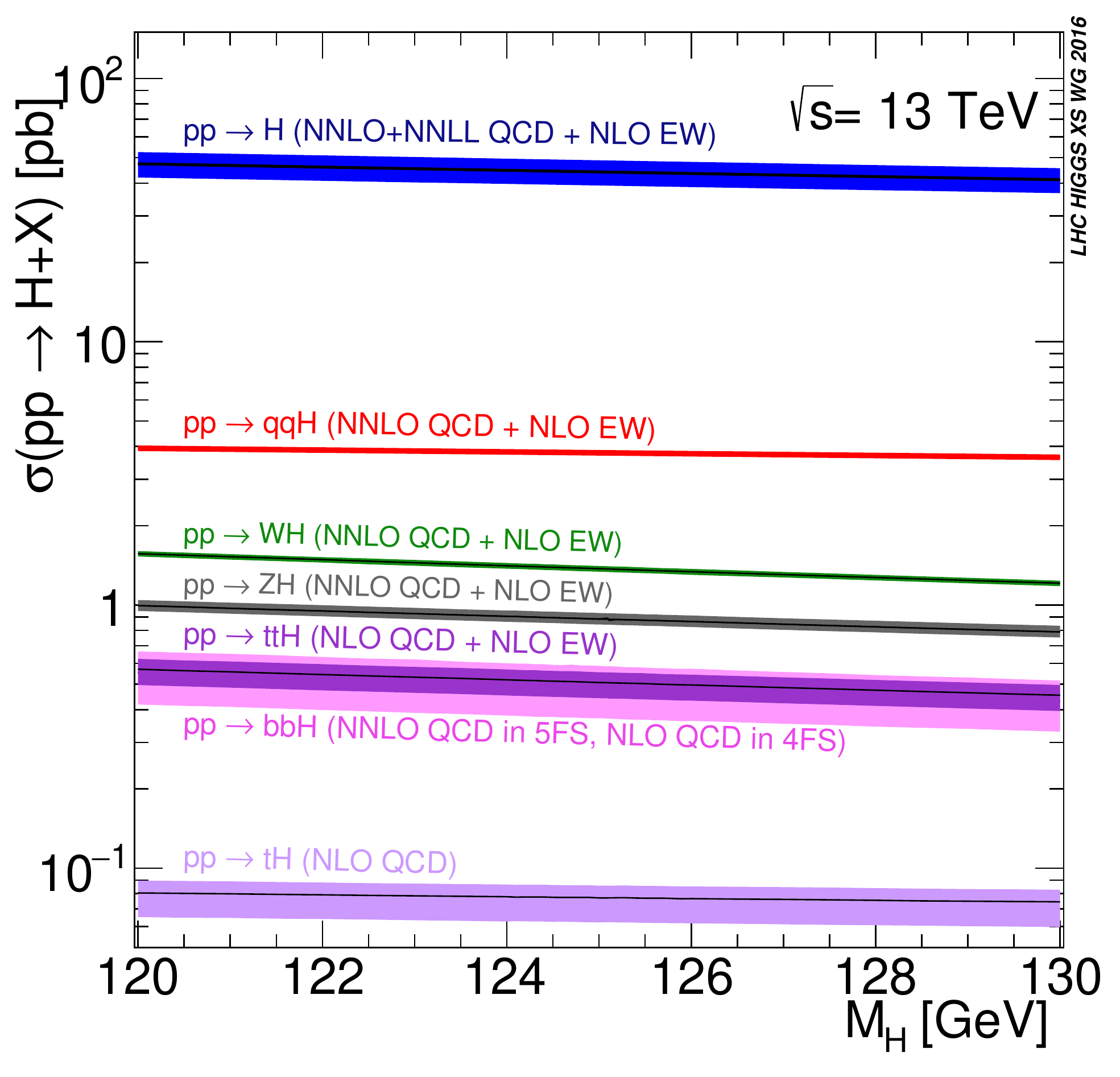}
\includegraphics[width=0.35\textwidth]{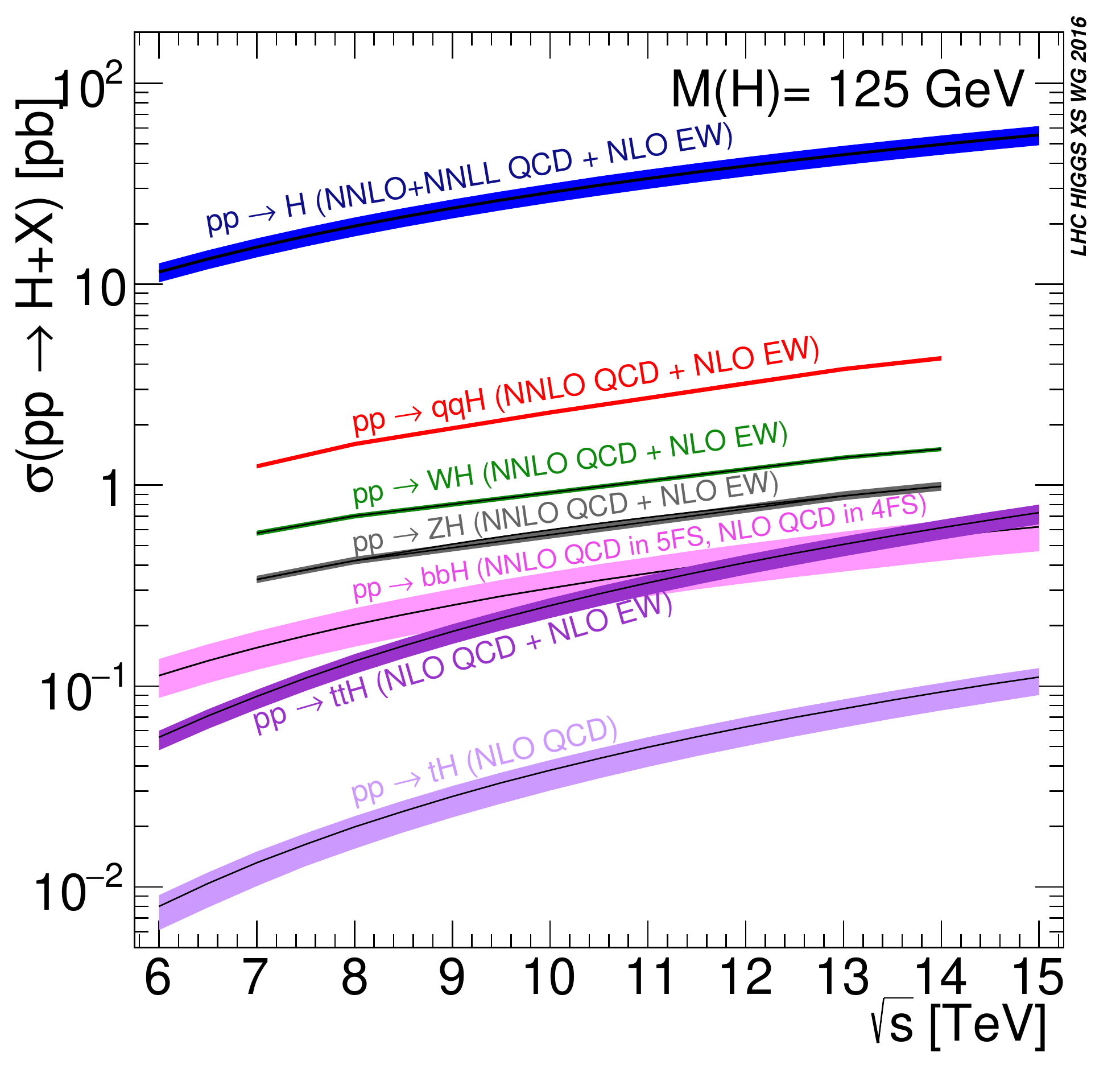}
\par\end{centering}
\caption{\label{fig:higgs_sig}
Total Higgs  production cross sections\cite{deFlorian:2016spz}. In this figure, $H$ is the SM Higgs boson. }
\end{figure}

The total rates for Higgs production in various channels are shown on the LHS of Fig. \ref{fig:higgs_sig} for arbitrary 
Higgs mass at $13~TeV$ (LHS) and as a function of center-of-mass energy (RHS) for the physics Higgs mass. 
 The curves include the most up-to-date
theoretical calculations, and the width of the curves represents an estimate of the uncertainties.  We will discuss
each production channel in turn in this section\cite{deFlorian:2016spz}. 
\subsubsection {$gg\rightarrow \hsm$}
The primary production mechanism for a Higgs boson
in hadronic collisions is through the couplings to heavy fermions, $g g \rightarrow \hsm$,
which is shown  on the LHS of Fig. \ref{fig:prodgg}. This process is dominated by the top quark loop and the
loop with a bottom quark contributes roughly $-5\%$ to the SM cross section. 

The  lowest order (LO)  amplitude for $g^{A,\mu}(p)+g^{B,\nu}(q)\rightarrow \hsm$ from a quark
of mass $m_q$ in the loop is, 
\begin{eqnarray}
A^{\mu\nu}(g^Ag^B\rightarrow \hsm)&=&
-{\alpha_s m_q^2 \over \pi v} \delta_{AB}
\biggl(g^{\mu\nu}{\mhsm^2\over 2}-p^\nu q^\mu\biggr)
\nonumber \\
&&\cdot \int_0^1 dx\int_0^{1-x} dy \biggl({1-4xy\over m_q^2-\mhsm^2 x y}\biggr)\epsilon_\mu(p)\epsilon_\nu(q)
\nonumber \\
&=&{\alpha_s  \over 4 \pi v} \delta_{AB}
\biggl(g^{\mu\nu}{\mhsm^2\over 2}-p^\nu q^\mu\biggr)F_{1/2}(\tau_q)\epsilon_\mu(p)\epsilon_\nu(q)
\nonumber \\
&\rightarrow &-{\alpha_s  \over 3 \pi v} \delta_{AB}
\biggl(g^{\mu\nu}{\mhsm^2\over 2}-p^\nu q^\mu\biggr)\epsilon_\mu(p)\epsilon_\nu(q)\qquad {\hbox{if}}~m_q>>\mhsm
\, .
\label{eq:ggans}
\end{eqnarray}

The partonic cross section can be found from the general resonance formula,
\begin{equation}
{\hat{\sigma}}(gg\rightarrow \hsm)={16\pi^2\over\mhsm}(2J+1){1\over 64}\cdot{1\over 4}\cdot 2 \Gamma(\hsm\rightarrow gg)
\delta(s-\mhsm^2)\, ,
\end{equation}
where the factors of ${1\over 64}$ and ${1\over 4}$ are the color and spin averages, $J=0$ is the 
Higgs spin, $s$ is the $gg$ partonic sub-energy, and the factor of $2$ undoes
the identical particle factor of ${1\over 2}$ in the decay width $\Gamma(\hsm\rightarrow gg)$.
 The lowest order partonic  cross section
for $gg\rightarrow \hsm$ is,
\beqn
{\hat \sigma}(gg\rightarrow \hsm)&= &
{\alpha_s^2\over 1024 \pi v^2}
\mid \sum_q  F_{1/2}(\tau_q)\mid^2\delta    
\biggl(1-{s\over \mhsm^2}\biggr)
\nonumber \\
&\equiv &{\hat \sigma}_0(gg\rightarrow h)\delta
\biggl(1-{s\over \mhsm^2}\biggr)
\, .
\label{eq:sigdef}
\eeqn     
In the heavy quark
limit, the cross
section is independent of the top quark mass and  becomes a constant,
\beq
{\hat \sigma}_0(gg\rightarrow h)\sim {\alpha_s^2\over
576 \pi v^2}      \,.
\label{eq:siginf}
\eeq
The heavy fermions do not decouple at high energy and the gluon fusion rate essentially counts
the number of SM-like  chiral quarks. 

The Higgs boson production cross section at a hadron collider can be 
found by integrating the partonic cross section,
$\sigma_0(pp\rightarrow h)$,
 with the gluon
parton distribution functions, $g(x,\mu)$,
\beq
\sigma(pp\rightarrow h)={\hat \sigma_0}z\int_z^1
{dx\over x} g(x,\mu)g\biggl({z\over x},\mu\biggr),
\label{logh}
\eeq
where $\sigma_0$ is given in Eq. \ref{eq:sigdef},
 $z\equiv \mhsm^2/S$, $\mu$ is the factorization scale and $S$ is
the hadronic center of mass energy.   It is particularly interesting to consider the theoretical accuracy 
at $N^3LO$\cite{Anastasiou:2016cez},
\begin{equation}
\sigma(pp\rightarrow \hsm)[13~\tev]=48.58^{+4.6\%}_{-6.7\%} (theory) \pm 3.2\% (PDF+\alpha_s)\, ,
\end{equation} 
where the theory uncertainty arises predominantly from the scale choice and the PDF+$\alpha_s$
uncertainty is the PDF and correlated uncertainty on $\alpha_s$. 

\begin{figure}
\begin{centering}
\includegraphics[width=0.5\textwidth]{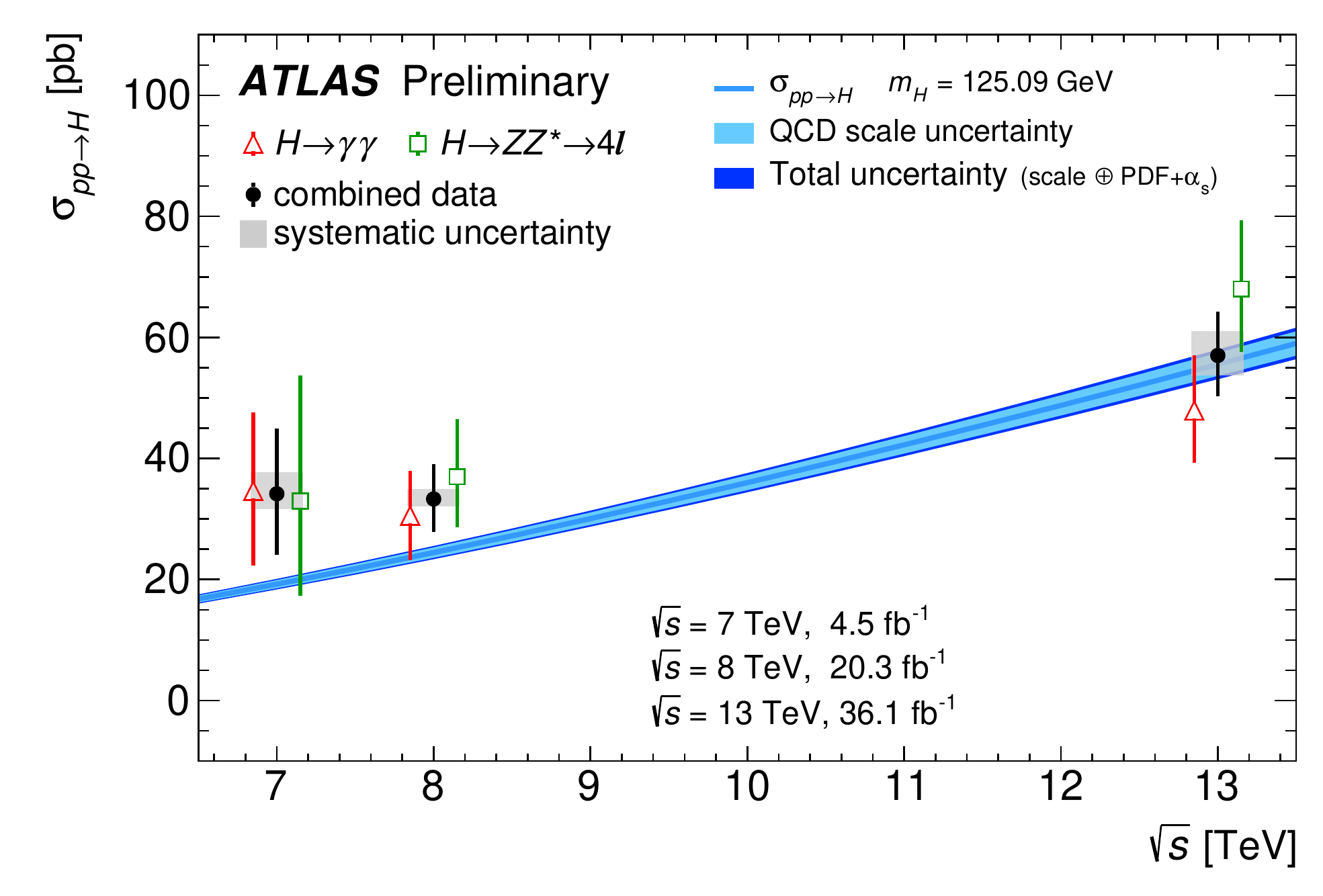}
\par\end{centering}
\caption{\label{fig:atsig} ATLAS measurements of the gluon fusion Higgs cross section, compared
to theory predictions\cite{ATLAS-CONF-2017-047} In this figure, $H$ is the SM Higgs boson mass. }
\end{figure}

The measured Higgs rate immediately rules out the possibility of a $4^{th}$ generation of SM chiral fermions.  Imagine that
there are heavy fermions, ${\cal{T}}$ and ${\cal {B}}$, with  identical quantum numbers as the SM top and bottom
quarks .  The new fermions would
contribute to Higgs production from gluon fusion as  on the LHS of  Fig. \ref{fig:prodgg}.
From Eq. \ref{eq:siginf}, we would have,
\beqn
{\hat \sigma}_0(gg\rightarrow h)&\rightarrow& {\alpha_s^2\over
576 \pi v^2}\biggl[1+1+1\biggr]^2 \nonumber \\
&\rightarrow& 9{\hat\sigma_0}(SM)\, ,
\eeqn
where the factors in the square bracket represent the contributions of  the SM $t,{\cal{T}}$ and 
${\cal{B}}$.  This is obviously excluded by
the measured rate for gluon fusion Higgs production, which is in good agreement with the SM prediction,
shown in Fig. \ref{fig:atsig}.

The tensor structure of Eq. \ref{eq:ggans} is exactly that required for the production of a spin-$0$ particle
from $2$-gluons with momentum, $g(k_1)$ and $g(k_2)$.
Starting from  a $G_{\mu\nu}G^{\mu\nu}$ term in the Lagrangian and considering only the Abelian
contributions for now,
\beqn
G_{\mu\nu} G^{\mu\nu}&\rightarrow & (\partial_\mu G_\nu-\partial _\nu G_\mu)
(\partial^\mu G^\nu-\partial ^\nu G^\mu)\, .
\eeqn
Making the replacement $\partial _\mu\rightarrow i k_\mu$,
\beqn
G_{\mu\nu} G^{\mu\nu}&\rightarrow &-(k_{1\mu} G_{1\nu}-k_{1\nu}G_{1\mu})
(k_{2}^{\mu} G_{2}^{\nu}-k_{2}^{\nu}G_{2}^{\mu})\nonumber \\
&=&-2\biggl(k_1\cdot k_2 G_1\cdot G_2-k_1\cdot G_2 k_2\cdot G_1\biggr)\nonumber \\
&=&-2 k_1\cdot k_2
G_{1\mu}G_{2\nu}\biggl[g^{\mu\nu}-{k_1^\nu k_2^\mu\over k_1\cdot k_2}\biggr]\, .
\label{eq:exp}
\eeqn
Comparing Eqs. \ref{eq:ggans} and \ref{eq:exp}\footnote{The extra factor of ${1\over 2}$ comes from 
the neglected color factor, $Tr(T^A T^B)={1\over 2}\delta_{AB}$.} suggests that the heavy quark limit for  the gluon fusion
production of a Higgs boson  can
be obtained from the effective dimension-$5$ Lagrangian
\beq
L_{EFT}={\alpha_s\over 12\pi} {\hsm\over v} G_{\mu\nu}^A G^{\mu\nu A}\, .
\label{eq:eft}
\eeq

\begin{figure}
\begin{centering}
\includegraphics[width=0.45\textwidth]{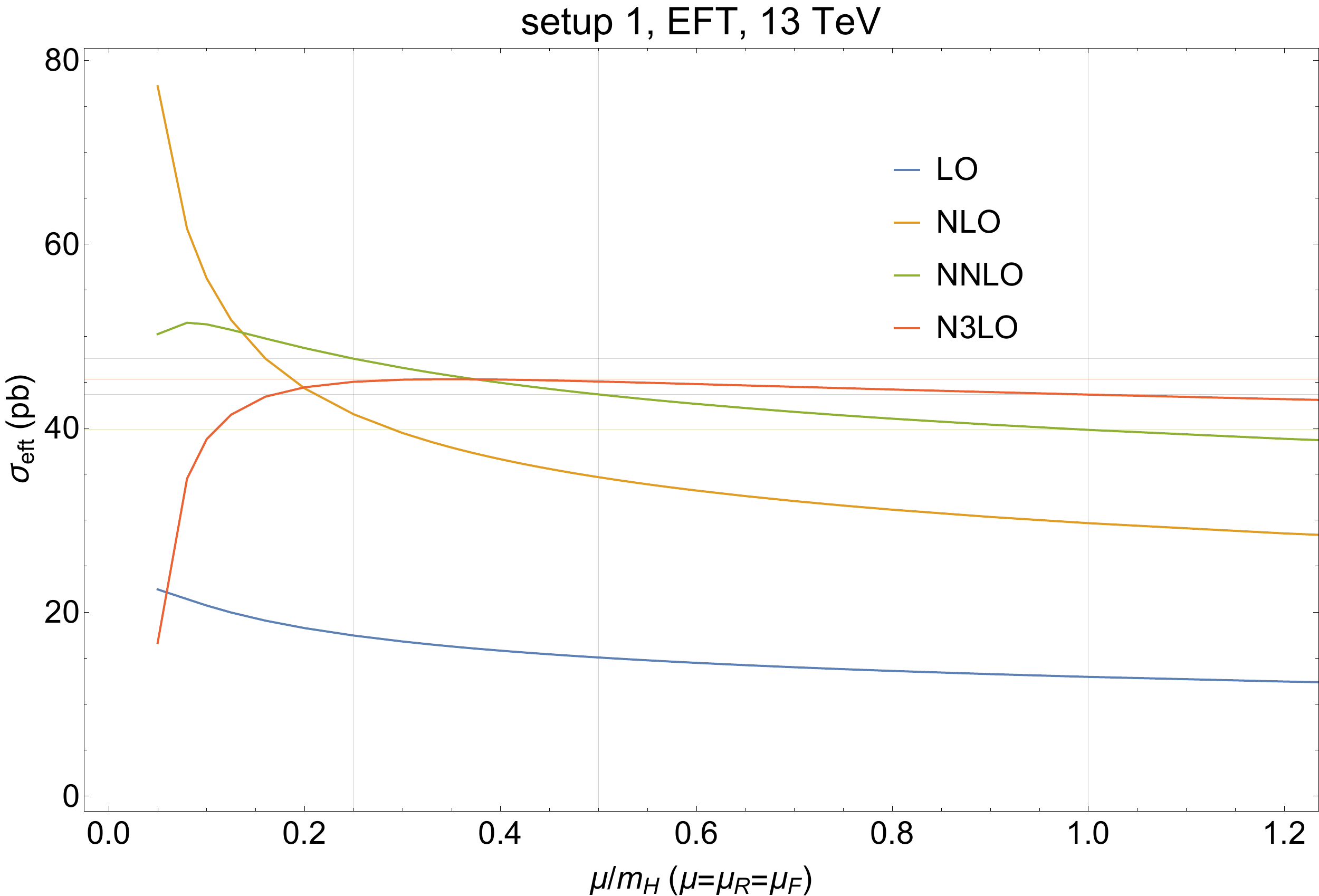}
\par\end{centering}
\caption{\label{fig:ggqcd}QCD corrected rate for gluon fusion as a function of the factorization and renormalization scale\cite{Anastasiou:2016cez}. }
\end{figure}

The effective Lagrangian of Eq. \ref{eq:eft} has been used to calculate the QCD corrections to gluon fusion
to NLO, NNLO, and N$^3$LO\cite{Anastasiou:2016cez}.  The result is shown in Fig. \ref{fig:ggqcd}.  Note that there
is a large correction (approximately a factor of $2$) going from LO to NLO.  The corrections at each order remain
sizable and the dependence on the factorization scale, $\mu$ is reduced at higher order. 

\noindent {\bf{\it{\bf Aside:}} {\it{Vector-like Fermions and Gluon Fusion}}}

The agreement of the measured rate for gluon fusion with the SM rate does not mean that all heavy fermions are excluded. 
A vector-like fermion is defined to have identical $SU(2)_L \times U(1)_Y$ transformation properties for both the left-and right-handed components.
 The simplest possibility is to add a fermionic top partner, $T$, for which both the left- and right- handed components
are weak singlets and color triplets\footnote{Recall that left- and right-handed fermions  contribute with opposite signs to anomalies and so the contributions cancel.  Therefore, it is not required to have a full generation of vector-like fermions to cancel
anomalies.}. In this scenario, the top partner can have a Dirac mass, (which has nothing to do with electroweak
symmetry breaking), and can mix with the SM top quark, $t$.   
The most general Yukawa interaction for a top partner
singlet is\cite{AguilarSaavedra:2002kr,Dawson:2012di}
\beqn
-L_Y&\sim & \lambda_t {\overline q}_L {\tilde \Phi} t_R +\lambda_2{\overline q}_L {\tilde \Phi} T_R+\lambda_3
{\overline T}_L t_R+\lambda_4 {\overline T}_L T_R\, ,
\eeqn
corresponding to the fermion mass matrix,
\begin{equation}
M^t=\left(
\begin{matrix}\lambda_t{v\over\sqrt{2}}&\lambda_2{v\over\sqrt{2}}\\
\lambda_3&\lambda_4\end{matrix}
\right)
\, .
\label{eq:vecmas}
\end{equation}

The mass eigenstates of charge ${2\over 3}$ ($t_1$ and $t_2$, with masses $m_{t1}$ and $m_{t2}$) are found through the
rotations,
\begin{equation}
\left( \begin{matrix}
t_{1,L,R}\\
t_{2,L,R}\end{matrix}\right)
\equiv U_{L,R}
\left(\begin{matrix}
t_{L,R}\\T_{L,R}
\end{matrix}
\right)\, .
\label{fg:chit}
\end{equation}
The matrices $U_{L,R}$ are unitary and can be parameterized as 
\begin{eqnarray}
U_L&=&
\left(\begin{matrix}
\cos\theta_L& -\sin\theta_L\\
\sin\theta_L& \cos\theta_L\end{matrix}
\right),\quad
U_R=
\left(\begin{matrix}
\cos\theta_R & -\sin\theta_R\\
\sin\theta_R & \cos\theta_R\end{matrix}
\right) \, .
\end{eqnarray}

The physical charge ${2\over 3}$ particles, $t_1$ and $t_2$, are therefore mixtures of $t$ and $T$,
\beqn
t_{1(L,R)}&=&\cos\theta_{L,R} t_{L,R}-\sin\theta_{L,R} T_{L,R}\nonumber \\
t_{2(L,R)}&=&\sin\theta_{L,R} t_{L,R}+\cos\theta_{L,R} T_{L,R}\, ,
\eeqn
The important point is that the couplings to the Higgs boson are changed in models with vector-like fermions,
\beqn
L_h&\rightarrow&-{m_{t1} \over v}\cos^2\theta_L {\overline{t}}_{1,L} t_{1,R}\hsm
-{m_{t2} \over v}\sin^2\theta_L{\overline{t}}_{2,L}t_{2,R}\hsm
\nonumber \\
&&
-
{m_{t2}\over v}\cos\theta_L\sin\theta_L{\overline{t}}_{1,L} t_{2,R}\hsm
 -
{m_{t1}\over v}\cos\theta_L\sin\theta_L{\overline{t}}_{2,L}t_{1,R}\hsm+h.c.
\eeqn
In the large mass limit ($m_{t1}$, $m_{t2}>>\mhsm$), the top and top partner contributions to gluon fusion yield
\beqn
{\hat \sigma}_0(gg\rightarrow h)&\rightarrow & {\alpha_s^2\over
576 \pi v^2}\biggl[\cos^2\theta_L+\sin^2\theta_L\biggr]^2 +{\cal{O}}\biggl({\mhsm^2\over m_{t1}^2},{\mhsm^2\over m_{t2}^2}
\biggr) \nonumber \\
&\rightarrow &{\hat\sigma_0}(SM)\, .
\eeqn
This equivalence with the SM gluon fusion rate in models with heavy vector- like fermions is a general 
feature\cite{Azatov:2011qy,Dawson:2012mk}. Observing the effects of top partners in single Higgs rates will be difficult, 
and instead  models with vector-like fermions are best probed by searches
for direct production of the new heavy quarks.  

\noindent {\it{\bf Aside:} Low Energy Theorems}
\label{sec:lowenergy}

\begin{figure}
\centering
\includegraphics[width=.5\textwidth]{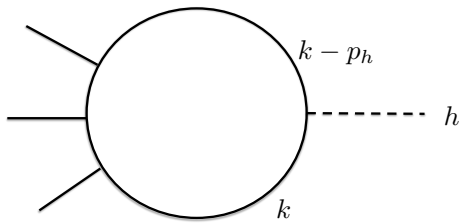}
\caption{General Higgs coupling to fermions. On the left-hand side is any initial or final state $X$.}
\label{figlet}
\end{figure}

We have seen that both in the SM and in the vector-like top partner singlet model, the gluon fusion contribution
to Higgs production takes a simple form in the heavy fermion mass limit.  
The idea that the Higgs gluon interactions due to heavy particles can  be derived from an effective Lagrangian 
as in Eq. \ref{eq:eft}  gives
rise to low energy theorems for Higgs-gluon couplings\cite{Kniehl:1995tn,Low:2010mr,Falkowski:2007hz}.  
Consider the Higgs coupling to a heavy fermion with mass $m$ as part of a complicated Feynman diagram as shown in 
Fig. \ref{figlet}.
The  sub-amplitude from the fermion-Higgs coupling can be written in the limit $p_\hsm\rightarrow 0$ as, 
\beqn
(...){i\over k-m}{-im\over v} {i\over k-p_h-m}(...)&\rightarrow &(...){i m\over v}\biggl({1\over k-m}\biggr)^2(....)
\nonumber \\
&=&(...){i m\over v} {\partial \over m} \biggl({1\over k-m}\biggr)(...)\, .
\end{eqnarray}
This  observation has been formalized to a theorem,
\beq
lim(p_h\rightarrow 0)A(\hsm X)={m\over v}{\partial \over \partial m}A(X)\, ,
\label{eq:let}
\eeq
where $A(X)$ is the amplitude for creating a state $X$. 
That is, adding a Higgs boson to a diagram is equivalent to taking the derivative with respect to the
heavy fermion 
mass\footnote{Identical reasoning holds for the coupling of a Higgs boson to gauge bosons.}.  It is 
important to note that the derivative is with respect to the unrenormalized mass.  At higher orders, there
are contributions from $\partial m_R/\partial m$, where $m_R$ is the renormalized mass. 

Let us apply Eq. \ref{eq:let} to the gluon 2-point function,
\beq
L=-{1\over 4 g_s^2}G_{\mu\nu}^{\prime A} G^{\prime \mu\nu A}\, ,
\eeq
where we have factored all coupling constant dependence out of the gluon field strength, 
$G^{\prime~A}_{\mu\nu}g_s\equiv G^A_{\mu\nu}$, for convenience.
Applying the low energy theorem of Eq. \ref{eq:let},
\beq
L_{EFT}=-{1\over 4}{\hsm\over v} \biggl[m{\partial\over \partial m}{1\over g_s^2}\biggr]
G_{\mu\nu}^{\prime A} G^{\prime \mu\nu A}\, . 
\label{eq:efttop}
\eeq
The dependence of $g_s$ on scale is given by the  QCD $\beta$ function,
\beq
\beta\equiv \mu{\partial g_s\over \partial \mu}\, .
\eeq
where only the heavy top quark contributes to the $\beta$ function here,
\beq
{\beta\over g_s}\rightarrow {\alpha_s\over 6 \pi}\, .
\eeq
The effective Lagrangian of Eq. \ref{eq:eft} follows immediatley.   

The low energy theorem is more than just a curiosity.  The effective field theory (EFT)  of Eq. \ref{eq:eft} has been used to calculate
 radiative corrections to Higgs production  in the large $\mt$ limit at NLO\cite{Dawson:1990zj,Graudenz:1992pv}, 
 NNLO\cite{Anastasiou:2002yz}, and N$^3$LO\cite{Anastasiou:2016cez}.  For the NLO corrections to $gg\rightarrow\hsm$, the $2$-loop virtual corrections in
the full theory  become 
$1$-loop calculations using the  EFT and so on.  This greatly reduces the complexity of the problem.  At NNLO, the validity of the EFT
has been checked numerically in the exact (top mass dependent)
 theory  by expanding the propagators in the large  top quark mass limit, and the agreement is within a few percent\cite{Harlander:2009mq,Pak:2009bx}. 
Practically speaking, the higher order results obtained using the EFT are typically used to rescale   the LO (or NLO)  kinematic
distributions obtained by  including the
full top quark mass dependence.    

The low energy theorem is particularly useful for estimating the effects of BSM physics on the gluon
fusion rate\cite{Low:2010mr}.
Consider, for example, a model with
multiple heavy fermions, ${\cal F}_i$, where  the interactions in the mass basis are,
\beq
L\sim \Sigma_i {\overline {\cal{F}}}_i{\tilde Y}_i(\hsm+v){\cal {F}}_i\,,
\eeq
with the fermion masses given by $m_i=v{\tilde Y}_i$
Then the obvious generalization of the results of the previous section is,
\beq
L_{EFT}={\alpha_s\over 12 \pi}{\hsm}\Sigma_i
{{\tilde{Y}}_i\over m_i}G_{\mu\nu}^AG^{\mu\nu A}\, .
\label{eq:fermlet}
\eeq
In general, however, the fermion-Higgs couplings are specified in the gauge basis, and it is quite a 
bit of work to obtain the results in the  mass basis.  This step can be eliminated using the low energy theorems.  We can start
from the general interactions of the fermions in the gauge basis, and diagonalize the  fermion mass matrix, $M$, using a unitary matrix, $U$,
where
$M_D=U^\dagger M U$,
\beqn
L&\sim& \Sigma_{ij}{\overline f}_iY_{ij}(\hsm+v)f_j\nonumber \\
&=&{\overline f}U U^\dagger Y U U^\dagger f (\hsm+v)\, ,
\eeqn
where the diagonal mass matrix is $M_D=vU^\dagger Y U$ and $M=Yv$ ($M$, $M_D$, ${\tilde {Y}}$ and $Y$ are now
all interpreted as matrices).  The Higgs couplings to gluons are determined,
\beqn
R_g&\equiv& \Sigma_i{{\tilde Y}_i\over m_i}
\nonumber \\
&=& \Sigma_i\biggl({U^\dagger Y U\over m}\biggr)_{ii} =Tr(U^\dagger Y U M_D^{-1})=Tr(YUM_D^{-1}U^\dagger)\, .
\eeqn
Using the matrix identity $M^{-1}M=M^{-1}UM_DU^\dagger=1$, 
\beqn
R_g&=&Tr(Y[M^{-1}UM_DU^\dagger][UM_D^{-1}U^\dagger])\nonumber \\
&=&Tr(YM^{-1})\nonumber\\
&=& Tr\biggl({\partial M\over \partial v}M^{-1}\biggl)\nonumber\\
&=&{\partial\over \partial v}\log\biggl(det (M)\biggr)\nonumber\\
&=& {\partial\over \partial v}Tr\biggl(log(M)\biggr)\, .
\end{eqnarray}
The effective Lagrangian is finally  given by\cite{Falkowski:2007hz,Low:2010mr,Dawson:2012mk},
\beq
L_{EFT}={\alpha_s\over 12 \pi}{h\over v}\biggl({\partial\over \partial \log(v)}Tr(log M)\biggr)
G_{\mu\nu}^AG^{\mu\nu A}\, ,
\eeq
and there is no need to diagonalize the mass matrix.

\subsubsection{$p_T$ distribution of Higgs Bosons}
\label{sec:higgspt}

\begin{figure}
\begin{centering}
\includegraphics[width=0.5\textwidth]{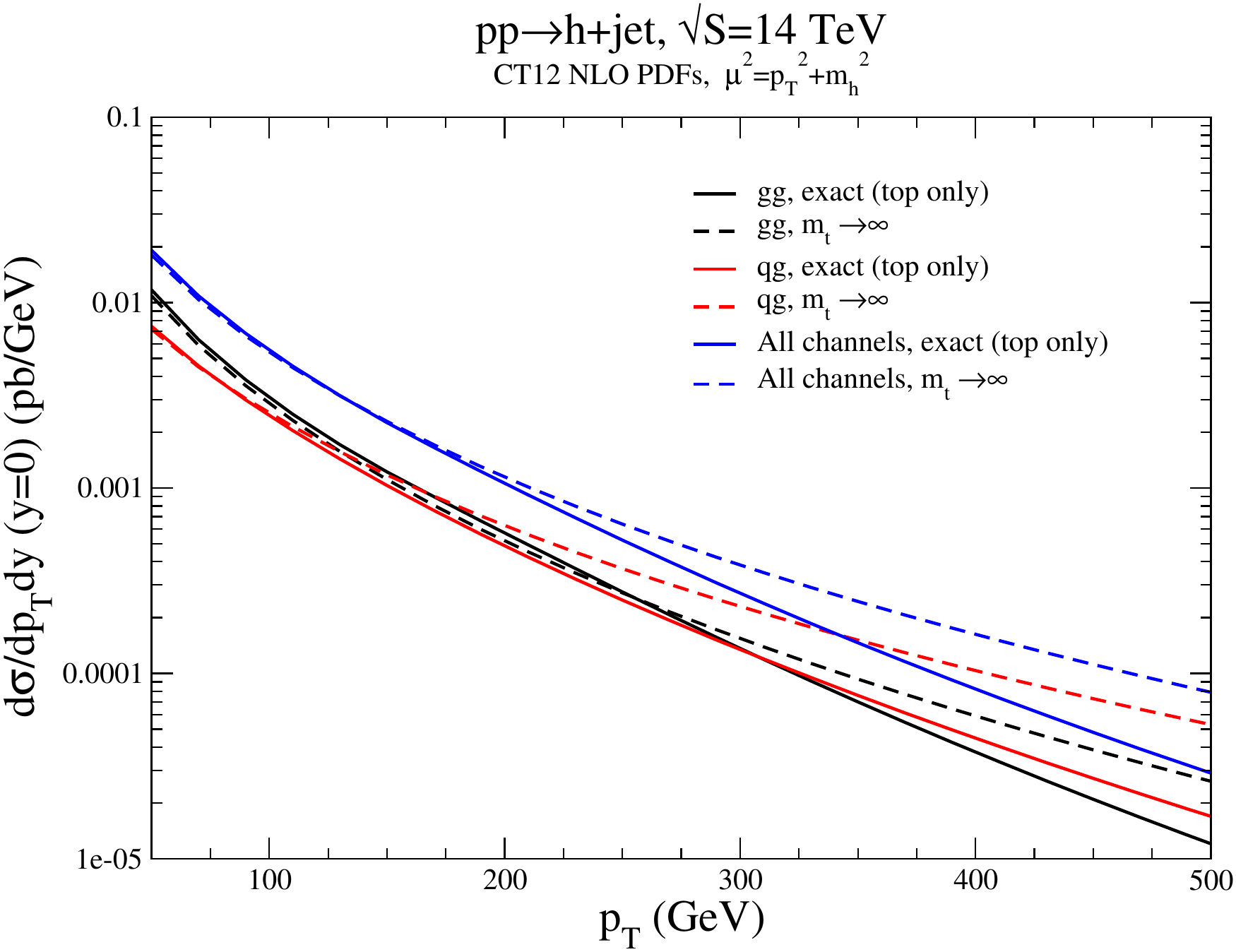}
\par\end{centering}
\caption{\label{fig:higgjet} Lowest order $p_T$ spectrum for Higgs plus jet production from Eq. \ref{eq:pth} and
the large $m_t$ approximation of Eq. \ref{eq:eft}. }
\end{figure}

At LO, the Higgs boson has no $p_T$ and a transverse momentum spectrum for the Higgs is first generated by the process, $gg\rightarrow g \hsm$,
which is an NLO contribution to the gluon fusion process\cite{Ellis:1987xu}.    As $p_T\rightarrow 0$,
the partonic cross section  for Higgs plus jet production diverges as $1/p_T^2$,
\beqn
{d{\hat\sigma}\over d { t}}(gg\rightarrow g \hsm)
&=&{\hat \sigma}_0{3 \alpha_s\over 2 \pi}\biggl\{
{1\over p_T^2}\biggl[
\biggl(1-{\mhsm^2\over {s}}\biggr)^4 + 1
+\biggl( {\mhsm^2\over { s}}\biggr)^4\biggr]
\nonumber \\
&& - {4\over { s}}\biggl(1-{\mhsm^2\over { s}}
\biggr)^2+{2 p_T^2\over {s}}\biggr\}\, ,
\label{eq:pth}
\eeqn
where ${\hat{\sigma}}_0$ is the LO $gg\rightarrow\hsm$ cross section given in Eq. \ref{eq:sigdef}, and $s, t$ and $u$ are the partonic 
Mandelstam invariants.    The $p_T$ spectrum for  Higgs plus jet at LO is shown in Fig. \ref{fig:higgjet}, where the
contributions from the $gg$ and $qg, {\overline{q}}g$ initial states are shown separately.  Also shown is the $m_t\rightarrow
\infty$ limit of the spectrum that is derived from the effective Lagrangian of Eq. \ref{eq:eft} .  The effective Lagrangian
approximation fails around $p_T\sim 2 m_t$. 
In this process, there are several distinct momentum scales ($p_T, \mhsm,\mt$), as opposed to gluon
fusion where there is only a single scale  ($\mhsm/m_t$) at LO.  The
expansion in ${\mhsm\over \mt}$  for $gg\rightarrow g\hsm$ receives corrections  of ${\cal{O}}({s\over \mt^2},{p_T^2\over \mt^2}) $
and for $p_T\gsim 2 \mt$, the EFT large top quark mass expansion cannot be used to obtain reliable distributions.

\begin{figure}
\begin{centering}
\includegraphics[width=0.5\textwidth]{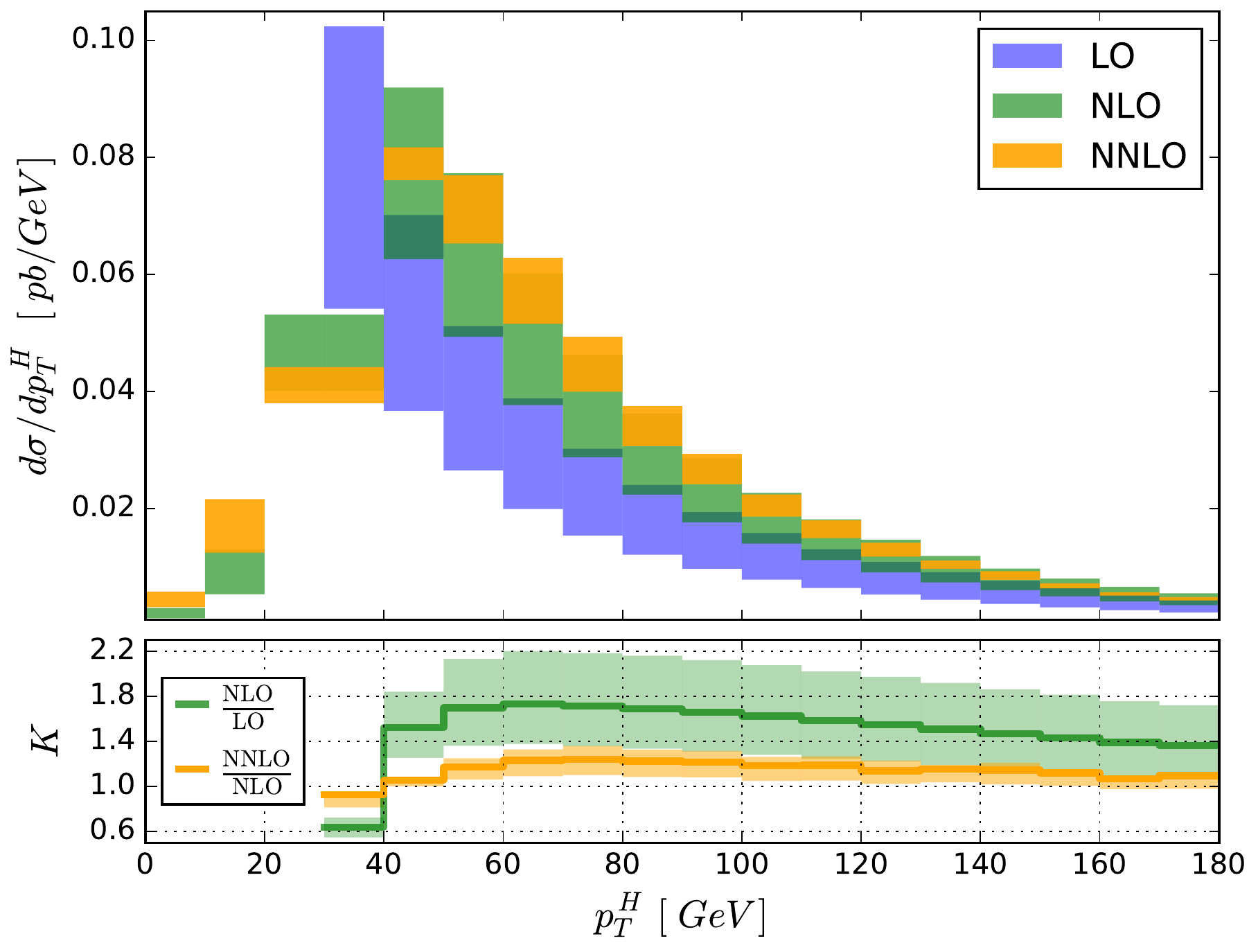}
\par\end{centering}
\caption{\label{fig:hpt}QCD corrected $p_T$ spectrum for Higgs plus jet production at $\sqrt{S}=8~TeV$\cite{Boughezal:2015aha,Boughezal:2015dra}. 
In this figure, $H$ is the SM Higgs boson. }
\end{figure}

NLO, NNLO, and N$^3$LO radiative corrections to Higgs plus jet production have been 
calculated\cite{Boughezal:2015aha,Boughezal:2015dra,Dulat:2017prg,Chen:2016vqn} using the 
$\mt\rightarrow \infty$ approximation.  The lowest order result of Eq. 
\ref{eq:pth} is then reweighted by a $K$ factor derived in the $\mt\rightarrow \infty$ limit for each
kinematic bin.  The effects of the higher order
corrections are significant  and increase the rate by a factor of around $1.8$ as shown in Fig. \ref{fig:hpt}.   The
 singularity of the LO result at $p_T=0$ is clearly visible in Fig. \ref{fig:hpt} and  we note that after the inclusion
of the NLO corrections, the $p_T$ spectrum no longer diverges as $p_T\rightarrow 0$.

\begin{figure}
\begin{centering}
\includegraphics[width=0.65\textwidth]{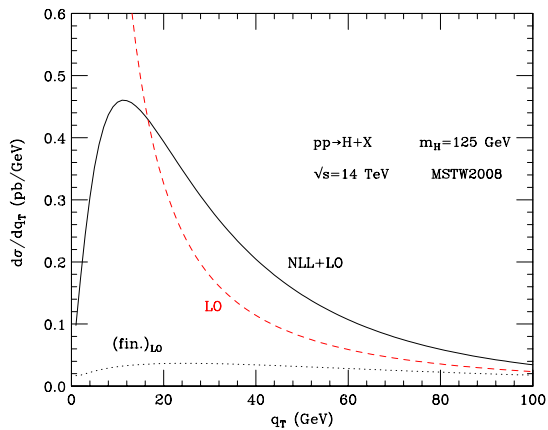}
\par\end{centering}
\vskip -1.5in
\caption{\label{fig:hptre} QCD NLL resummed $p_T$ spectrum for Higgs plus jet production at $\sqrt{S}=14~TeV$\cite{deFlorian:2011xf}.  In this
figure, $H$ is the SM Higgs boson.}
\end{figure}

\begin{figure}
\begin{centering}
\hskip -.2in
\includegraphics[width=0.4\textwidth]{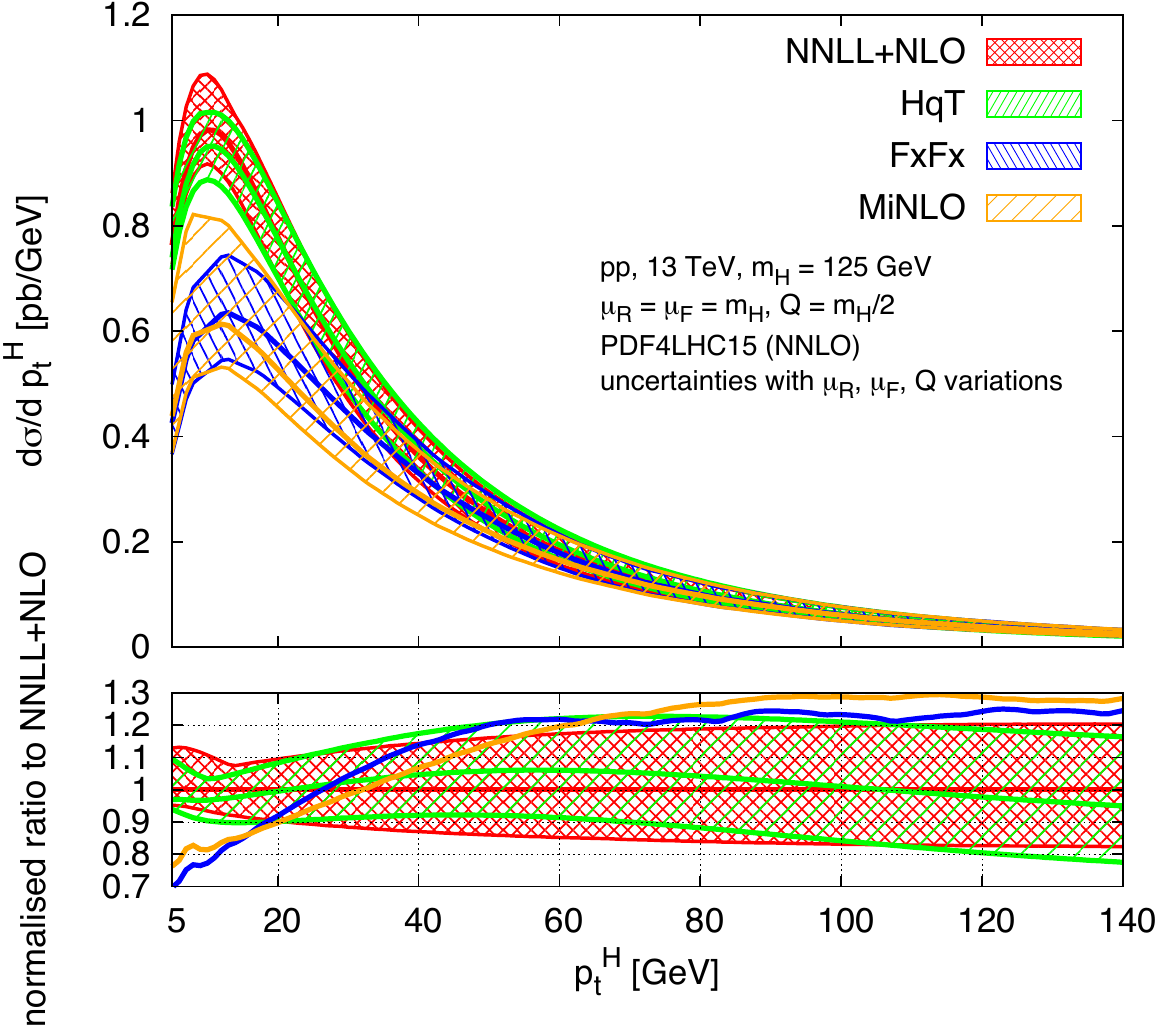}
\par\end{centering}
\caption{\label{fig:hptr2} NNLL QCD resummed $p_T$ spectrum  for Higgs plus jet production\cite{Monni:2016ktx}. In this figure, $H$ is the SM Higgs boson. }
\end{figure}

The terms which are singular as $p_T\rightarrow 0$ can
be isolated and the integrals performed explicitly.
Considering only the $gg$
initial state\cite{deFlorian:2011xf},
\beq
{d \sigma\over d p_T^2 d y} (pp\rightarrow g \hsm)\mid _{
p_T^2\rightarrow 0}\sim
{\hat \sigma}_0 { 3\alpha_s\over 2 \pi} {1\over p_T^2}
\biggl[
6 \log \biggl({\mhsm^2\over p_T^2}\biggr) - 2 \beta_0
\biggr]
g(z e^y)g (z e^{-y})+...
\eeq
where $z\equiv\mhsm^2/S$, $\beta_0=(33-2 n_{lf})/6$, and $n_{lf}=5$ is
the number of light flavors.  Clearly when $p_T << \mhsm$, the terms
containing the logarithms resulting from soft gluon emission can give a large 
numerical contribution.  The logarithms of the form $\alpha_s^n \log^m(\mhsm^2/p_T^2)$ 
can be resummed\cite{deFlorian:2011xf,Kauffman:1991cx} to  improve the theoretical
accuracy in the regime $p_T\rightarrow 0$, 
as can be seen in the curve labelled NLL+LO in  Fig. \ref{fig:hptre}.   Additional  logarithms can also
be resummed\cite{Monni:2016ktx}, as shown in the curve labelled NNLL+NLO  in  Fig.  \ref{fig:hptr2}.

\subsubsection{Measuring the Higgs width with $gg\rightarrow \hsm\rightarrow ZZ$}

\begin{figure}
\begin{centering}
\includegraphics[width=0.4\textwidth]{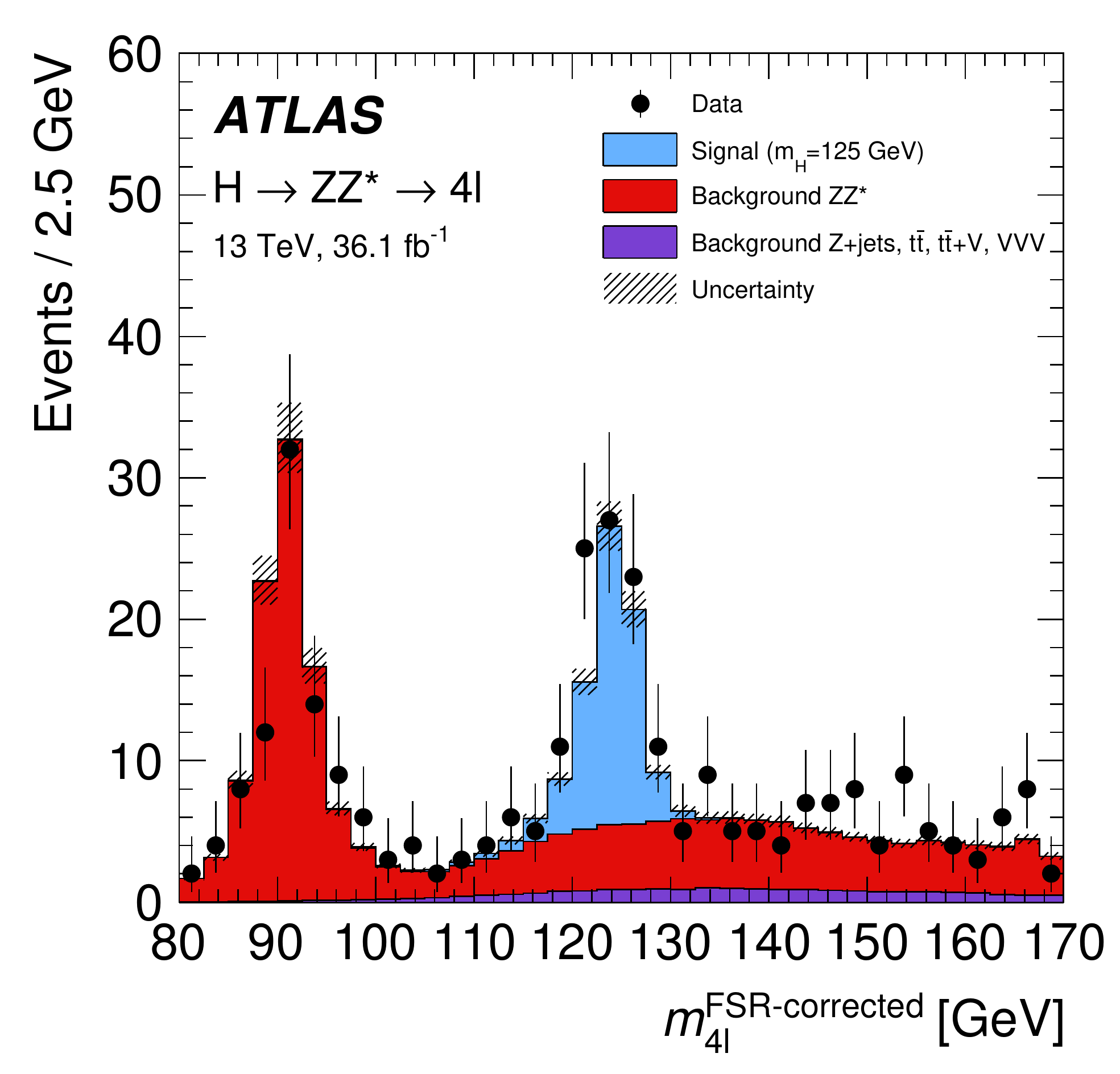}
\includegraphics[width=0.4\textwidth]{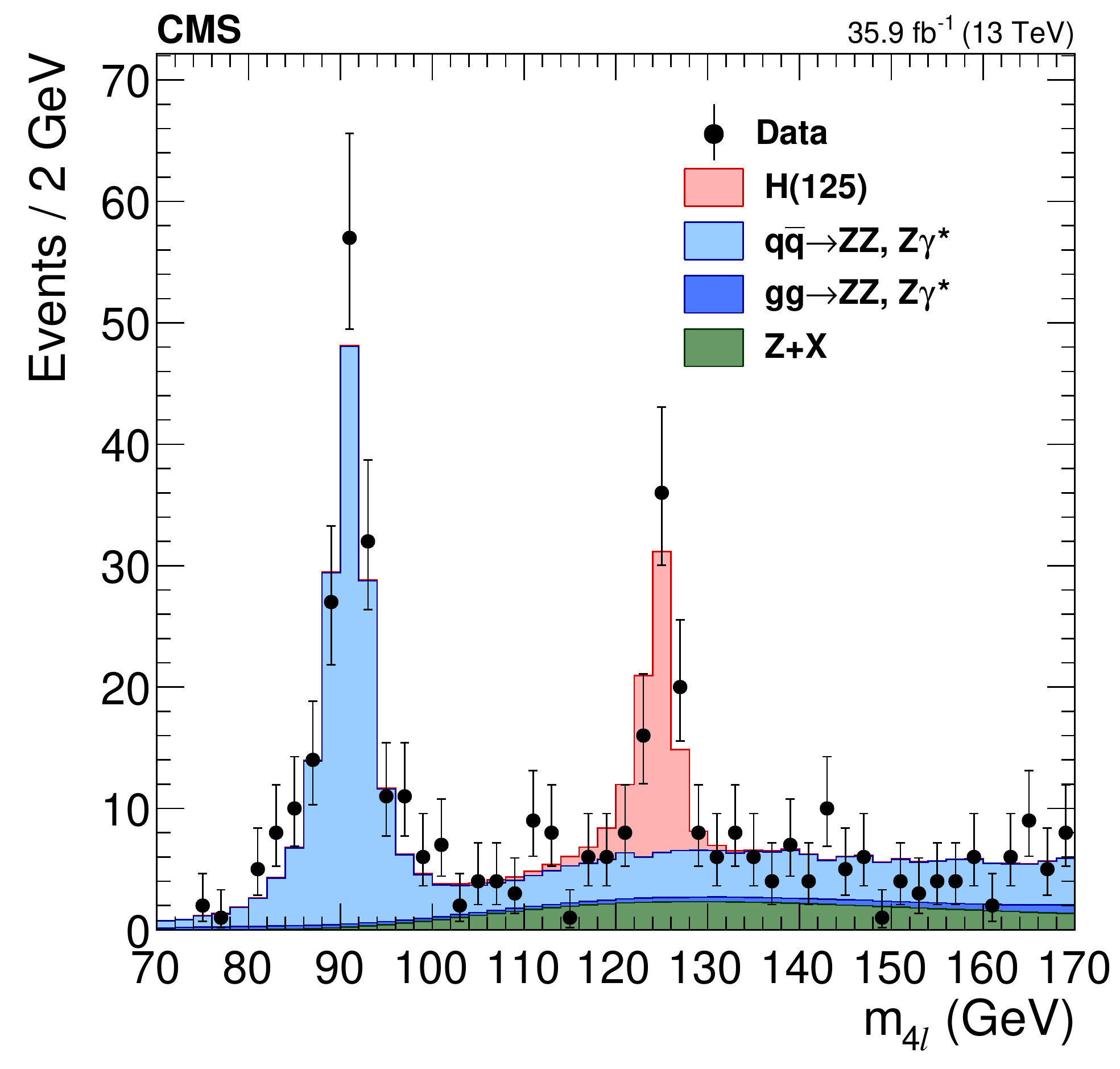}
\par\end{centering}
\caption{\label{fig:hzzplot}$h\rightarrow ZZ\rightarrow$ 4 lepton signal at $13~TeV$\cite{Aaboud:2017oem,Sirunyan:2017exp}.}
 \end{figure}

Gluon fusion with the subsequent Higgs decay to $ZZ\rightarrow 4$ leptons or $\gamma\gamma$ were
the Higgs discovery channels. The $h\rightarrow ZZ\rightarrow 4$ lepton signals
at 1$3~TeV$ are shown in Fig. \ref{fig:hzzplot} \cite{Aaboud:2017oem,Sirunyan:2017exp} and the Higgs resonance is clearly visible.  
Making a direct measurement  of the Higgs width  by fitting a Breit-Wigner  function to  the resonance shape
 is not possible since the detector resolution is ${\cal{O}}(1-2)~GeV$,
much larger than the Higgs width, $\Gamma_h\sim~4~MeV$ .

\begin{figure}
\centering
\hskip3in
\includegraphics[width=.8\textwidth]{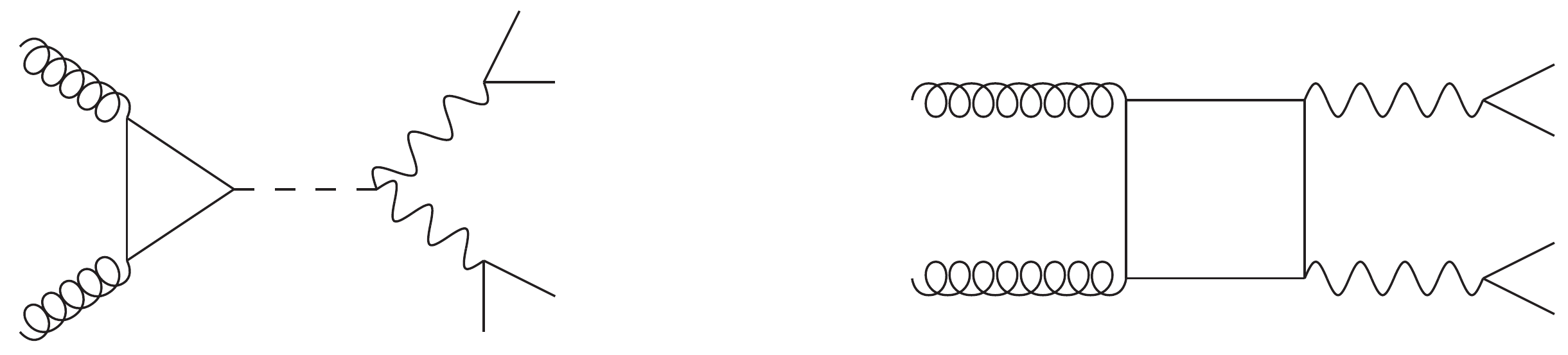}
\caption{Contributions to $gg\rightarrow ZZ\rightarrow 4l$. The dominant contributions to the triangle and box diagrams are from the top
quark. }
\label{fig:ggzzfig}
\end{figure}

 A clever
idea uses the properties of the longitudinal $Z$ polarizations\cite{Caola:2013yja,Kauer:2013qba}.   Consider the process $gg\rightarrow ZZ\rightarrow 4l$ 
shown in Fig. \ref{fig:ggzzfig}.
The Higgs contribution is shown  on the LHS of Fig. \ref{fig:ggzzfig} and  the partonic cross section from the
Higgs contribution alone is generically
given by,
\beq
{\hat \sigma}(gg\rightarrow\hsm\rightarrow 
ZZ)\sim 
\int  ds {\mid A(gg\rightarrow\hsm) \mid^2 \mid A(\hsm  \rightarrow ZZ)\mid^2\over (s-\mhsm^2)^2+\Gamma_h^2\mhsm^2}\, .
\label{eq:hzzdef}
\eeq
We allow the effective $gg\rightarrow \hsm$ and $\hsm\rightarrow ZZ\rightarrow $ couplings to be scaled  from
the SM values by arbitrary factors $\kappa_g(s)$ and $\kappa_Z(s)$, where
we explicitly note that the $\kappa$ factors can in principle depend on scale, 
\beq
\mid A(gg\rightarrow\hsm)\mid^2 \mid A(\hsm \rightarrow ZZ)\mid^2\sim \kappa_g^2(s)\kappa_Z^2(s)\mid \epsilon_{Z1}\cdot \epsilon_{Z2}\mid^2\, ,
\eeq 
where $\epsilon_{Zi}^\mu$ are the $Z$ polarization vectors. 

The interesting observation is that Eq. \ref{eq:hzzdef} behaves very differently above the Higgs resonance and near the resonance.
Above the resonance, $s>>\mhsm^2$, Eq. \ref{eq:hzzdef} becomes,
\beq
{\hat \sigma}(gg\rightarrow\hsm\rightarrow 
ZZ)^{above}\sim \int ds {\kappa_g^2(s)\kappa_Z^2(s) \mid \epsilon_{Z1}\cdot \epsilon_{Z2}\mid^2
\over s^2} 
\, .
\label{eq:hzzabove}
\eeq
For transverse polarizations, nothing particularly interesting happens, but because of the electroweak symmetry breaking
the longitudinally polarized $Z$ bosons have a novel feature.  Defining the momenta of the outgoing $Z$ bosons as $p_{Z1}$
and $p_{Z2}$ and remembering that the longitudinal polarization is approximately given by,
\beq
\epsilon_L^\mu (p_Z)\sim {p_Z^\mu\over M_Z}+{\cal O}\biggl({M_Z^2\over s}\biggr)\, ,
\eeq
we observe that  $\epsilon_L\cdot\epsilon_L\sim {p_{Z1}\cdot p_{Z2}\over M_Z^2}\sim{s\over M_Z^2}$. 
 Eq. \ref{eq:hzzabove} has the approximate form for $s>>m_h^2$,
\beq
{\hat \sigma}(gg\rightarrow\hsm\rightarrow 
Z_L Z_L)^{above}\sim \int ds {\kappa_g^2(s)\kappa_Z^2(s) \over M_Z^4}
\, .
\label{eq:hzzabove_final}
\eeq
We note that Eq.~\ref{eq:hzzabove_final} exhibits no dependence on the Higgs width.

Near the Higgs resonance, we can use the narrow width approximation, which amounts to the replacement,
\begin{equation}
{1\over
(s-\mhsm^2)^2+(\mhsm\Gamma_\hsm)^2}\rightarrow {\pi\over \mhsm\Gamma_\hsm}\delta(s-\mhsm^2)
\end{equation}
and Eq. \ref{eq:hzzdef} is approximately,
\beq
{\hat \sigma}(gg\rightarrow\hsm\rightarrow 
ZZ)^{on}\sim {\kappa_g^2(\mhsm^2)\kappa_Z^2(\mhsm^2) \over \mhsm\Gamma_h}\, .
\label{eq:hzzon}
\eeq

The idea is that by measuring the $gg\rightarrow 
\hsm\rightarrow ZZ$ rate above and on the resonance, information can be extracted about the Higgs width.
Assuming the $\kappa$ factors do not depend on scale,  
\beq
\Gamma_h\sim{{\hat\sigma}^{above}\over {\hat\sigma}^{on}}\, .
\label{eq:kap}
\eeq

At $8~TeV$, approximately $15\%$ of the cross section has 
$m_{4l}>140~GeV$, so this is a promising idea.  If the $\kappa$ factors have an energy dependence, they do 
not cancel in Eq. \ref{eq:kap} and the interpretation of the measurement becomes more complicated. 

\begin{figure}
\centering
\includegraphics[scale=0.3,angle=270]{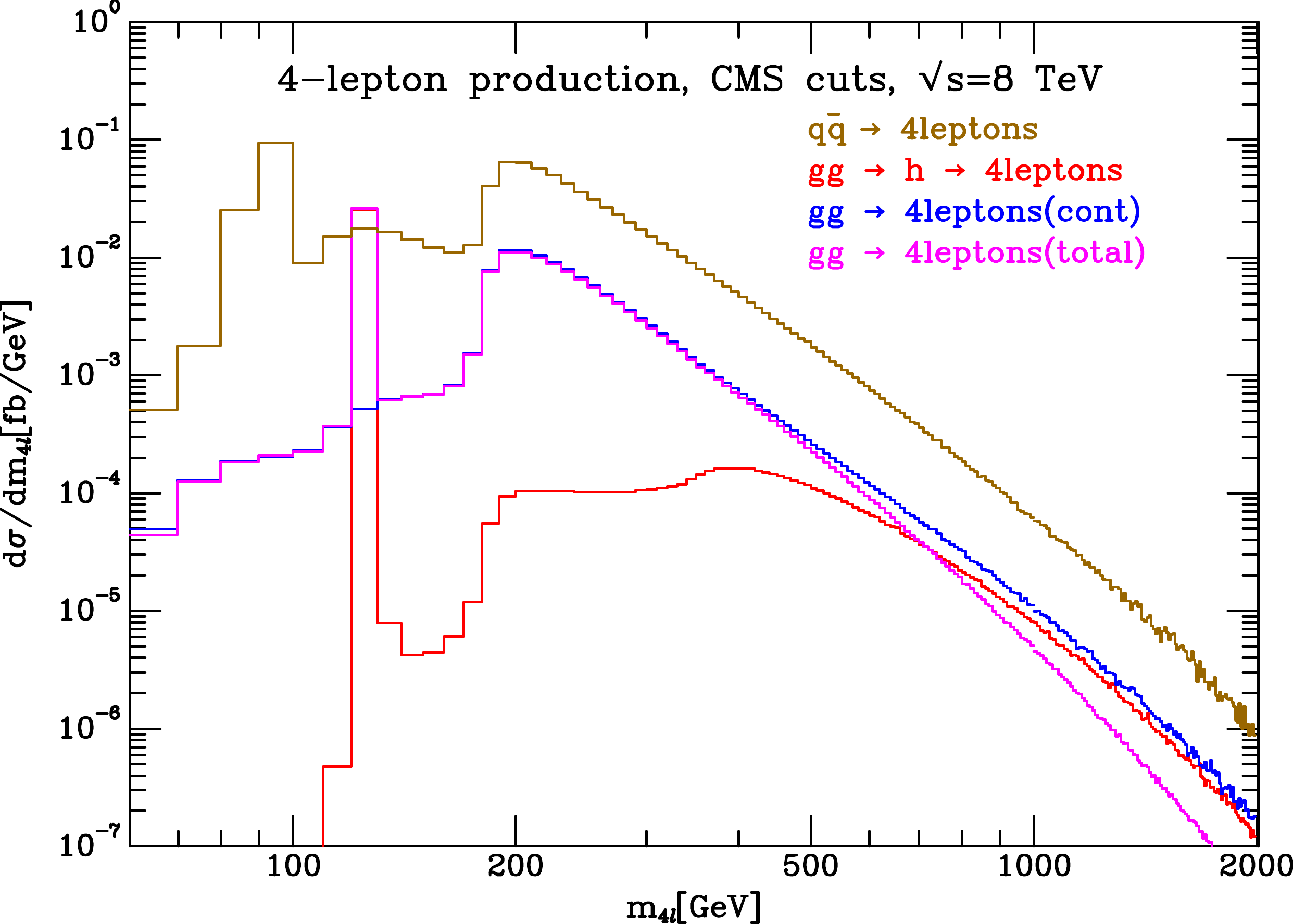}
\caption{Contributions to $gg\rightarrow ZZ\rightarrow 4l$ at $8~TeV$.  The Higgs contributions are shown in red, while
the total rate from gluon fusion including interference is given in magenta\cite{Campbell:2013una}.}
\label{fig:ggzz_width}
\end{figure}

Of course, a real calculation needs to include both the diagrams of Fig. \ref{fig:ggzzfig}, along with the interference, and this
has been done by several groups with results shown in Fig. \ref{fig:ggzz_width}.  The importance of including the interference
terms is apparent, but the long tail at high $m_{4l}$ (shown in red) is clear.  ATLAS and CMS have used this technique
to place limits on the Higgs width\cite{Khachatryan:2014iha,Aad:2015xua},
\beq
\Gamma_h\lsim (4-5)\Gamma_h^{SM}\, .
\eeq

There are some big assumptions in this extraction of the Higgs width, the most obvious of which is the assumption
that the $\kappa$ factors are the same on and off the Higgs resonance peak.  This is clearly a false assumption, since in a quantum field theory
all couplings run.  If there are anomalous $\hsm ZZ$ (or $\hsm gg$) couplings, than the running could be changed 
significantly\cite{Azatov:2014jga,Gainer:2014hha}.
For example, a contribution to the EFT of the form,
\beq
L\sim {c_Z\over \Lambda^2}{\hsm\over v}Z_{\mu\nu}Z^{\mu\nu}
\eeq
would give contributions of ${\cal O}\biggl( {s\over \Lambda^2}\biggr)$ and would cause $m_{4l}$ to grow above the peak, and would
invalidate the extraction of $\Gamma_h$. Additional colored particles in the $ggh$ loop would also change the interpretation of the $gg\rightarrow ZZ
\rightarrow $
4 lepton result as a measurement of the Higgs width\cite{Englert:2014ffa}.

It is worth noting that an $e^+e^-$ collider with an energy of $\sqrt{s}=500~GeV$ can make a $5\%$ measurement of $\Gamma_h$ 
with an integrated luminosity of $500~GeV$\cite{Dawson:2013bba}.  First the measurement of $e^+e^-\rightarrow Z\hsm$ is made by tagging the $Z\hsm$ events
where the recoil mass is consistent with a Higgs boson.  This is done using conservation of momenta and determines $\sigma(Z\hsm)$. Next we can measure the $\hsm\rightarrow ZZ$ rate to determine $BR(\hsm\rightarrow ZZ)$.  The Higgs width is then determined in a model independent fashion,
\beqn
\Gamma_h&=& \Gamma(\hsm\rightarrow ZZ)BR(\hsm\rightarrow ZZ)\nonumber \\
&\sim & {\sigma(Z\hsm)\over BR(\hsm\rightarrow ZZ)}\, . 
\eeqn

\subsubsection{Vector Boson Scattering}
The vector boson scattering (VBS) process is shown  on the RHS of Fig. \ref{fig:prodgg}.  It can be thought of as $2$ incoming quarks 
each radiating a $W$ or $Z$ boson, which then form a Higgs. 
Vector boson fusion also offers the opportunity to observe the $2\rightarrow 2$ scattering process, $VV\rightarrow VV$,
($V=Z,W$), 
which is extremely sensitive to new physics in the electroweak sector.  The $VV\rightarrow VV$ sub-process
plays a special role in Higgs physics since the  Higgs exchange contributions  unitarize the scattering amplitude,
as discussed in Sec. \ref{sec:unitsec}.

VBS production of a Higgs occurs through the purely electroweak process $q {\overline q}^\prime \rightarrow 
q {\overline q}^\prime h$ which has a distinctive experimental signature and vanishes
in the limit $v=0$.  The outgoing jets are peaked in the forward and backward regions and can be used to tag the VBF event.  This can easily be seen by considering
the top leg of the RHS of Fig. \ref{fig:prodgg}:
\begin{equation}
q(p)\rightarrow q^\prime(p^\prime) V(k)\, .
\end{equation}
In the lab frame,
\begin{eqnarray}
p&\equiv& E(1,0,0,1)\nonumber \\
p^\prime &\equiv& E^\prime(1,0,\sin\theta, \cos\theta)\,.
\end{eqnarray}
The integral over the final state phase space for the VBS scattering cross section has a generic
contribution,
\begin{equation}
\sigma\sim \int 
{(\hbox{Phase~ Space})\over [(p-p^{\prime~2})^2-M_V^2]^2}
\sim \int {\theta d\theta
\over [2E E^\prime (1-\cos \theta)-M_V^2]^2}\sim \int {\theta d\theta
\over [\theta^2-M_V^2/E E^\prime)^2}
\end{equation}
which is enhanced in the $\theta\rightarrow 0$ region for $E, E^\prime >> M_V^2$.  
In addition, these forward tagging jets have
a large invariant mass and small $p_T$. 
Typical cuts on the jets are,
\begin{equation}
p_{T_j}> 20~GeV, \,\,\mid y_j\mid < 5 \, , \mid y_{j_1}-y_{j_2}\mid > 3\, , M_{jj}> 130~ GeV\, .
\end{equation}
  The decay products from the intermediate $VV$ scattering
are mostly contained in the central rapidity region.  These characteristics can be used to separate
VBS scattering from QCD gluon initiated events and the non-VBS contributions
can be suppressed to $\sim 1-2\%$\cite{DelDuca:2001fn}. The ability to separate  the Higgs signal
into gluon initiated events and VBF events is crucial for the extraction of Higgs coupling constants.

\subsubsection{Associated Production}
At the LHC
the process $q {\overline q}\rightarrow V\hsm$ offers the hope
of being able to tag the Higgs boson by the
$V$ boson decay products\cite{Stange:1994bb}, although as shown in Fig. \ref{fig:higgs_sig} the rate is 
significantly smaller than the dominant $gg\rightarrow h$ production mechanism.
The cross section for $Wh$ production is, 
\beq
{\hat \sigma}(q_i {\overline q}_j\rightarrow W^\pm h)
={G_F^2 M_W^6 \mid V_{ij}\mid ^2\over 6\pi {s}^2
(1-M_W^2/{ s})^2}\lambda_{Wh}^{1/2}\biggl[
1+{{ s}\lambda_{Wh}\over 12 M_W^2}\biggr]
\quad ,
\eeq
where $\lambda_{Wh}                                                             
=1-2(M_W^2+m_h^2)/{ s}+(M_W^2-m_h^2)^2/{ s}^2$
and $V_{ij}$ is the CKM angle associated with the
$q_i {\overline q}_j W$ vertex.
The rate for $Zh$ is about a factor of $3$ smaller than that for $Wh$ and analytic results can be found in 
Ref. \cite{Djouadi:2005gi}.
The NNLO QCD and NLO electroweak corrections are known, so there is relatively little uncertainty on
the prediction\cite{Brein:2012ne,Denner:2011id}.

The $Vh$ associated channel has recently been used to observe the decay 
$h\rightarrow b {\overline {b}}$\cite{Sirunyan:2017dgc,Aaboud:2017xsd}, using
the jet substructure techniques first proposed in Ref. \cite{Butterworth:2008iy}.  The idea is that by  going to high transverse
momentum for the Higgs, the backgrounds can be significantly reduced.   Jet substructure techniques are discussed
in the lectures of Schwartz at this school\cite{Schwartz:2017hep}.

\subsubsection{$t {\overline t}\hsm$ Production}

The top quark Yukawa coupling, $Y_t$, can be directly measured in the $t {\overline t}\hsm$ process shown 
on the RHS of Fig. \ref{fig:prodtt}.
Recall that the gluon fusion production of the Higgs is also proportional to the top quark Yukawa, but in addition it
can
 receive enhanced contributions from the bottom quark Yukawa interactions in some BSM scenarios, along with contributions
from new colored scalars.  
The NLO QCD\cite{Dawson:2003zu,Dawson:2002tg,Beenakker:2002nc,Beenakker:2001rj}  and electroweak
 corrections\cite{Frixione:2014qaa,Frixione:2015zaa}  for $t {\overline {t}}h$ production are known and contribute to very precise 
predictions\cite{deFlorian:2016spz}:
\begin{eqnarray}
\sqrt{S}&=8~TeV\qquad \qquad & \sigma_{tt\hsm}=.133~pb^{+4\%}_{-9\%}(scale)\pm 4.3\%(PDF+\alpha_s)\nonumber\\
\sqrt{S}&=13~TeV\qquad \qquad & \sigma_{tt\hsm}=.507~pb^{+5.8\%} (scale)_{-9.2\%}\pm 3.6\%
(PDF+\alpha_s)\, .
\end{eqnarray}
Although numerically small, electroweak corrections  spoil the direct proportionality of the lowest
order cross section to $Y_t^2$.  

This process has large backgrounds from $t {\overline t}b {\overline b}$ and
$t {\overline t}jj$.
In order to suppress the backgrounds, many $t {\overline t}\hsm$ searches are done 
in the boosted regime, where the electroweak Sudakov logarithms become relevant.  A definitive
measurement of this channel has not yet been made, and 
will be one of the important
milestones of the coming LHC run.

The associated production  of $b{\overline b}\hsm$ is not relevant in the SM, but can be important in models
with enhanced $b$ Yukawa couplings.

\subsubsection{Double Higgs Production}
\label{sec:ddhh}
Finally, we  need to measure the parameters of the Higgs potential, Eq. \ref{eq:potfin},
 to determine if electroweak symmetry
breaking really proceeds as in the SM. 
In the SM, the Higgs potential from Eq. \ref{eq:smpot} is, 
\beqn
V&=&
{\mhsm^2\over 2}\hsm^2+\lambda_3 \hsm^3+\lambda_4 \hsm^4\, ,
\label{eq:lamdef}
\eeqn
where $\lambda_3^{SM}=\mhsm^2/(2v)$ and $\lambda_4^{SM}=\hsm^2/(8v^2)$.
It is apparent that the Higgs self- couplings are weak,
\begin{equation}
\lambda_3^{SM}=.13 v,\qquad \lambda_4^{SM}=.03\, .
\end{equation}
\begin{figure}
\centering
\includegraphics[width=.3\textwidth]{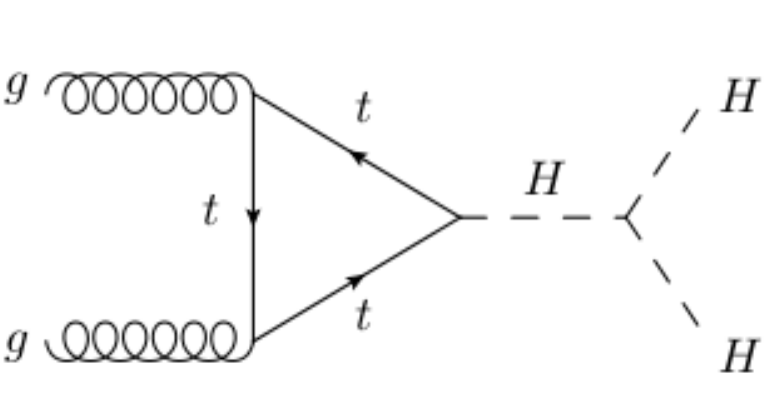}
\includegraphics[width=.3\textwidth]{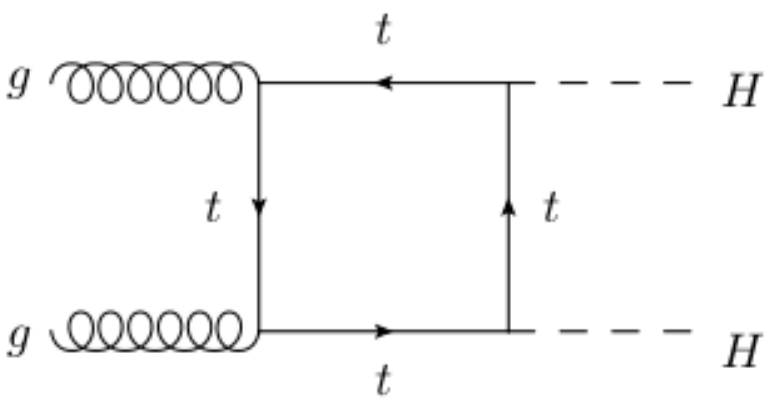}
\includegraphics[width=.3\textwidth]{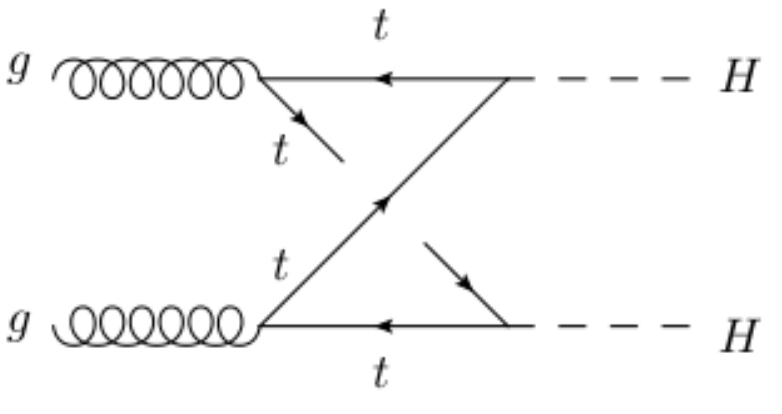}
\caption{\label{fg:hhfig}Contributions to $gg\rightarrow \hsm\hsm$ in the SM. The dominant contribution to the triangle and box diagrams are from the top
quark.  In this figure, $H$ is the SM Higgs boson. }
\end{figure}
The only way to directly probe the $\hsm^3$ coupling is by double Higgs production and the dominant production mechanism is gluon
fusion as shown in
Fig. \ref{fg:hhfig}.  
The result is sensitive to new colored particles running in the loops, along with modifications to  the 
Higgs tri-linear self-coupling and the top quark Yukawa coupling (Eqs. \ref{eq:lamdef} and \ref{eq:uyuk}).

The amplitude for $g^{A,\mu}(p_1)g^{B,\nu}(p_2)\rightarrow h(p_3)h(p_4)$ is
\begin{equation}
	A^{\mu\nu}_{AB}={\alpha_s\over 8 \pi v^2}\delta_{AB}\biggl[P_0^{\mu\nu}(p_1,p_2){\hat F}_1(s,t,u,m_t^2)+
	P_2^{\mu\nu}(p_1,p_2,p_3){\hat F}_2(s,t,u,m_t^2)\biggr]\; ,
\label{eq:amp}
\end{equation}
where $P_0$ and $P_2$ are the orthogonal  projectors onto the spin-$0$ and spin-$2$ states respectively,
\begin{eqnarray}
	P_0^{\mu\nu}(p_1,p_2)&=&
	g^{\mu\nu}-{p_1^\nu p_2^\mu\over p_1\cdot p_2} \; , \nonumber \\
	P_2^{\mu\nu}(p_1,p_2,p_3)&=&g^{\mu\nu}+{2\over s p_T^2 } \left(
		\mhsm^2 p_1^\nu p_2^\mu 
		- 2 p_1.p_3 \, p_2^\mu p_3^\nu - 2 p_2.p_3 \, p_1^\nu p_3^\mu + s \, p_3^\mu p_3^\nu
\right)\, ,
\end{eqnarray}
	$s = (p_1+p_2)^2 ,
	t = (p_1-p_3)^2 ,
	u = (p_2-p_3)^2 $,
and
$p_T$ is the transverse momentum of the Higgs boson,
\beq 
	p_T^2={ut-\mhsm^4\over s} \, .
\eeq
The functions ${\hat F}_1$ and ${\hat F}_2$ are known analytically~\cite{Glover:1987nx,Plehn:1996wb}. 

In the SM, the largest contributions come from top quark loops and 
in the limit $m_t^2 >> s$, the leading terms are,
\begin{eqnarray}
\label{eq:gghh_SM_expansion}
	{\hat F}_1(s,t,u,m_t^2)&\equiv &{\hat F}_1^{tri}(s,t,u,m_t^2)+{\hat F}_1^{box}(s,t,u,m_t^2) \; \nonumber\\
	{\hat F}_1^{tri}(s,t,u,m_t^2)&=& {4 \mhsm^2\over s - \mhsm^2} s\biggl({\lambda_3\over \lambda_3^{SM}}\biggr)
		\nonumber \\
	{\hat F}_1^{box} (s,t,u,m_t^2)&=&-{4\over 3}s \nonumber \\
	{\hat F}_2(s,t,u,m_t^2) & = & - {11\over 45} s {p_T^2\over m_t^2} \label{eq:fdef}\, ,
\end{eqnarray}
where we have allowed an arbitrary rescaling of the Higgs tri-linear coupling. 
It is important to remember that in the SM, there is no freedom to rescale $\lambda_3$, making this a BSM effect.

\begin{figure}
\centering
\includegraphics[width=.5\textwidth]{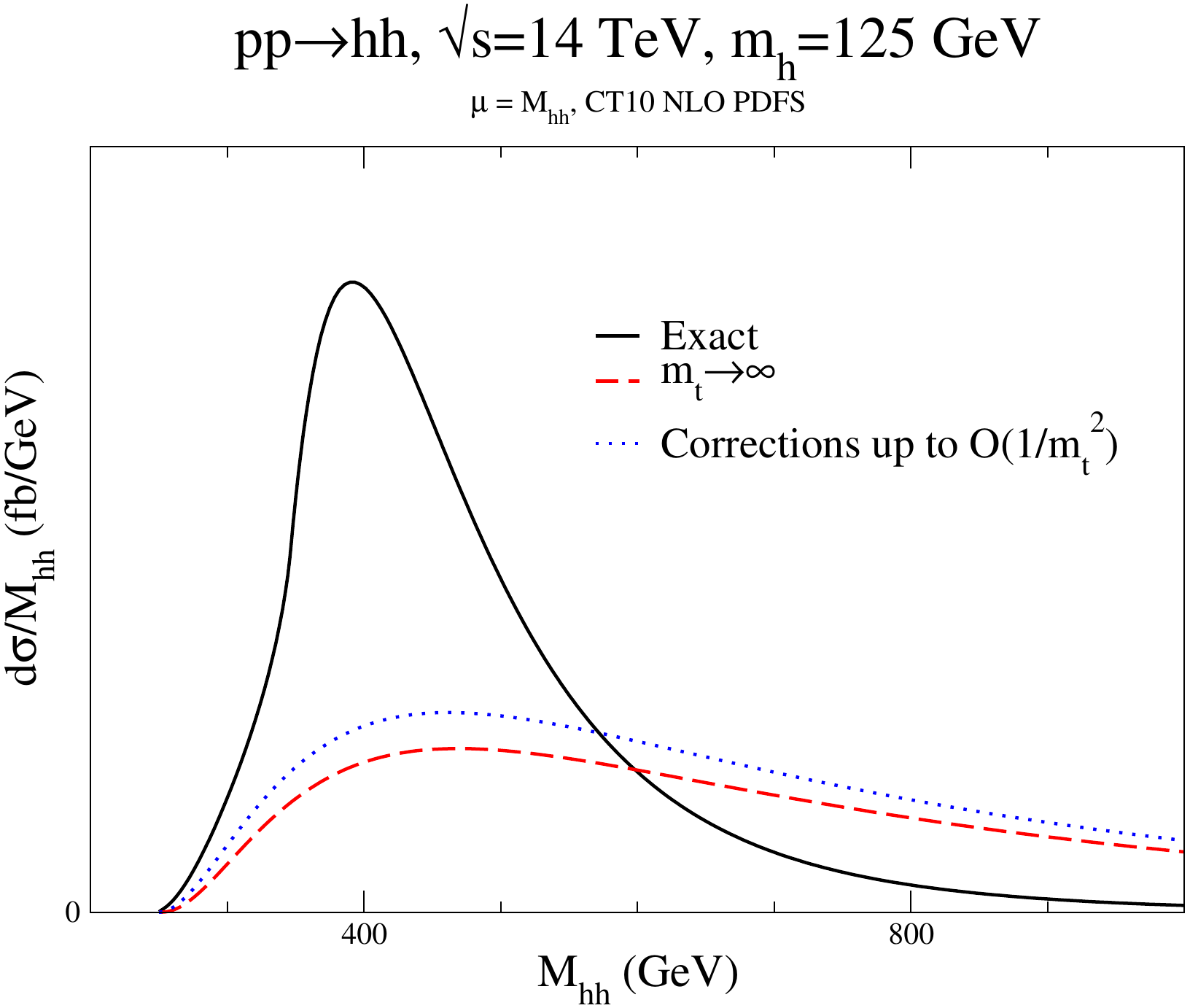}
\caption{LO transverse momentum distribution for double Higgs production in the SM, compared with the large $\mt$
limit, along with the first correction  of ${\cal {O}}(s/m_t^2)$. }
\label{fig:hhpt}
\end{figure}

We see that the amplitude vanishes at threshold in the SM  in the large $m_t$ limit, reducing the sensitivity to $\lambda_3$.
The expansion in powers of $1/\mt$ poorly reproduces kinematic distributions, due to the presence 
of contributions proportional to $s/m_t^2$, as is obvious in Fig. \ref{fig:hhpt}\cite{Dolan:2012ac,Dawson:2012mk}.  

\begin{figure}
\centering
\includegraphics[width=.5\textwidth]{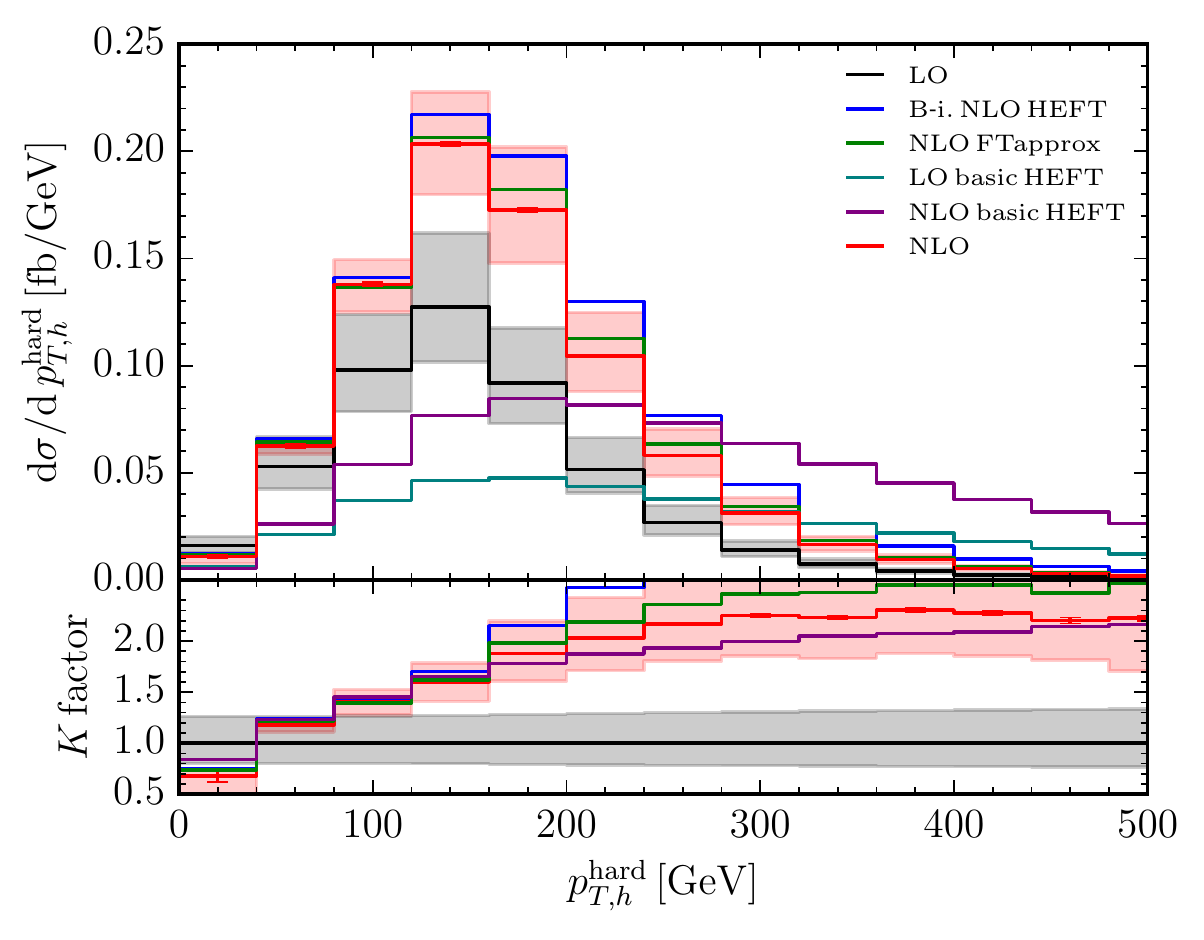}
\caption{Transverse momentum distribution for double Higgs production in the SM, including various approximations for 
the QCD corrections. The curve labelled NLO includes all finite $m_t$ effects\cite{Heinrich:2017kxx}.\label{fig:hhpt_NLO}}
\end{figure}

The large $\mt$ limit has  been
used to compute QCD corrections to NLO \cite{Dawson:1998py} and  NNLO\cite{deFlorian:2015moa}.  In this approach, a $K$ factor is computed: 
\begin{equation}
K\equiv{d\sigma_{NNLO}\over d\sigma_{LO}}\, ,
\label{eq:kdef}
\end{equation}
where the distributions in Eq. \ref{eq:kdef} are computed in the $\mt\rightarrow\infty$ limit and are then used to
rescale the lowest order distributions computed with finite $m_t$\footnote{This is termed the B.i. NLO HEFT in Fig. \ref{fig:hhpt_NLO}.}
\cite{deFlorian:2016uhr,deFlorian:2017qfk,deFlorian:2013jea,Grigo:2014jma}. 
The exact NLO result for double Higgs production including all top mass effects  is now known and can be used to obtain  distributions\cite{Borowka:2016ypz,Heinrich:2017kxx}.  The effects of including the top quark mass
exactly at NLO are significant and reduce the total cross section by $\sim 14\%$ at $14~TeV$ from  the B.i. NLO HEFT
 limit. Including the top quark mass effects also has significant effects on distributions, as demonstrated in 
Fig. \ref{fig:hhpt_NLO}.

  The dependence of $\hsm\hsm$ production on $\lambda_3$ from various production mechanisms is shown
in Fig. \ref{fg:hhfig}\cite{Frederix:2014hta} as a function of $\delta_3\equiv {\lambda_3\over \lambda_3^{SM}}$. 
\footnote{The curve labelled EFT loop-improved is identical to  the  B.i. NLO HEFT approximation.}
\begin{figure}
\centering
\includegraphics[width=.5\textwidth]{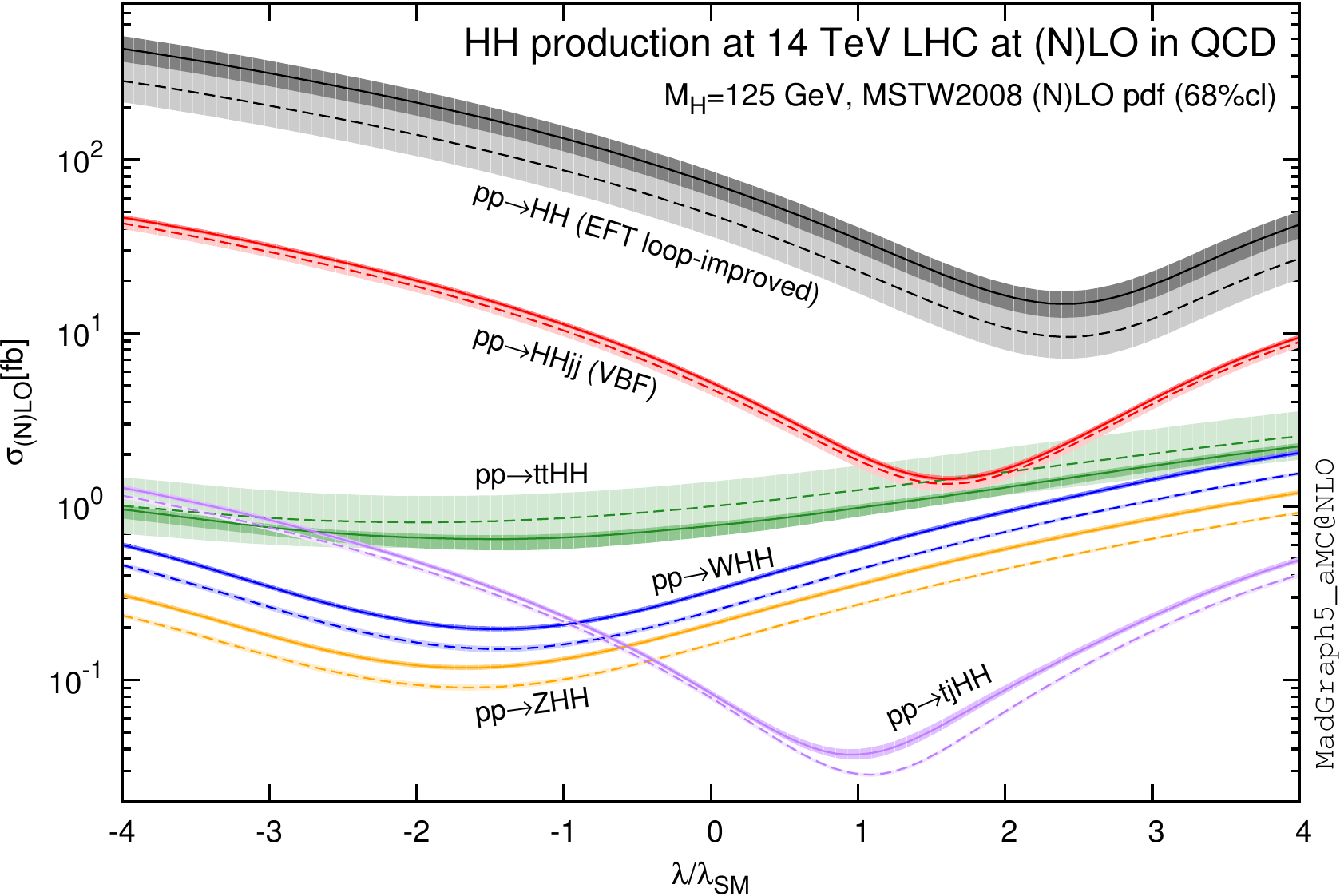}
\caption{Dependence of  double Higgs production rates on the Higgs tri-linear self-coupling\label{fg:hhfig}\cite{Frederix:2014hta}.}
\end{figure}

The best current  limits from the $8~TeV$ data on double Higgs production are,
 \begin{eqnarray}
 {\sigma(pp\rightarrow \hsm\hsm)\over \sigma(pp\rightarrow \hsm\hsm)\mid_{SM}}&<29\qquad &
 {\text{ATLAS}}\, , \nonumber \\
 {\sigma(pp\rightarrow \hsm\hsm)\over \sigma(pp\rightarrow \hsm\hsm)\mid_{SM}}&<19\qquad &
 {\text{CMS}}\, ,
 \end{eqnarray}
  which still leaves a way to go before we get to an interesting regime. 
   The ATLAS limit is from the $b{\overline b} b {\overline b}$ final state\cite{ATLAS-CONF-2016-049},
   while the CMS limit
  is from the $b {\overline b}\gamma \gamma$ final state\cite{CMS-PAS-HIG-17-008}. 
 ATLAS estimates that
a luminosity of $3~ab^{-1}$ will be sensitive  to $\delta_3 > 8.7$ and $\delta_3<-1.3$\cite{ATL-PHYS-PUB-2015-046}.  This is clearly not the precision measurement we desire and
the need to measure the Higgs tri-linear coupling
is one of the major motivations for a $100~TeV$ collider.  

\begin{figure}
\centering
\includegraphics[width=.5\textwidth]{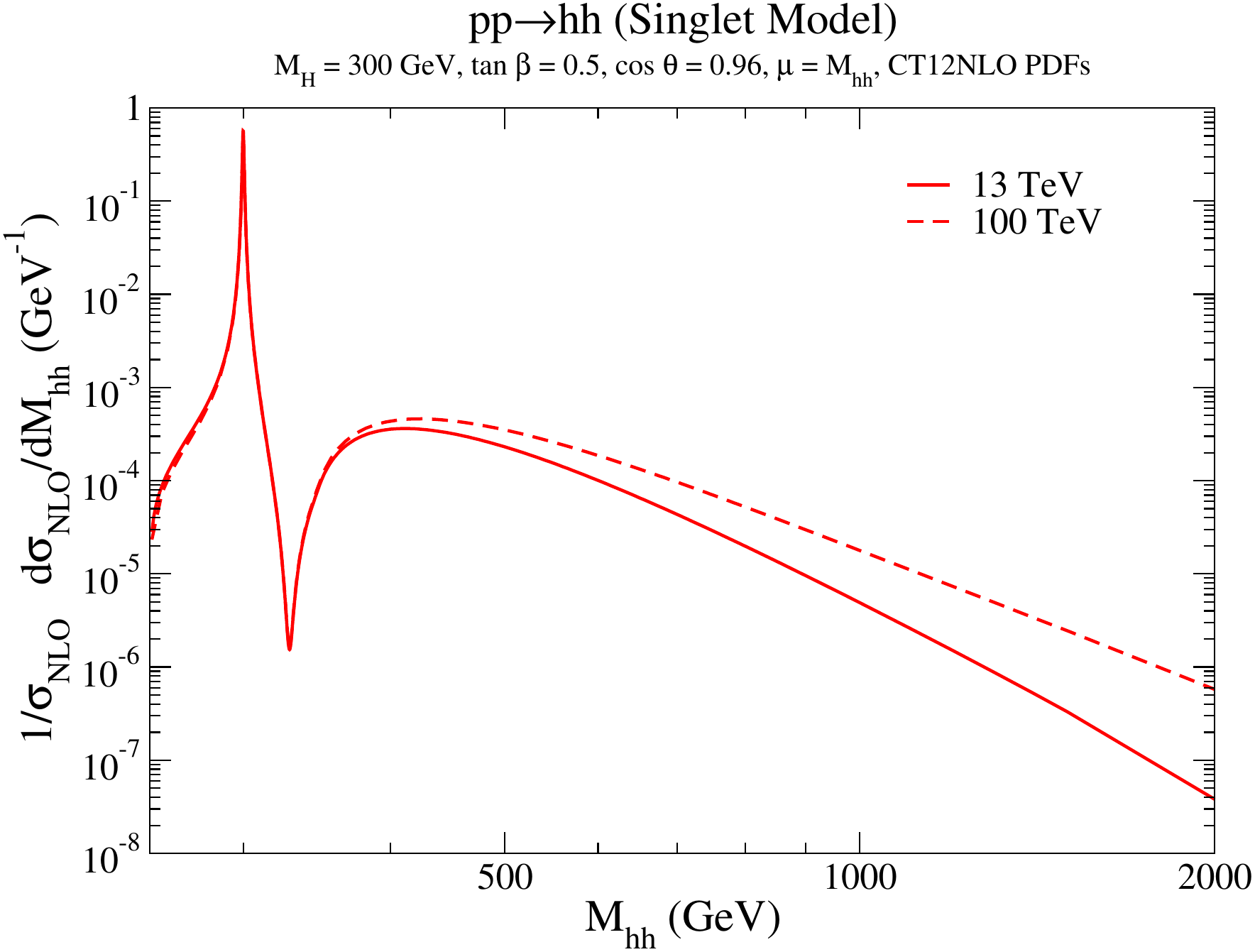}
\caption{Double Higgs production in the $Z_2$ symmetric singlet model
with a heavy neutral scalar of mass $M_H=300~GeV$\cite{Chen:2014ask}.\label{fg:hhsing}}
\end{figure}

\begin{figure}
\centering
\includegraphics[width=.5\textwidth]{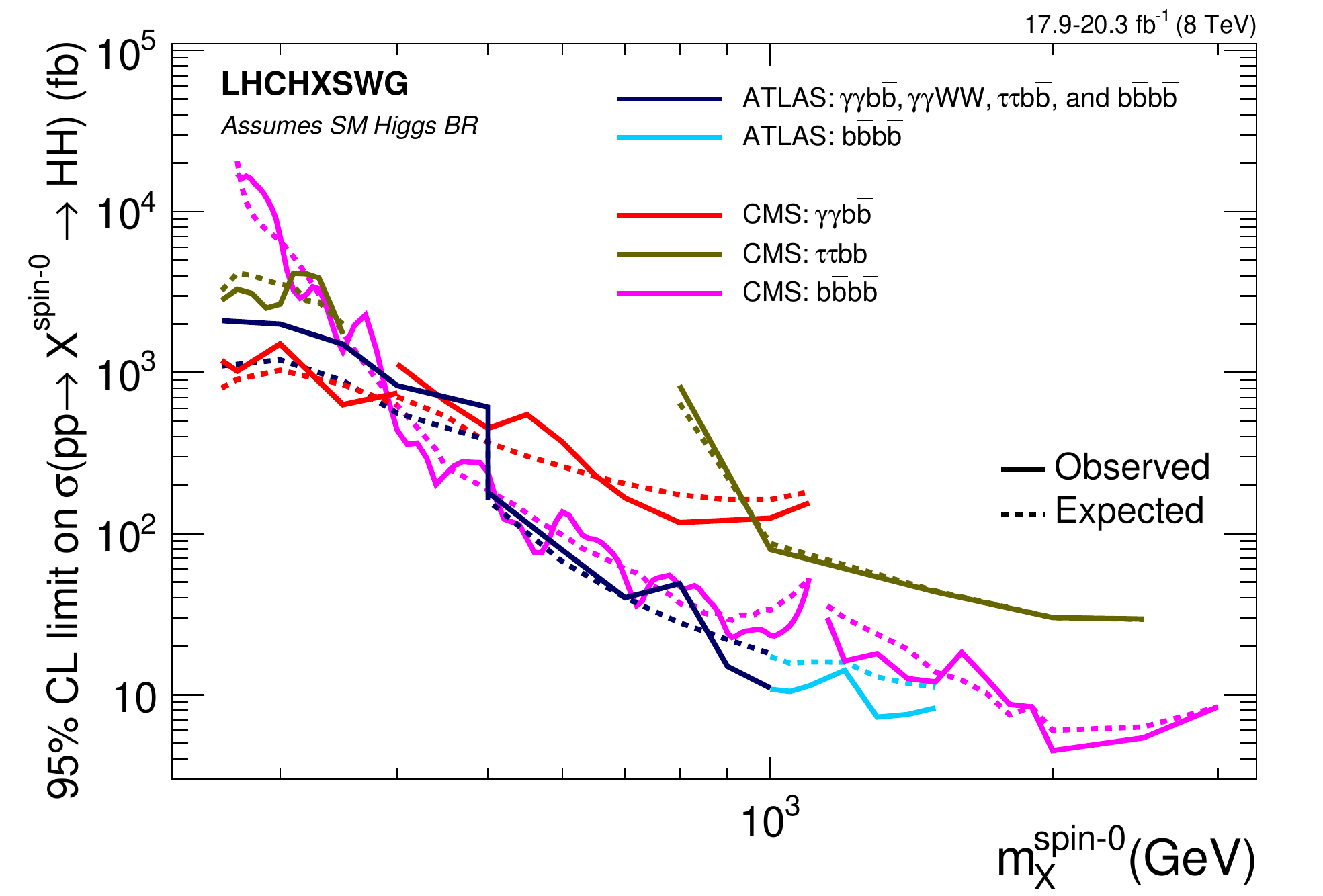}
\caption{Experimental limits from the LHC on $\hsm\hsm$ production in a BSM theory containing an $s-$ channel scalar 
resonance with mass $M_X$\cite{deFlorian:2016spz}.\label{fg:hhlims}}
\end{figure}

The fact that the SM rate for double Higgs production is quite small makes it an ideal place to search for new physics.  Many models (singlet, 2HDM, MSSM, NMSSM, etc)\cite{Chen:2014ask,Barr:2014sga,Costa:2015llh,Dawson:2017jja,Dawson:2015haa} 
contain heavy neutral scalars that can decay into 2 SM Higgs bosons with a significant  ($\sim 30\%$) branching ratio.  
In these models, there is an $s-$ channel resonance
 from the heavy Higgs particle, and there will be interference between this new scalar and the SM Higgs giving the classic dip structure shown in Fig. \ref{fg:hhsing} for the example of the singlet model.
Limits on  resonant decays in the generic BSM process, $gg\rightarrow X\rightarrow \hsm\hsm$  for various
final states are shown in Fig. \ref{fg:hhlims}, where for heavy resonances, the most
important search channel is the $4b$ final state.  

It has been proposed that indirect limits on $\lambda_3$ may be extracted from the dependence of electroweak 
radiative corrections  to single Higgs production on the Higgs tri-linear coupling.  This coupling enters the rate for $gg\rightarrow h$ at $2-$ loops and 
contributes to the $t{\overline {t}}h$, $Vh$, and VBS processes at $1-$ loop.  Of course $\lambda_3$ is not a free 
parameter in the SM, and some care must be taken with the renormalization prescription.  Ref. \cite{Degrassi:2016wml}
obtains the allowed $2\sigma$ region from a fit to single Higgs production,
\beq
-9.4 < \delta_3 < 16\, .
\eeq
Similar allowed regions  are obtained in Refs. \cite{DiVita:2017eyz,Kribs:2017znd,Bizon:2016wgr,Degrassi:2017ucl}.  The allowed parameter 
space from current fits to single Higgs production are not significantly different from the expected limits on $\lambda_3$ with $3~ab^{-1}$ at the LHC.

\section{Effective Field Theory and the Higgs Boson}
\label{sec:eftch}

\subsection{Higgs Boson Coupling measurements}

The production of the Higgs boson in Run-I at the LHC produced results which basically agree with the SM
predictions at the $10-20\%$ level\cite{Khachatryan:2016vau}.   Preliminary Higgs coupling results 
at $13~   TeV$ \cite{ATLAS-CONF-2017-043,ATLAS-CONF-2017-045,CMS-PAS-HIG-16-044,CMS-PAS-HIG-16-021,CMS-PAS-HIG-16-041,Aaboud:2017xsd}, are also in reasonable agreement with expectations. The rates are as predicted, and there are no non-SM like light (EW scale) particles observed. 

What we need is a way to quantify small
deviations from the SM predictions.  The simplest way is to introduce an arbitrary scaling into the SM interactions,
\begin{eqnarray}
L_{\kappa}&=& \Sigma_f{\kappa_f}{m_f\over v} {\overline f} f \hsm+\kappa_Wg M_W W^{+\mu}W^-_{\mu}\hsm
+\kappa_Z g {M_Z\over c_W} Z^{\mu}Z_{\mu}\hsm \, .
\label{eq:kapdef}
\end{eqnarray}
In the SM, all $\kappa$ parameters are $1$, so a deviation would indicate some physics not contained in the SM.  Of course,
Eq. \ref{eq:kapdef} is not $SU(2)_L\times U(1)_Y$ gauge invariant, but it serves as a starting point for study.   

For a given production and decay channel, $i\rightarrow \hsm\rightarrow j$,
\begin{eqnarray}
\kappa_i^2&=& {\sigma(i\rightarrow \hsm)\over \sigma(i\rightarrow \hsm)_{SM}}\nonumber \\
\kappa_j^2&=& {\Gamma(\hsm\rightarrow j)\over \Gamma(\hsm\rightarrow j)_{SM}}\, .
\end{eqnarray} 
The $\kappa$ formalism also
rescales the total width,
\begin{eqnarray}
\kappa_h & \equiv & {\Gamma_\hsm\over \Gamma_{\hsm}^{SM}}\nonumber \\
\Gamma_h&=& \Sigma_X\kappa_X^2\Gamma(\hsm\rightarrow XX)+\Gamma(\hsm\rightarrow
{\hbox{invisible}})\, , 
\end{eqnarray}
where $\Gamma(\hsm\rightarrow
{\hbox{invisible})}$ is any unobserved decay. 
This approach assumes that there are no new light resonances, no new tensor structures in the Higgs interactions
beyond those of the SM, that
the narrow width approximation for Higgs decays is valid, and is based on rescaling total rates (that is, no new
dynamics is included).   

\begin{figure}
\begin{centering}
\includegraphics[width=0.7\textwidth]{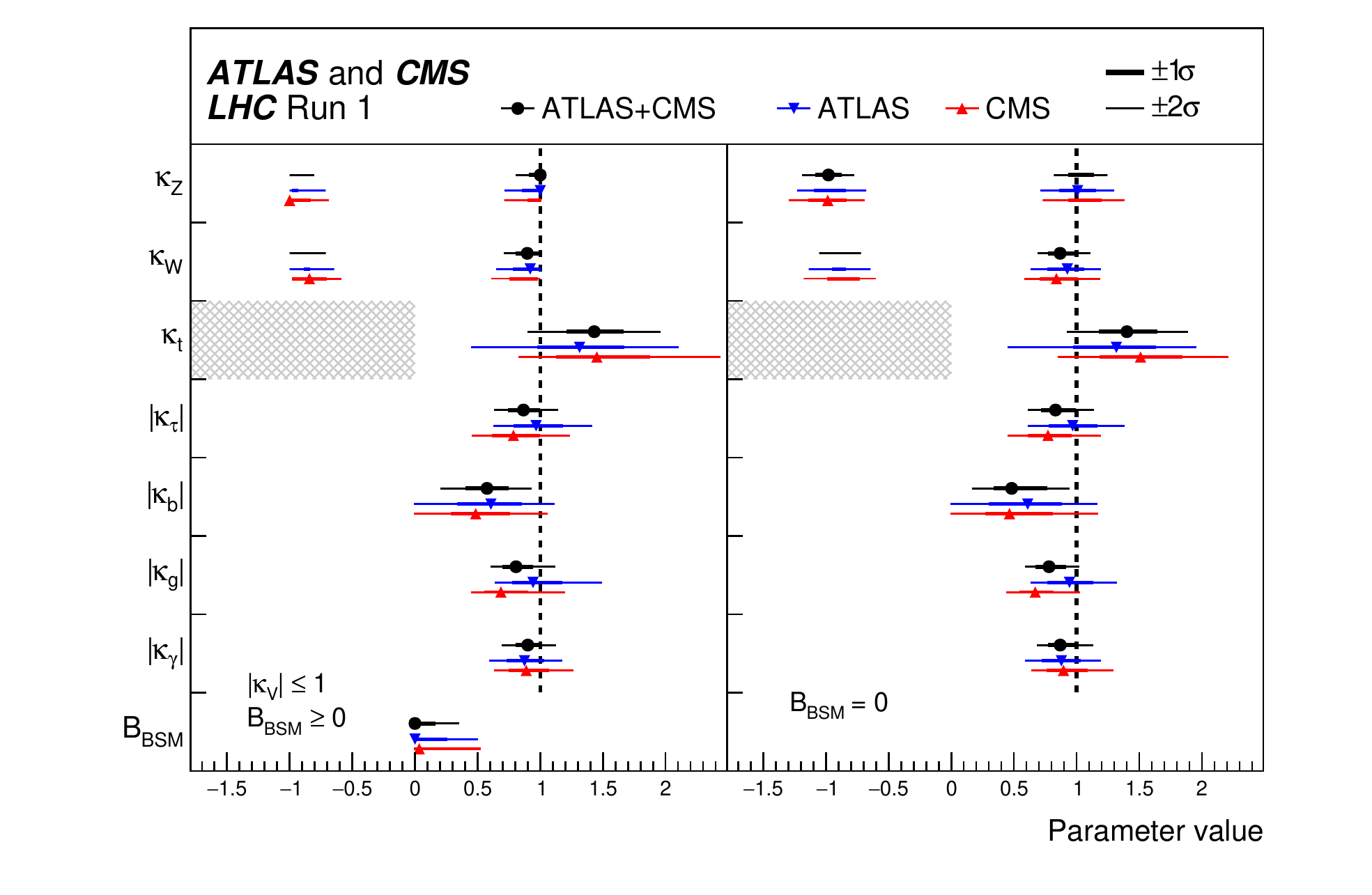}
\end{centering}
\caption{Combined ATLAS/CMS $\kappa$ fits to Run-1 data\cite{Khachatryan:2016vau}.\label{fg:kapfits}}
\end{figure}

\begin{figure}
\begin{centering}
\includegraphics[width=0.4\textwidth]{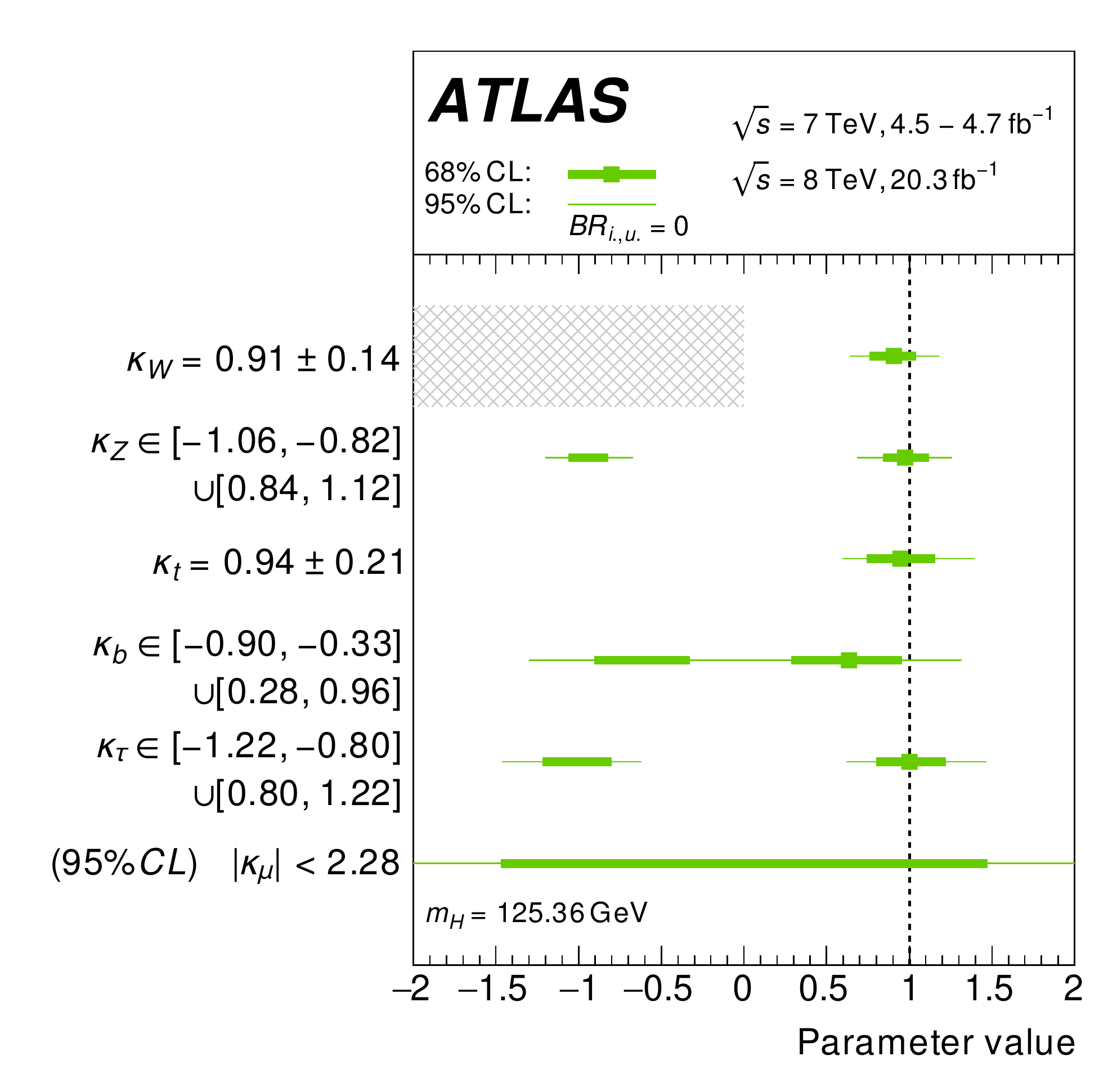}
\includegraphics[width=0.4\textwidth]{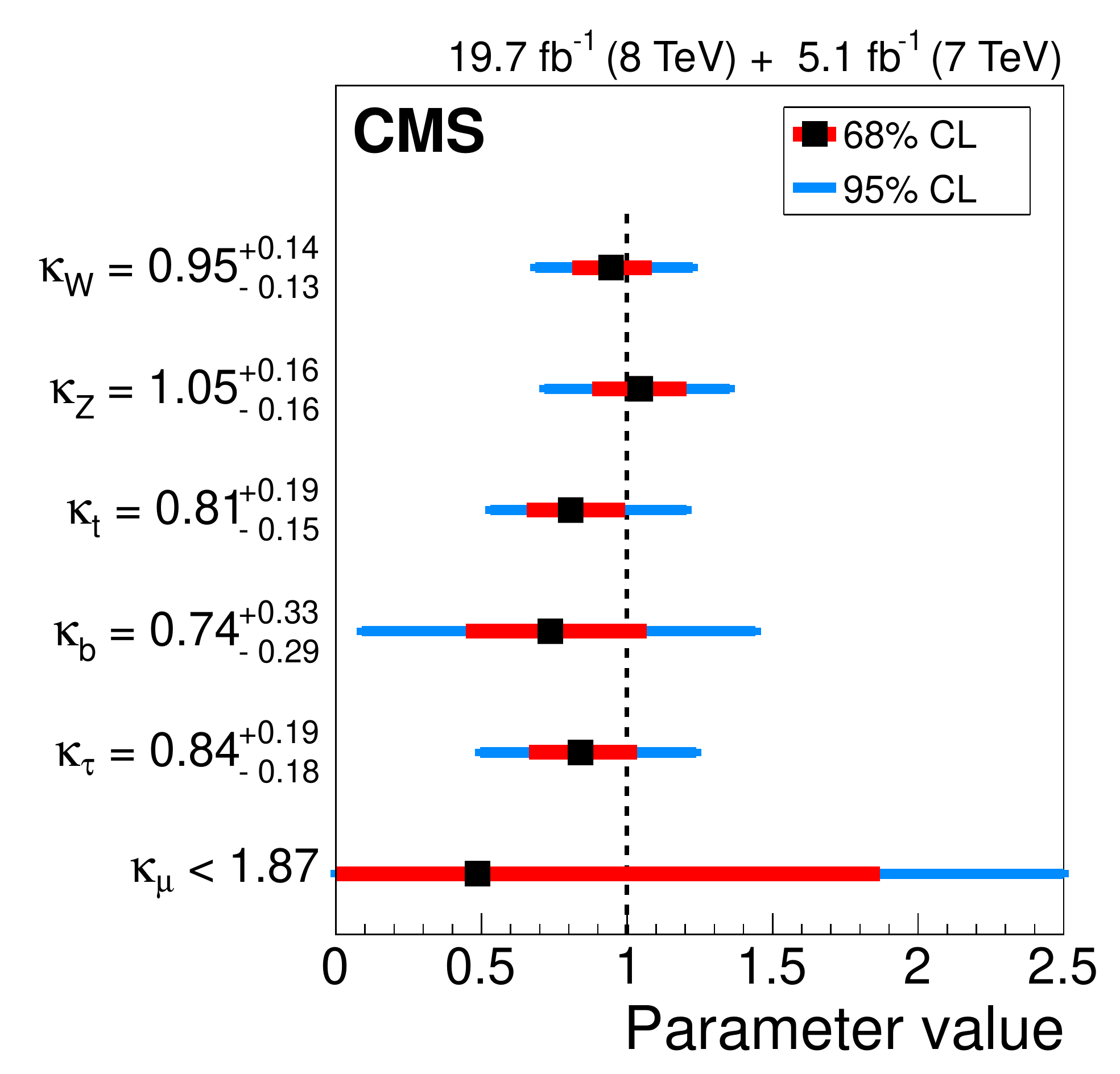}
\end{centering}
\caption{ ATLAS $\kappa$ fits to Run-1 data\cite{Aad:2015gba} (LHS) and CMS 
$\kappa$ fits to Run-1 data\cite{Khachatryan:2014jba} (RHS) .\label{fg:kap}}
\end{figure}

   A combined CMS/ATLAS fit is shown in Fig. \ref{fg:kapfits}.   This particular
    fit does not allow for new physics in the $gg\rightarrow \hsm$ and $\hsm\rightarrow \gamma\gamma$ channels, but instead parameterizes  the effective couplings in terms of the SM interactions of the Higgs with the top and bottom
    ($\kappa_g$) and  with the $W$ and top ($\kappa_\gamma)$ as,
  \begin{eqnarray}
  \kappa_g^2 &\sim & 1.06\kappa_t^2+.01\kappa_b^2-.07\kappa_t\kappa_b\nonumber \\
  \kappa_\gamma^2&\sim& 1.59 \kappa_W^2+.07 \kappa_t^2-.66\kappa_W\kappa_t\, .
  \label{eq:kaps}
\end{eqnarray}
Similar results are shown in Fig. \ref{fg:kap}, and again the results are in general agreement with the SM predictions. 
With the addition of $13~TeV$ data, the Higgs  couplings should become even more constrained.  In particular, the $tth$ 
and $bbh$ coupling measurements have been significantly updated from Fig. \ref{fg:kap}.

ATLAS and CMS have various types of fits. 
In some fits, they separate Higgs bosons from different production  and decay channels. 
Other
fits allow for unobserved decay channels, or new contributions to gluon fusion or the 
decay to $\gamma \gamma$.   None of
the fits show any significant deviation from the SM predictions.  

Finally, a fit to all Higgs production  and decay channels yields the combined ATLAS/CMS result\cite{Khachatryan:2016vau},
\begin{eqnarray}
\mu& \equiv & {\sigma_h\over \sigma_h(SM)}=1.0\pm 0.07 (stat)\pm 0.04 (syst)\pm 0.03 (theory)\, .
\label{eq:finsig}
\end{eqnarray}
From Eq. \ref{eq:finsig}, it is clear that the accuracy of the theoretical predictions will soon be the limiting
factor in the interpretation of Higgs measurements. 

To improve on the fits to total rates,
 we need to construct an effective field theory, which is the topic of the next section. 

\subsection{Effective Field Theory Basics}

The  effective field theory (EFT)
 Lagrangian we use assumes that there are no new light degrees of freedom and is constructed by writing an $SU(2)_L\times U(1)_Y$
invariant Lagrangian as an expansion in powers of $v/\Lambda$, where $\Lambda$ is some high scale 
where we envision that there is a UV complete theory\cite{Contino:2013kra,Brivio:2017vri},
\begin{equation}
L_{EFT}=L_{SM}+\Sigma_i{c_i^5\over \Lambda}O_i^5+\Sigma_i{c_i^6\over\Lambda^2}O_i^6
+.....
\label{eq:smeft}
\end{equation}
and $O_i^n$ is a dimension-$n$ operator constructed from SM fields.
  The EFT allows for a systematic study of BSM physics effects in a gauge invariant fashion and radiative
corrections can be implemented order by order in ${v\over \Lambda}$. 

The only possible dimension-$5$ operator violates lepton number conservation
and is typically neglected in studies of Higgs physics.
There are many possible bases for constructing the dimension-$6$ operators, of which the most well-known are the Warsaw\cite{Buchmuller:1985jz}, HISZ\cite{Hagiwara:1993ck},
 and SILH\cite{Giudice:2007fh} bases.  By using the equations of motion, there is a mapping from one basis to the 
 next\cite{Falkowski:2015wza,Wells:2015uba}.  Note that the HISZ basis does not
contain fermion interactions.

There are several approaches to using the dimension-$6$ truncation of the 
EFT of Eq. \ref{eq:smeft}.  One could calculate an amplitude to ${\cal{O}}\biggl( {v^2\over \Lambda^2}\biggr)$,
\begin{equation}
A\sim A_{SM}+{A^{6}_{EFT}\over \Lambda^2}\, .
\end{equation}
Squaring the amplitude,
\begin{equation}
\mid A\mid^2\sim \mid A_{SM}+{A^6_{EFT}\over \Lambda^2}\mid^2\, ,
\label {eq:eftprobs}
\end{equation}
we obtain results that are guaranteed to be positive-definite.  The problem is
that Eq. \ref{eq:eftprobs} contains terms $\sim {(A^6_{EFT})^2\over \Lambda^4}$ that are of the same
order in $v^2/\Lambda^2$ as the neglected dimension-$8$ terms.  The expansion only makes sense if
\begin{equation}
\mid A^6_{EFT}\mid^2 << \mid A_{SM}^*A^8_{EFT}\mid \, ,
\end{equation} 
which can be arranged in some BSM models\cite{Contino:2016jqw} .

We begin by  considering a simple  EFT with just $2$  non-SM terms, 
\beqn
L\sim L_{SM}+{\alpha_s\over 4 \pi}{c_g\over \Lambda ^2}
(\Phi^\dagger \Phi) G_{\mu\nu}^AG^{\mu\nu A}
+\biggl({c_t Y_t\over \Lambda^2}{\overline {q}}_L{\tilde \Phi} q_R (\Phi^\dagger \Phi)+h.c.\biggr)\, .
\label{eq:eftbsm}
\eeqn
After spontaneous symmetry breaking, the top mass is shifted,
\begin{equation}
m_t={Y_t v\over\sqrt{2}}\biggl( 1-{v^2 c_t\over 2 \Lambda^2}\biggr)\, .
\end{equation}
The Higgs coupling to the top quark is no longer proportional to $m_t$ and Eq. ~\ref{eq:eftbsm}
becomes
\begin{equation}
L\rightarrow {\alpha_s\over 4 \pi}
{c_g\over \Lambda ^2}
h G_{\mu\nu}^AG^{\mu\nu A}
-m_t t {\overline{t}}\biggl[
1+{h\over v}\biggl(1-{v^2 c_t\over \Lambda^2}\biggr)
\biggr] +...
\label{eq:eftcg}
\end{equation}
When flavor indices are included in the fermion interactions, Eq. ~\ref{eq:eftbsm} can generate flavor violation in the Higgs sector\cite{Harnik:2012pb}. 

Both  $c_g$ and $c_t$ contribute to  $gg\rightarrow h$,\footnote{{\it{Caveat~ emptor}}:  Practically every EFT paper uses different normalization
and sign conventions for the EFT operators.  The only way to check results like Eq. ~\ref{eq:topeft} is to start from the definition of
the operators in the Lagrangian.}
\beq
\sigma(gg\rightarrow \hsm)=\sigma(gg\rightarrow \hsm)_{SM}  \biggl(1+2{v^2\over \Lambda^2}(3c_g-c_t)\biggr)
+{\cal{O}}\biggl({m_h^2\over m_t^2}, {v^4\over \Lambda^2}\biggr)\, ,
\label{eq:topeft}
\eeq
and so gluon fusion cannot distinguish between $c_g$ and $c_t$\cite{Azatov:2016xik,Contino:2013kra,Chen:2014xwa,Goertz:2014qta,Azatov:2015oxa,Gillioz:2012se}.
The $t {\overline{t}}h$ process  is independent of $c_g$ at leading order and can be used to obtain
a measurement of $c_t$.  Once radiative corrections (both QCD 
and electroweak) are included, however, the situation becomes murkier and the $t {\overline{t}}h$ rate is no longer  directly proportional
to $c_t$. 

\begin{figure}
\begin{centering}
\includegraphics[width=0.7\textwidth]{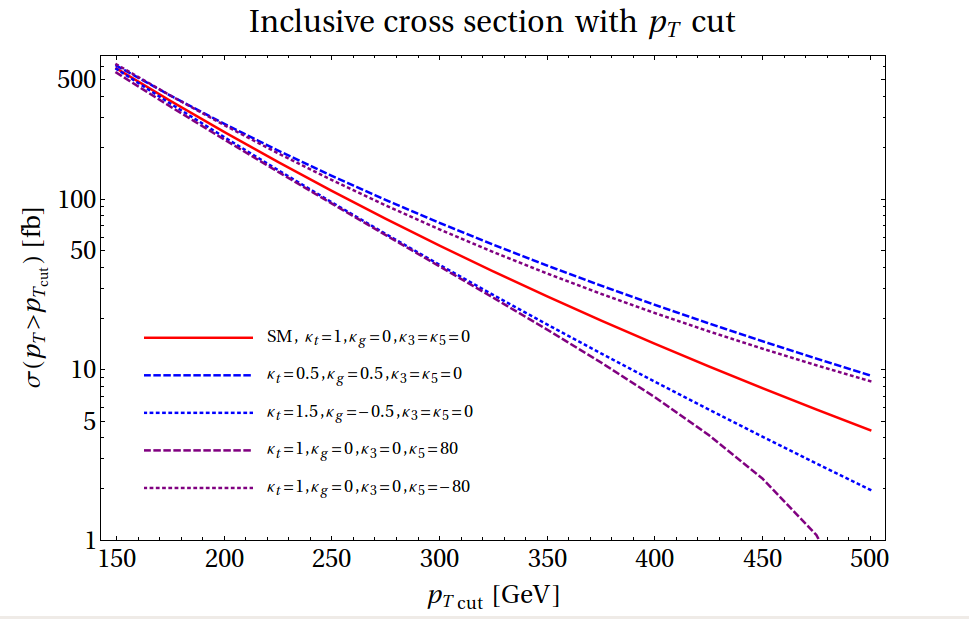}
\end{centering}
\caption{Effects of the dimension-$8$ operators of Eq. \ref{eq:op5} on the $p_T$ spectrum
of $gg\rightarrow gh$\cite{Dawson:2015gka}.
\label{fig:dim8}}
\end{figure}

At dimension-$6$, the unique operator contributing to gluon fusion of the Higgs is
\begin{equation}
O_1=G_{\mu\nu}^A G^{\mu\nu,A}\Phi^\dagger\Phi\, ,
\end{equation}
generating  the effective Lagrangian of Eq. ~\ref{eq:eftbsm}  and discussed in Sec. \ref{sec:lowenergy}.
The Higgs gluon effective interactions can be further altered at dimension-$8$ by the inclusion of the 
operators,\cite{Dawson:2015gka,Dawson:2014ora,Neill:2009tn,Harlander:2013oja,Buchmuller:1985jz,Grazzini:2016paz},
\begin{eqnarray}
O_2&=&  D_\sigma G^A_{\mu\nu}D^\sigma G^{A,\mu\nu}h \nonumber
\\
O_3&=&f_{ABC}G_\nu^{A,\mu} G_\sigma^{B,\nu}G_\mu^{C,\sigma} h \nonumber
\\
O_4&=&g_s^2h \Sigma_{i,j=1}^{n_{lf}} {\overline \psi}_i\gamma_\mu T^A \psi_i \,
 {\overline \psi}_j\gamma^\mu T^A \psi_j \nonumber 
\\
O_5&=& g_s h\Sigma_{i=1}^{n_{lf}} 
G_{\mu\nu}^A D^\mu\, {\overline \psi}_i\gamma^\nu T^A\psi_i\, .
\label{eq:op5}
\end{eqnarray}
The operator, $O_3$, of Eq. \ref{eq:op5} not only affects Higgs interactions, but also changes the kinematics of dijet production\cite{Dixon:1993xd}.  The Higgs
$p_T$ spectrum discussed in Sect. \ref{sec:higgspt}  can be significantly affected by the presence of the dimension-$8$ operators\cite{Dawson:2015gka,Dawson:2014ora,Harlander:2013oja,Grazzini:2016paz}.  In Fig. \ref{fig:dim8}\footnote{
$\kappa_t$ and $\kappa_g$ are defined in Eq. \ref{eq:kapdef} and $\kappa_5$ is the
scaling relative to the contribution of a $500~\gev$ scalar as discussed in Ref. \cite{Dawson:2015gka}.}, we show the effects on the Higgs $p_T$ spectrum for 
the lowest order rate for $gg\rightarrow g h$ with a cut implemented on the jet energy of $p_{T~cut}$.  For
$p_T\gsim 300~GeV$, the effects of the higher dimension operators can be numerically relevant. 
 This plot illustrates an important point about the EFT expansion.  For $2\rightarrow 2$ processes,
there are contributions of ${\cal{O}}\biggl({p_T^2\over\Lambda^2}\biggr)$, so care needs to be taken to stay in the region of validity of the 
expansion\footnote{This failure of the EFT also occurs in the $m_t\rightarrow\infty$ limit of the $gg\rightarrow hh$ process discussed in Sec.
\ref{sec:ddhh}.}  .

We turn now to a discussion of the effects of dimension-$6$ operators in the electroweak sector.  
As an example, we consider the SILH basis relevant for gauge-Higgs interactions\cite{Giudice:2007fh},
\begin{eqnarray}
L_{SILH}&=&{c_H\over 2\Lambda^2}\biggl(\partial ^\mu\mid\Phi\mid^2\biggr)^2 +{c_T\over 2 \Lambda^2}
\biggl(\Phi^\dagger\overleftrightarrow D^\mu\Phi\biggr)^2+\biggl({c_fy_f\over \Lambda^2}\mid\Phi\mid^2{\overline f}_L\Phi f_R+hc\biggr)-{c_6\lambda\over \Lambda^2}\mid\phi\mid^6\nonumber \\
\nonumber \\
&&+{ig c_W\over 2\Lambda^2}\biggr(\Phi^\dagger\sigma^I\overleftrightarrow D^\mu \Phi\biggr)\biggr(D^\nu W_{\mu\nu}^I\biggr)
+{ig^\prime c_B\over 2\Lambda^2}\biggr(\Phi^\dagger\overleftrightarrow D^\mu \Phi\biggr)\biggr(D^\nu B _{\mu\nu}\biggr)\nonumber\\
&&+{igc_{HW}\over 16\pi^2\Lambda^2}
\biggl(D^\mu\Phi\biggr)^\dagger \sigma^i \biggl(D^\nu \Phi\biggr)W_{\mu\nu}^i
+{ig^\prime c_{HB}\over 16\pi^2\Lambda^2}\biggl(D^\mu\Phi\biggr)^\dagger  \biggl(D^\nu \Phi\biggr)B_{\mu\nu}
\nonumber \\
&&+{c_\gamma g^{\prime ~2}g^2\over 16\pi^2\Lambda^2}
\mid \Phi\mid ^2 B_{\mu\nu}B^{\mu\nu} 
+{c_g g_s^2\over 16\pi^2\Lambda^2}\mid \Phi\mid ^2 G_{\mu\nu}^A G^{A,\mu\nu}
\, .
\label{eq:silhlag}
\end{eqnarray}
Note that the normalization of the operators is arbitrary and merely reflects a prejudice about the origins
of the new physics,    $I=1,2,3$ are $SU(2)$ indices and we have not written terms involving only fermions, or terms
that do not contain a Higgs field.  Many of the operators of Eq. \ref{eq:silhlag} introduce momentum dependence into
the Higgs couplings to SM fermions and so the kinematic distributions of the Higgs will be affected.

  We briefly discuss some of the phenomenological effects of Eq. \ref{eq:silhlag}. Three of the coefficients are strongly 
limited by precision electroweak measurements as parameterized by the oblique parameters,
\begin{eqnarray}
\Delta T&=& {v^2\over \Lambda^2} c_T\nonumber \\
\Delta S&=& {M_W^2\over \Lambda^2}(c_W+c_B)\, .
\end{eqnarray}
Using the fit from Ref. \cite{deBlas:2016ojx},  
$\mid c_T\mid\lsim {\cal{O}}(.03)$ and $\mid c_W+c_W\mid\lsim {\cal{O}}(.1)$  for $\Lambda\sim 1~TeV$.

The coefficient $c_H$ modifies the Higgs  boson kinetic energy.  The physical Higgs field needs to be rescaled,
\begin{equation}
\hsm\rightarrow \hsm \biggl( 1-{c_H v^2 \over 2 \Lambda^2}\biggr)\, ,
\label{eq:hshift}
\end{equation} 
in order to have canonically normalized kinetic energy.  This  shift introduces a dependence on $c_H$ into  all of the Higgs decay
widths.  The tree level Higgs decay widths to ${\cal O}({v^2\over \Lambda^2})$  in the SILH formalism are,
\begin{eqnarray}
{\Gamma(\hsm\rightarrow W W*)\over \Gamma(\hsm\rightarrow WW^*)\mid_{SM}}&=&1-{v^2\over \Lambda^2}
\biggl[
c_H-g^2\biggl(c_W+{c_{HW}\over 16\pi^2}\biggr)\biggr]\nonumber \\
{\Gamma(\hsm\rightarrow ZZ*)\over \Gamma(\hsm\rightarrow ZZ^*)\mid_{SM}}&=&
1-{v^2\over \Lambda^2}
\biggl[
c_H-g^2\biggl(c_W+\tan^2\theta_Wc_B+{c_{HW}+\tan^2\theta_2 c_{HB}\over 16\pi^2}\biggr)\biggr]
\nonumber \\
{\Gamma(\hsm\rightarrow f {\overline f})\over \Gamma(\hsm\rightarrow f {\overline f} )\mid_{SM}}&=& 1-{v^2\over \Lambda^2}(c_H+2c_f)
\, . \nonumber \\
\end{eqnarray}
The loop processes, $gg\rightarrow\hsm$ and $\hsm\rightarrow\gamma\gamma$, also receive corrections from the EFT 
operators.  
The expressions for Higgs decays in the SILH Lagrangian  have been implemented into an update of the HDECAY program, EDECAY\cite{Contino:2014aaa}.
In the Warsaw basis, they can be obtained using the SMEFTsim  code\cite{Brivio:2017btx}. 
Fits to the EFT coefficients can be performed using total Higgs rates  (as is done  in the $\kappa$ formalism)
or including information from distributions\cite{DiVita:2017eyz,Butter:2016cvz}.  The kinematic information provides a significant
improvement to the fits from using only the total rates.  

Some of the operators of Eq.~ \ref{eq:silhlag} not  only affect Higgs production, but they also change
the $WWZ$ and $WW\gamma$ vertices.  
 Assuming CP conservation, the most general Lorentz invariant $3-$gauge
boson couplings can be written
as~\cite{Gaemers:1978hg,Hagiwara:1986vm}
\begin{eqnarray}
 L_{V}&=&
-ig_{WWV}\biggl[g_1^V\left(W^+_{\mu\nu}W^{-\mu}V^\nu-W_{\mu\nu}^-W^{+\mu}V^\nu\right)+\kappa^VW^+_\mu
            W^-_\nu V^{\mu\nu} 
            \nonumber \\ && 
            +\frac{\lambda^V}{M^2_W}W^+_{\rho\mu}{W^{-\mu}}_\nu V^{\nu\rho}\biggr]\, ,
\label{eq:lagdefwwv}
\end{eqnarray}  
where $V=(Z,\gamma)$, $g_{WW\gamma}=e$, and $g_{WWZ}=g c_W$.  In the SM, $g_1^Z=g_1^\gamma=\kappa^Z
=\kappa^\gamma=1$, $\lambda^Z=\lambda^\gamma=0$ and $SU(2)$ gauge invariance implies,
\begin{eqnarray}
\lambda^\gamma&=&\lambda^Z\nonumber \\
g_1^Z&=&\kappa^Z+{s_W^2\over c_W^2}(\kappa^\gamma-1)\, .
\end{eqnarray} 

The fields in Eq.~\ref{eq:lagdefwwv} are the
canonically normalized mass eigenstate fields.  These coefficients can be mapped to EFT coefficients
in a straightforward manner and a subset of the dimension-$6$ coefficients contribute both to gauge boson pair
production and Higgs production\cite{Falkowski:2016cxu,Butter:2016cvz,Berthier:2016tkq}.  

\begin{figure}
\begin{centering}
\includegraphics[width=0.35\textwidth]{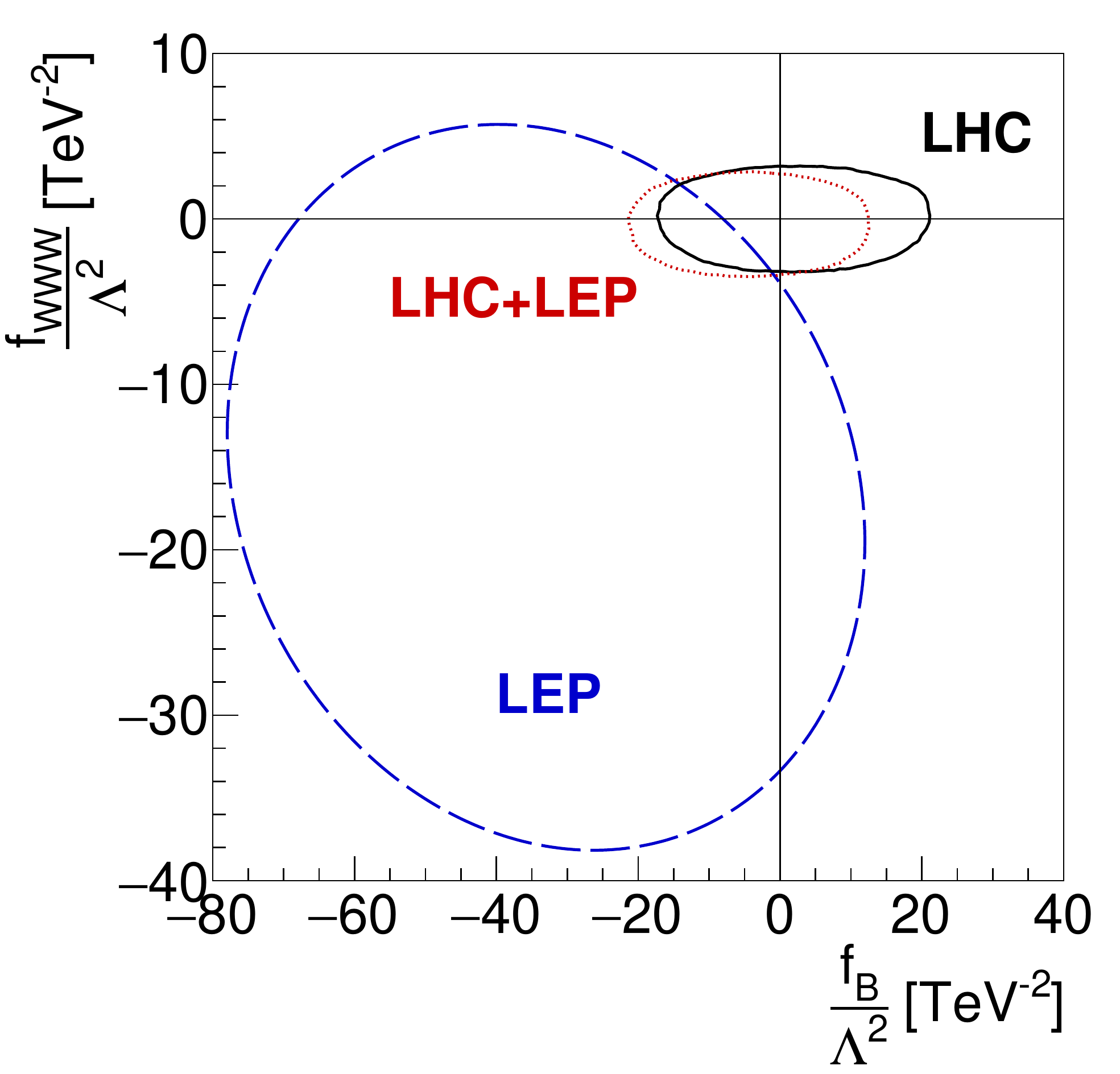}
\includegraphics[width=0.35\textwidth]{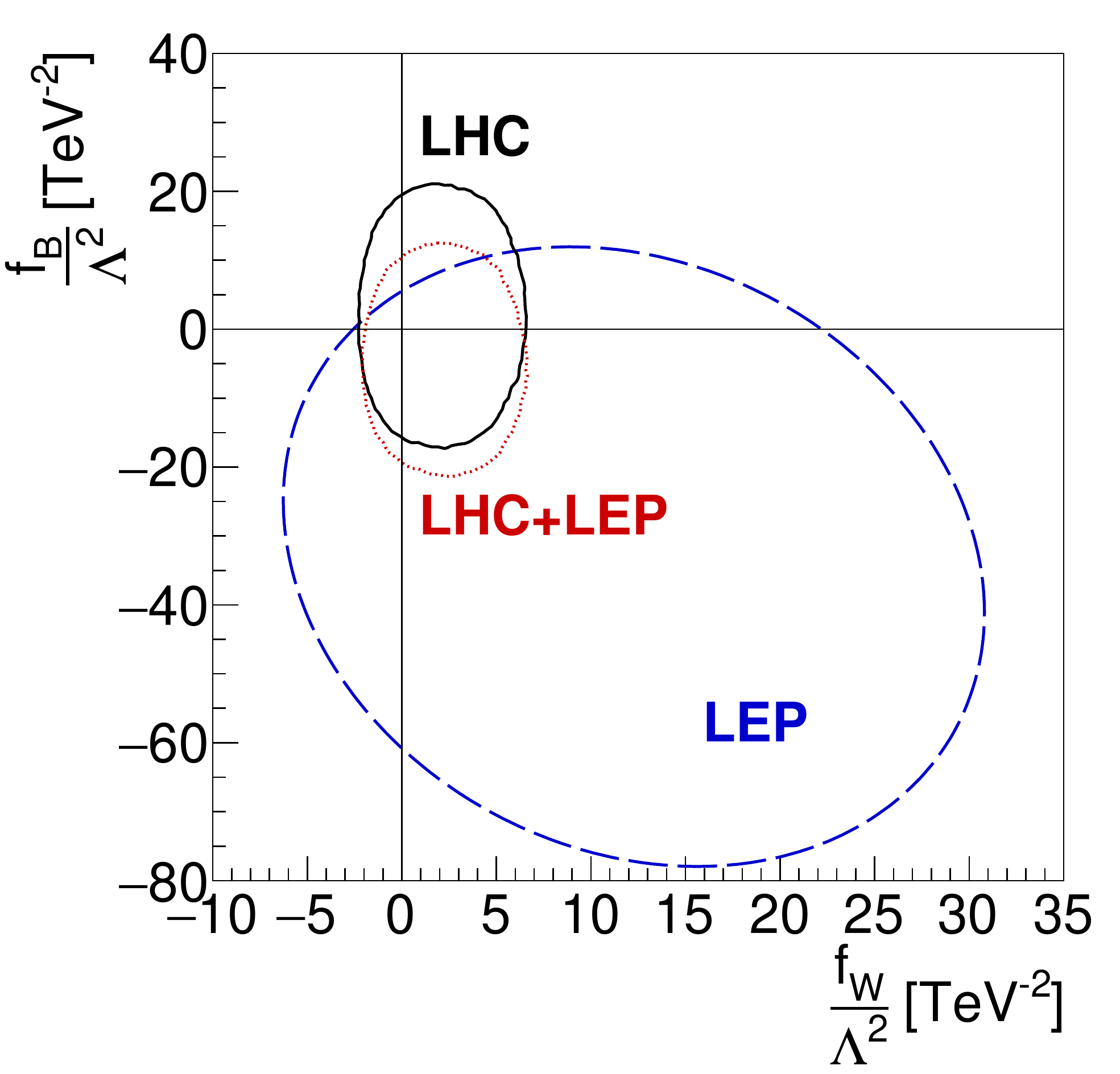}
\includegraphics[width=0.35\textwidth]{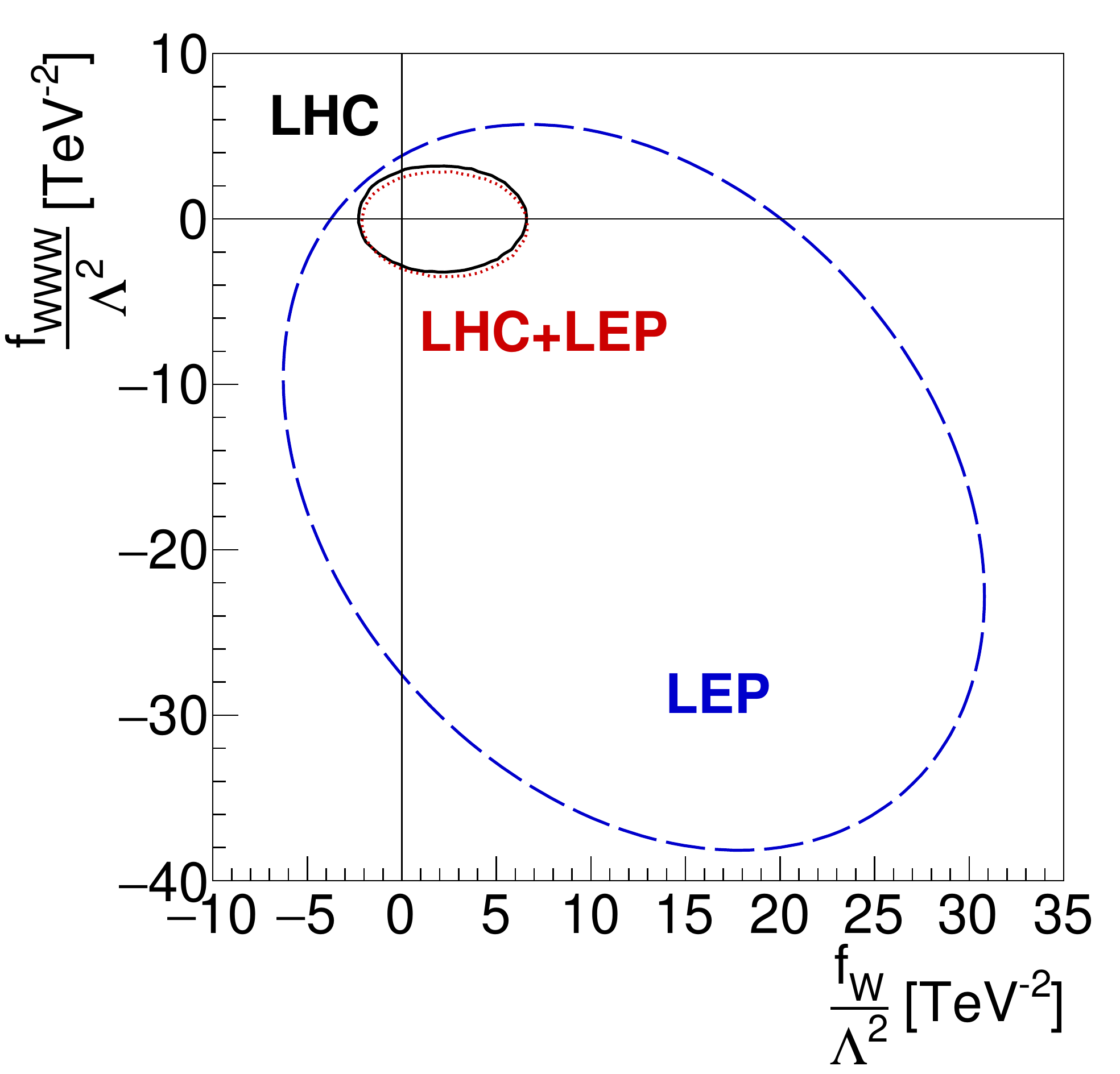}
\par\end{centering}
\caption{\label{fg:tilfig} Fits to LEP data, LHC data and the combination of both\cite{Butter:2016cvz}.}  \end{figure}

A consistent fit must include not only Higgs data, but also fits to anomalous gauge couplings. 
 In Fig. \ref{fg:tilfig}, we show  fits to $3$  of the EFT couplings that contribute to both $W^+W^-$ and Higgs production,
 including only LEP data on $W^+W^-$ pair production, 
 only LHC data on   $W^+W^-$  and Higgs production, and the resulting fit combining the two.
 The LHC results have now surpassed the LEP results in terms of precision\cite{Butter:2016cvz}.  
 This figure includes the full  set of dimension-$6$ squared contributions. In terms of the parameters of Eq. ~\ref{eq:lagdefwwv},
 \begin{eqnarray} 
 f_W&=&{2\Lambda^2\over M_Z^2}(g_1^Z-1)\nonumber \\
 f_B&=& {2\Lambda^2\over M_W^2}\biggl[(\kappa_\gamma-1)-c_W^2(g_1^Z-1)\biggr]\nonumber \\
 f_{WWW}&=&{4\Lambda^2\over 3 g^2M_W^2}\lambda^\gamma\, . 
 \end{eqnarray}

Global fits to EFT coefficients in the SILH basis can be found in Ref. \cite{Falkowski:2014tna,DiVita:2017eyz} and 
in the Warsaw basis in Ref. \cite{Berthier:2016tkq}. 
Many of the EFT coefficients are only weakly constrained. 
These results illustrate, however,  that fits performed to only a single operator typically significantly overestimate the sensitivity.
 As of this writing,
the experimental collaborations have not performed such global EFT  fits.

Finally, it is interesting to ask what the target precision is for measuring EFT coefficients.  In any given UV complete model, these
coefficients can be calculated, and the scale $\Lambda$ will be of the same order of magnitude as the mass of the new particles. 
This suggests that as direct searches for new particles get more and more precise, it is necessary to measure the EFT  coefficients
more and more precisely.   In a specific UV complete model, not all coefficients will be generated, and the pattern of non-zero coefficients
will be a guide to the underlying model.  The EFT coefficients for numerous models with heavy 
scalars\cite{Dawson:2017vgm,deBlas:2017xtg,Henning:2014wua,Brehmer:2015rna,Gorbahn:2015gxa} and heavy vector-like 
quarks\cite{Chen:2017hak,AguilarSaavedra:2009mx} are
known and suggest that measurements of ${\cal {O}}(2-3\%)$ 
will be necessary to probe models with new particles at the $2-3~TeV$ scale.

\section{Outlook}
\label{sec:conc}

The discovery of a SM-like Higgs boson opened a new era in particle physics.  We do not yet know if we have
discovered ${\it{\bf{a}}}$ Higgs boson or ${\it{\bf{the}}}$ Higgs boson.  To make this determination, the measurements of Higgs interactions need
to be improved to the few $\%$ level and the Higgs self-interactions need to be observed.  These  precision measurements 
will begin during the high luminosity run of the LHC, but will require a future 
high energy hadron collider or $e^+e^-$ collider to reach the desired accuracy.  A limiting factor will be the precision of
theoretical predictions--predictions accurate at the few $\%$ level will require a dedicated effort in the coming years and 
improvement of our knowledge of PDFs. 
I have not discussed models with extra scalar particles other than the singlet model.  One of the most important efforts 
of the Higgs program in the next few years will be the search for additional Higgs-like particles.  The observation of another scalar
would be the cleanest possible indication of new BSM physics in the scalar sector.
  \section*{Acknowledgements}
  I thank the TASI 2016 students for making the school an exciting place to discuss my favorite topic-- the Higgs boson--
  and the organizers, Rouven Essig and Ian Low, for encouraging me to finally write my lecture notes.  I am grateful to R.K. Ellis
  and T. Degrand for pointing out typos.  My research
  is supported  by the U.S. Department of Energy under grant DE-SC0012704. 
\newpage
\bibliographystyle{utphys}
\bibliography{tasi}
\end{document}